\documentclass{jfm}

\usepackage{graphicx,graphics,color}
\usepackage{natbib}
\usepackage{array} 
\usepackage{booktabs}
\usepackage{tabularx}
\usepackage{amsmath}
\usepackage{relsize}
\usepackage{etoolbox} 
\usepackage{scalerel}
\usepackage{pgfplots}
\usepackage{amssymb}
\usepackage{amsfonts}
\usepackage{upmath}

\usepackage{breqn}

\ifCUPmtlplainloaded \else
  \checkfont{eurm10}
  \iffontfound
    \IfFileExists{upmath.sty}
      {\typeout{^^JFound AMS Euler Roman fonts on the system,
                   using the 'upmath' package.^^J}%
       \usepackage{upmath}}
      {\typeout{^^JFound AMS Euler Roman fonts on the system, but you
                   dont seem to have the}%
       \typeout{'upmath' package installed. JFM.cls can take advantage
                 of these fonts,^^Jif you use 'upmath' package.^^J}%
       \providecommand\upi{\pi}%
      }
  \else
    \providecommand\upi{\pi}%
  \fi
\fi

\ifCUPmtlplainloaded \else
  \checkfont{msam10}
  \iffontfound
    \IfFileExists{amssymb.sty}
      {\typeout{^^JFound AMS Symbol fonts on the system, using the
                'amssymb' package.^^J}%
       \usepackage{amssymb}%

      }{}
  \fi
\fi

\ifCUPmtlplainloaded \else
  \IfFileExists{amsbsy.sty}
    {\typeout{^^JFound the 'amsbsy' package on the system, using it.^^J}%
     \usepackage{amsbsy}}
    {\providecommand\boldsymbol[1]{\mbox{\boldmath $##1$}}}
\fi

\newsavebox{\astrutbox}
\sbox{\astrutbox}{\rule[-5pt]{0pt}{20pt}}

\title[Rayleigh-B\'enard Stability]{Rayleigh-B\'enard stability and the validity of quasi-Boussinesq or quasi-anelastic liquid approximations }

\author[T. Alboussi\`ere and Y. Ricard]%
{Thierry Alboussi\`ere
  \thanks{Email address for correspondence: thierry.alboussiere@ens-lyon.fr}
and Yanick Ricard}

\affiliation{Univ Lyon, Universit\'e Lyon 1, Ens de Lyon, CNRS, UMR 5276 LGL-TPE, F-69622, Villeurbanne, France}

\pubyear{}
\volume{}
\pagerange{}

\date{\today}

\begin{document}

\maketitle

\begin{abstract}
The linear stability threshold of the Rayleigh-B\'enard configuration is analyzed with compressible effects taken into account. It is assumed that the fluid under investigation obeys a Newtonian rheology and Fourier's law of thermal transport with constant, uniform (dynamic) viscosity and thermal conductivity in a uniform gravity field. Top and bottom boundaries are maintained at different constant temperatures and we consider here mechanical boundary conditions of zero tangential stress and impermeable walls. Under these conditions, and with the Boussinesq approximation, \citet{rayleigh} first obtained analytically the critical value $27 \upi ^4 / 4$ for a dimensionless parameter, now known as the Rayleigh number, at the onset of convection. This manuscript describes the changes of the critical Rayleigh number due to the compressibility of the fluid, measured by the dimensionless dissipation parameter $\mathcal{D}$ and due to a finite temperature difference between the hot and cold boundaries, measured by a dimensionless temperature gradient $a$. Different equations of state are examined: ideal gas equation, Murnaghan's model (often used to describe the interiors of solid but convective planets) and a generic equation of state with adjustable parameters, which can represent any possible equation of state.  
In the perspective to assess approximations often made in convective models, we also consider two variations of this stability analysis. In a so-called quasi-Boussinesq model, we consider that density perturbations are solely due to temperature perturbations. In a so-called quasi-anelastic liquid  approximation model (quasi-ALA), we consider that entropy perturbations are solely due to temperature perturbations. In addition to the numerical Chebyshev-based stability analysis, an analytical approximation is obtained when temperature fluctuations are written as a combination of only two modes. This analytical expression allows us to show that the superadiabatic critical Rayleigh numbers depart quadratically in $a$ and $\mathcal{D}$ from $27 \upi ^4 / 4$. That quadratic departure is shown to involve the expansion of density up to the degree three in terms of pressure and temperature. 
\end{abstract}


\section{Introduction}
\label{intro}

Thermal, or natural, convection results from a complex interaction between dynamical principles and thermodynamics of a fluid. This complexity was an obstacle to the analysis of even the most idealized configurations. A great simplification, assumed to be valid when compressibility effects can be ignored, was put forward by \citet{oberbeck}, then \citet{boussinesq}, at the expense of thermodynamic coherence. Using Boussinesq's equations, \citet{rayleigh} was able to solve the problem of the stability of a fluid layer heated from below, and obtained a critical value, now expressed as a dimensionless number named after him, the Rayleigh number. For boundary conditions of no shear stress with imposed temperatures, the critical Rayleigh number is $27 \upi ^4 / 4$. Thanks to the Oberbeck-Boussinesq model, this stability analysis can be done analytically, with a simple eigenvector spatial structure for temperature perturbations of the form of plane waves, with lateral wavenumber equal to $\upi / \sqrt{2}$ and a cosine dependence along the vertical direction. 

Meanwhile, \citet{schwarzschild} proved that a sufficient condition for stability in a compressible fluid was obtained when the temperature gradient does not exceed the adiabatic gradient, which can equivalently be stated as the non-decrease of entropy with height. Then \citet{jeffreys} showed that, in the limit of small compressibility effects, the critical threshold for convection instability was identical to the original critical Rayleigh number, as long as the temperature difference is replaced by the excess temperature difference above the adiabatic temperature difference (usually called the super-adiabatic temperature difference). 
 
Since these pioneering works, stability of compressible convection has continued to be an active subject of research. \citet{spielgel} has been studying the convective instability of a layer of ideal gas. A single small parameter was identified, equivalent to the dissipation number. It was found that the critical superadiabatic Rayleigh number does not depend on that parameter at order $1$ (when evaluated in the middle of the layer), so that the first deviation is of order $2$. \citet{gs1970} and, more recently, \citet{bormann} argue essentially that \citet{jeffreys} is correct and the superadiabatic critical Rayleigh number has small deviations from its Boussinesq value $27 \upi ^4 / 4$. 
Another series of papers have attempted to evaluate the change in critical superadiabatic Rayleigh number, when compressibility effects are negligible but when the temperature difference is large \citep{busse1967,pc1987,flp1992}. They show that the deviation from the Boussinesq value scales as the square of the  dimensionless temperature difference between the bottom and top boundaries $(T_{bottom}-T_{top})/T_0$ (where $T_0$ is the average temperature $(T_{bottom}+T_{top})/2$). 

A category of research works are related to the formal derivation of the Boussinesq equations from the general equations. \citet{sv1960} use one small parameter $\Delta \rho / \rho$, \citet{mihaljan1962} uses two small parameters, $\alpha T$ and the ratio between the dissipation number and the dimensionless temperature difference, while \citet{malkus1964} considers the vanishing limit of the dissipation parameter and of the dimensionless temperature difference: we shall here choose the same small parameters as Malkus. Another type of research is highly relevant to the present study, namely the derivation of intermediate models between the exact and Boussinesq models. A number of `sound-proof' models have been proposed whose first motivation was to remove sound waves from the set of solutions to the convection equations. Otherwise, one would like the anelastic models to be able to model accurately convective phenomena. The anelastic model was derived first for atmospheric studies by \citet{op1961}, then for the Earth's core by \citet{br1995} and for stellar convection by \citet{lf1999}. The anelastic model is basically a linear expansion of the general equations around an isentropic profile. This is in complete correspondence with \citet{jeffreys}, as the reference takes into account the adiabatic profile already and only superadiabatic quantities are computed. The anelastic liquid approximation (ALA) was proposed later by \citet{ajs05}, where the contribution of pressure fluctuations are neglected compared to that of entropy fluctuations. In the present work, we shall test one aspect only of these models, their ability to provide a good approximation of the critical superadiabatic Rayleigh number. It should be noted however that we will have to make changes to these approximation models in order to study their stability: essentially, instead of an adiabatic base profile, we will need to take a conductive base profile. The adiabatic profile is indeed unconditionally stable. Other sound-proof models \citep{durran1989,lipps1990}, used preferentially in stratified cases, will not be considered in this paper.

The structure of the present work is the following. Section \ref{configuration} will be devoted to the geometry, notations, governing equations and boundary conditions. Dimensional scales and dimensionless equations will be presented in section \ref{nondim}, base profile solutions in section \ref{initial}. In section \ref{eigenvalue}, we present the linear stability analysis and the determination of eigenvalues using the tau-Chebyshev expansion. An approximate stability analysis is performed in section \ref{estimate} using two modes only for temperature disturbances (with vertical dependence in $\cos ( \upi z) $ and $\sin (2 \upi z)$, where $-1/2 < z < 1/2$ is the range of the dimensionless vertical coordinate $z$), allowing us to obtain analytical equations for the critical superadiabatic Rayleigh number up to degree $2$ in the dissipation number and in the dimensionless temperature difference. In section \ref{themodels} we introduce the approximation models which will be tested compared to the exact stability analysis: the quasi-Boussinesq and quasi-ALA (quasi-Anelastic Liquid Approximation): they have the same features as the Boussinesq and ALA models, but the base profile is the conduction profile with compressibility taken into account (for the determination of the profile of density, pressure, entropy...). In section \ref{eos}, we consider different equations of state (ideal gas, Murnaghan's equation for condensed matter, and a generic equation of state) and solve the linear stability analysis. We compare the numerical Chebyshev results to the analytical expressions obtained from the two-modes analysis. Those expressions allow us to predict, for each equation of state, the accuracy achieved by the approximation models considered, as far as the critical superadiabatic Rayleigh number is concerned (see section \ref{discussion}). In the same section, we discuss the validity of the approximation models in geophysical objects. 
In section \ref{conclusion}, the current state of our knowledge is summarized.

\section{Rayleigh-B\'enard configuration and governing equations}
\label{configuration}
 
\begin{figure}
\begin{center}
\includegraphics{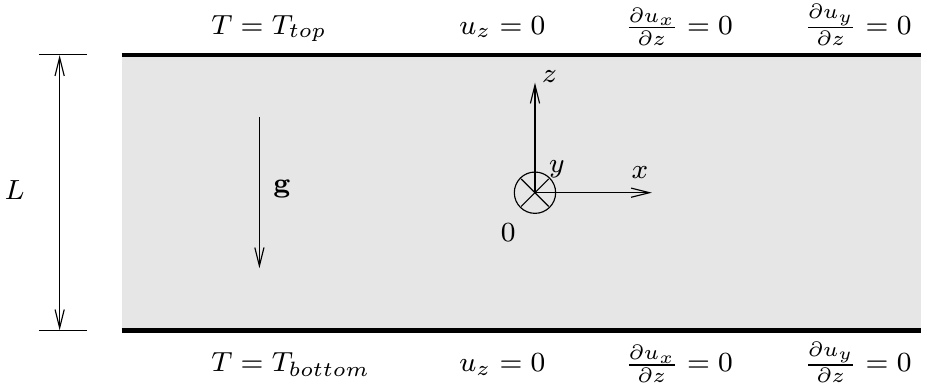}
\caption{Rayleigh-B\'enard configuration, with imposed temperatures and tangential stress-free boundary conditions}
\label{config}
\end{center}
\end{figure}

A horizontal fluid layer of thickness $L$, in a uniform gravity field ${\bf g } = - g {\bf e}_z$, is heated from below: the lower and upper boundaries are maintained at $T_{bottom}$ and $T_{top}$ respectively. The fluid is a Newtonian fluid and obeys the Fourier law of heat conduction. Its dynamic viscosity $\mu$ and thermal conductivity $k$ are taken to be uniform, 
independent of pressure and temperature, for simplicity. The mechanical boundary conditions are stress-free, impermeable, on the upper and lower planar boundaries.  The governing equations for convection consist in the equations of continuity, momentum conservation (Navier-Stokes with no bulk viscosity), entropy balance and an equation of state:
\begin{eqnarray}
\frac{\partial \rho }{\partial t} + \nabla \cdot \left( \rho {\bf u} \right) &=& 0, \label{continuity} \\
\rho \frac{\mathrm{D} {\bf u} }{\mathrm{D} t} &=& - {\bf \nabla } p + \rho {\bf g} + \mu \left( {\bf \nabla}^2 {\bf u} + \frac{1}{3} {\bf \nabla } \left( {\bf \nabla} \cdot {\bf u}  \right) \right) \label{NS} \\
\rho c_p \frac{\mathrm{D} T }{\mathrm{D} t } - \alpha T \frac{{\mathrm D} p }{{\mathrm D} t } &=& \dot{\epsilon}  : \tau + k {\bf \nabla}^2  T, \label{entropy} \\
\rho &=& \rho (p, T), \label{EoS}
\end{eqnarray}
where $t, \rho, {\bf u}, p, T, c_p, \alpha$ are the time, density, velocity vector, pressure, temperature, heat capacity at constant pressure and expansion coefficient respectively. 
A vertical coordinate axis $z$ is defined with its origin on the mid-plane of the layer (see Fig.~\ref{config}). Horizontal coordinates $x$ and $y$ form an orthogonal unit reference frame.
The boundary conditions associated with the governing equations are the following:
\begin{eqnarray}
u_z \left( z = \pm \frac{L}{2} \right) = 0, & \label{no-penetrative} \\
T \left( z = \frac{L}{2} \right) = T_{top}, &\hspace*{1 cm} & T \left( z = - \frac{L}{2} \right) =T_{bottom}, \label{fixedtemperature} \\
\frac{\partial u_x }{ \partial z} \left( z = \pm \frac{L}{2} \right) = 0, & \label{no-stress_x} \\
\frac{ \partial u_y }{ \partial z } \left( z = \pm \frac{L}{2} \right) = 0, & \label{no-stress_y} 
\end{eqnarray}
The initial condition considered will be a quiescent state and will be described in section \ref{initial}. 
The mass of fluid per horizontal unit surface area is set when the density of the base profile $\rho _0$ is specified at $z=0$. 

\section{Dimensionless formulation}
\label{nondim}

The dimensional quantities will be made dimensionless with the help of the quiescent base solution. Density, thermal expansion coefficient and specific heat capacity at constant pressure of the base solution at $z=0$, $\rho _0$, $\alpha _0$ and $c_{p0}$ will be the scales for density, temperature, thermal expansion coefficient and specific heat capacity at constant pressure, and $T_0 = \left( T_{top} + T_{bottom} \right) / 2$ will be the scale for temperature. Pressure $p$, velocity ${\bf u}$, time $t$, and spatial coordinates ${\bf x}$ are made dimensionless using $\rho _0 g L$, $k /(\rho _0 c_{p0} L)$, $L^2 \rho _0 c_{p0} / k$ and $L$ respectively. 
The governing equations take the following dimensionless form:
\begin{eqnarray}
\frac{\partial \rho}{\partial t} + {\bf \nabla} \cdot \left( \rho {\bf u} \right) &=& 0, \label{continuitya} \\
Pr^{-1} \rho \frac{\mathrm{D} {\bf u}}{\mathrm{D} t} &=& - Ra_{th} {\bf \nabla } p - Ra_{th}  \rho  {\bf e}_z +   {\bf \nabla }^2 {\bf u} + \frac{1}{3} {\bf \nabla }  \left( {\bf \nabla} \cdot  {\bf u} \right)  , \label{NSa} \\
\rho c_p \frac{\mathrm{D} T }{\mathrm{D} t } - {\cal{D}} \alpha T \frac{\mathrm{D} p }{\mathrm{D} t } &=& \dot{\epsilon}  : \tau + {\bf \nabla} ^2  T , \label{entropya} 
\end{eqnarray}
where $Pr = \mu c_{p0} / k$ is the Prandtl number, $Ra_{th} = \rho _0^2 g c_{p0} L^3 / (\mu k)$ is called here the thermodynamic Rayleigh number (the classical Rayleigh number in the Boussinesq approximation
is $Ra_{th}\alpha_0(T_{bottom}-T_{top})$) and ${\cal{D}} = \alpha _0 g L / c_{p0} $ is the dissipation number. The thermal boundary conditions necessitate an additional dimensionless parameter and we choose the ratio of 
the temperature difference to the average temperature $a = 2 \left( T_{bottom} - T_{top} \right) / \left( T_{bottom} + T_{top} \right) $ so that the boundary conditions (\ref{fixedtemperature}) become: 
\begin{equation}
T \left( z=\frac{1}{2} \right) = 1 - \frac{a}{2}, \hspace*{1 cm}  T \left( z=-\frac{1}{2} \right) = 1 + \frac{a}{2} . \label{fixedtemperaturea} 
\end{equation}
From our choice of dimensional scales, another dimensionless number is obtained from the product $\alpha _0 T_0$. 
The equations of state will also be made dimensionless when they are considered in section \ref{eos}. Depending on the equation of state, dimensionless parameters other than the four numbers listed above may be necessary or not. We have not specified how the viscous dissipation term $\dot{\epsilon}  : \tau $ was made dimensionless because this term is quadratic in terms of velocity disturbances, hence will play no role in the linear stability analysis. 

\section{Motionless base solution}
\label{initial}

The base solution is a pure conduction, hydrostatic state. The dynamic and thermal equations (\ref{NSa}) and (\ref{entropya}) lead to the following equations for $p_b$, $\rho _b$ and $T_b$, the base pressure, density and temperature solutions which are functions of $z$ only:
\begin{eqnarray}
\frac{\mathrm{d} p_b}{\mathrm{d} z} &=& - \rho _b \label{hydrostatic} \\
 \frac{\mathrm{d}^2 T_b }{\mathrm{d} z^2}    &=& 0. \label{conduction}
\end{eqnarray}
The boundary condition (\ref{fixedtemperaturea}) for temperature needs to be satisfied. The conduction solution can be expressed as 
\begin{equation}
T_b=1 - a z. \label{basetemperature}
\end{equation}
The opposite of the temperature gradient is $a$ and the bottom to top temperture ratio $T_{bottom} / T_{top}$ is $r  = (2+a) / (2-a)$.

\section{Eigenvalue equations for infinitesimal disturbances}
\label{eigenvalue}

Infinitesimal disturbances, denoted by primes, are added to the base solution and the temporal linear stability is analyzed. 
The governing equations are linearized around the base solution and the resulting problem can be written:
\begin{eqnarray}
\frac{\partial {\rho '}}{\partial t} &=& - {\bf \nabla} \cdot \left( \rho _b {\bf u}' \right), \label{cont_eig} \\
Pr^{-1} \rho _b \frac{\partial {\bf u}'}{\partial t} &=& - Ra_{th} {\bf \nabla } p' - Ra_{th}  { \rho ' }  {\bf e}_z +   {\bf \nabla }^2 {\bf u}' + \frac{1}{3} {\bf \nabla } \left( {\bf \nabla} \cdot  {\bf u}' \right)  , \label{NS_eig} \\
\rho _b c_{pb} \frac{\partial T'}{\partial t} - {\cal{D}} \alpha _b T_b \frac{\partial p' }{\partial t} &=& - \rho _b c_{pb} u'_z \frac{\mathrm{d} T_b }{\mathrm{d} z} + {\cal{D}} \alpha _b T_b u'_z \frac{\mathrm{d} p_b}{\mathrm{d} z} + {\bf \nabla} ^2 T'  , \hspace*{1 cm} \label{entropy_eig} 
\end{eqnarray}
where $c_{pb}$ and $\alpha_b$ are the heat capacity and thermal expansivity along the base profile. 
The problem does not explicitly depend on time, $t$, nor on the horizontal directions $x$ and $y$. Thus general solutions can be searched in the linear space of plane waves:
\begin{equation}
T' = {\widetilde{T}}(z) \exp \left( \sigma t + \mathrm{i} k_x x + \mathrm{i} k_y y \right), \label{vectpropre}
\end{equation} 
where $\sigma$ is the growth rate of the disturbance, $k_x$ and $k_y$ its horizontal wavenumbers. As rotation along a vertical axis leaves the problem unchanged, we can restrict the analysis to $k_y=0$ without any loss of generality. 
Equations (\ref{cont_eig}) , (\ref{NS_eig}) and (\ref{entropy_eig})  are then changed into the following eigenvalue problem:
\begin{eqnarray}
\sigma \widetilde{\rho } &=& - \mathrm{i} k_x \rho _b \widetilde{u}_x - \rho _b \frac{\mathrm{d} \widetilde{u}_z }{\mathrm{d} z}  - \frac{\mathrm{d} \rho _b}{\mathrm{d} z} \widetilde{u}_z , \label{cont_eig2} \\
\sigma Pr^{-1} \rho _b \widetilde{ u }_x &=& - Ra_{th} \mathrm{i} k_x \widetilde{p} - \frac{4}{3} k_x^2 \widetilde{u}_x + \frac{\mathrm{i} k_x }{3} \frac{\mathrm{d} \widetilde{u}_z }{\mathrm{d} z} + \frac{\mathrm{d}^2 \widetilde{u}_x }{\mathrm{d} z^2}   , \label{NSx_eig2} \\
\sigma Pr^{-1} \rho _b \widetilde{ u}_z &=& - Ra_{th} \frac{\mathrm{d} \widetilde{p}}{\mathrm{d} z} - Ra_{th}  \widetilde{ \rho }  - k_x^2 \widetilde{u}_z + \frac{\mathrm{i} k_x }{3} \frac{\mathrm{d} \widetilde{u}_x }{\mathrm{d} z} + \frac{4}{3} \frac{\mathrm{d}^2 \widetilde{u}_z }{\mathrm{d} z^2} , \hspace*{1 cm} \label{NSz_eig2} \\
\sigma \rho _b c_{pb} \widetilde{T} -\sigma {\cal{D}} \alpha _b T_b \widetilde{p} &=& - \rho _b c_{pb} \widetilde{u}_z \frac{\mathrm{d} T_b }{\mathrm{d} z} + {\cal{D}} \alpha _b T_b \widetilde{u}_z \frac{\mathrm{d} p_b}{\mathrm{d} z} -k_x^2 \widetilde{T} + \frac{\mathrm{d}^2 \widetilde{T} }{\mathrm{d} z^2} . \label{entropy_eig2} 
\end{eqnarray}
Finally, the density disturbance $\widetilde{ \rho }$ is expanded linearly in terms of temperature $\widetilde{ T }$ and pressure $\widetilde{ p }$ disturbances in equation (\ref{NSz_eig2}) when a particular equation of state will be considered: 
\begin{equation}
\widetilde{ \rho } = \left. \frac{\partial \rho}{\partial T}\right| _p \widetilde{ T } + \left. \frac{\partial \rho}{\partial p} \right| _T \widetilde{ p }, \label{eos_rho}
\end{equation}

Our objective is to obtain the critical value of the thermodynamic Rayleigh number $Ra_{th}$ as a function of the other dimensionless numbers.  
We restrict our analysis to the critical threshold, $\Re (\sigma ) = 0$. The eigenvalue problem is not self-adjoint in general, unlike the classical Boussinesq problem, however the imaginary part of the critical eigenvalue is always found to be zero in our numerical calculations. The first instability takes the form of a stationary pattern, not a travelling wave.  
A consequence is that the Prandtl number is irrelevant in our study, since it appears only as the product $\sigma \, Pr^{-1}$ in the eigenvalue problem, in equations (\ref{NSx_eig2}) and (\ref{NSz_eig2}). 

The eigenvalue problem is solved and the critical Rayleigh number for neutral stability is obtained. The method is that of Chebyshev collocation expansion and we use the differentiation matrices provided by the DIFFMAT suite \citep{wr2000}. The computations are run in GNU Octave on a laptop.  
The critical Rayleigh number is expressed using the superadiabatic Rayleigh number:
\begin{equation}
Ra_{SA} = Ra_{th} \alpha _0 T_0 \left( \frac{T_{bottom}-T_{top}}{T_0} - \left[ T_a \left( -\frac{1}{2} \right) - T_a \left( \frac{1}{2} \right) \right] \right), \label{RaSA}
\end{equation}
where $T_a$ denotes the dimensionless adiabatic (isentropic) temperature profile anchored at $z=0$ to $1$ ($T_0$ in dimensional terms), {\it i.e.} the steady state hydrostatic profile solution of (\ref{entropya}) neglecting dissipation and diffusion, obeying
\begin{equation}
\frac{\mathrm{d} T_a}{\mathrm{d} z} = - \frac{\mathcal{D} \alpha _a}{c_{pa}} T_a, \label{adiabgrad}
\end{equation} 
where $\alpha _a$ and $c_{pa}$ are themselves computed along the adiabatic and hydrostatic profile with $T_a (z=0) = 1$.
The results of the numerical stability analysis will be presented in section \ref{eos}.

When an equation of state is specified, in order to carry out the linear analysis above, we first need to determine the $z$-profile of the following quantities associated with the base solution: $\rho _b (z)$, $\mathrm{d} \rho _b / \mathrm{d} z (z)$, $ \left. \partial \rho / \partial T \right| _p (z)$, $\left. \partial \rho / \partial p \right| _T (z)$, $c_{pb} (z)$, $\alpha _b T_b (z)$.

\section{An approximate analysis with two modes}
\label{estimate}

We assume that the imaginary part of the eigenvalue is zero at critical conditions, $\sigma =0$, and equations (\ref{cont_eig2}), (\ref{NSx_eig2}), (\ref{NSz_eig2}) and (\ref{entropy_eig2}) take the form:
\begin{eqnarray}
0 &=& - \mathrm{i} k_x \rho _b \widetilde{u}_x - \rho _b \mathrm{D} \widetilde{u}_z  - \rho' _b \widetilde{u}_z, \label{cont_eig3} \\
0 &=& - Ra_{th} \mathrm{i} k_x \widetilde{p} - \frac{4}{3} k_x^2 \widetilde{u}_x + \frac{\mathrm{i} k_x }{3} \mathrm{D} \widetilde{u}_z  + \mathrm{D}^2 \widetilde{u}_x \\
0 &=& - Ra_{th} \mathrm{D} \widetilde{p} -Ra_{th} \left. \frac{ \partial \rho }{\partial T} \right| _P \widetilde{T} - Ra_{th} \left. \frac{ \partial \rho }{\partial p} \right| _T \widetilde{p} - k_x^2 \widetilde{u}_z + \frac{\mathrm{i} k_x }{3} \mathrm{D} \widetilde{u}_x + \frac{4}{3} \mathrm{D}^2 \widetilde{u}_z \\
0 &=& \left( -\rho _b c_{pb} T'_b + {\cal{D}} \alpha _b T_b p'_b \right) \widetilde{u}_z - k_x^2 \widetilde{T} + \mathrm{D}^2 \widetilde{T} 
\end{eqnarray}
The primes denotes $z$-derivatives 
of the base solution profiles, while the symbol $\mathrm{D}$ (resp. $\mathrm{D}^2$, $\mathrm{D}^3$...) denotes $z$-derivatives (resp. second, third derivatives...) of the perturbation variables. Then $\widetilde{u}_x$ is substituted using the first equation, and a function of $z$ is introduced $g(z) = \left( \rho _b c_{pb} T'_b - {\cal{D}} \alpha _b T_b p'_b \right) ^{-1}$ in order to simplify the fourth equation, which takes the form $\widetilde{u}_z = g(z) \left( \mathrm{D}^2 - k_x^2 \right)  \widetilde{T} $. Note that $g(0) = \left( T'_b (0) - \mathcal{D} p'_b (0)  \right) ^{-1} = (-a + \mathcal{D} ) ^{-1}$. The pressure term is substituted using the second equation into the third one and $\widetilde{u}_z$ is expressed in terms of $\widetilde{T}$ using the fourth equation. Finally, we get a single differential equation for the perturbation $\widetilde{T}$: 
\begin{eqnarray}
0&=&- (\mathrm{D}^2-k_x^2) \mathrm{D} \left( \mathrm{D}+ \frac{\rho '_b}{\rho _b} \right) g (\mathrm{D}^2 -k_x^2) \widetilde{T} - k_x^2 Ra_{th} \left. \frac{ \partial \rho }{\partial T} \right| _p \widetilde{T} + k_x^2 (\mathrm{D}^2-k_x^2) g (\mathrm{D}^2- k_x^2) \widetilde{T} \nonumber \\
&& + \left. \frac{ \partial \rho }{\partial p} \right| _T \left[ \frac{1}{3} k_x^2 \frac{\rho '_b}{\rho _b} g (\mathrm{D}^2 -k_x^2) - (\mathrm{D}^2-k_x^2) \left( \mathrm{D}+\frac{\rho '_b}{\rho _b} \right) g (\mathrm{D}^2 -k_x^2)  \right] \widetilde{T} . 
\end{eqnarray}
We now introduce $f(z) = \Delta T_{SA}\  g(z) = \Delta T_{SA} / \left( \rho _b c_{pb} T'_b - {\cal{D}} \alpha _b T_b p'_b \right) $, where the superadiabatic temperature difference $\Delta T_{SA}$ is estimated to the order $3$ in terms of $a$ and $\mathcal{D}$:
\begin{equation}
 \Delta T_{SA} = a - \mathcal{D} + \frac{1}{24} \frac{\mathrm{d}^3 T_a }{ \mathrm{d} z^3} (z=0) \label{deltaTSA}
\end{equation}
The temperature difference of the conduction solution is exactly $a$ while the adiabatic temperature profile is not necessarily linear: the adiabatic temperature gradient at $z=0$ is $\mathcal{D}$, the second derivative does not contribute to the difference between symmetric altitudes at $z=\pm 1/2$ and the third derivative provides a cubic contribution in terms of $\mathcal{D}$ which must be determined when an equation of state is specified. Because the denominator of the function $f$ above is $-a+\mathcal{D}$, linear in terms of $a$ and $\mathcal{D}$, 
our choice for $\Delta T_{SA}$ implies that the function $f$ can be evaluated correctly up to the order $2$ in $a$ and $\mathcal{D}$. In particular, its value at $z=0$, is
\begin{equation}
 \left. f \right| _0  \simeq  -1 - \frac{1}{24 (a - \mathcal{D})} \frac{\mathrm{d}^3 T_a }{ \mathrm{d} z^3} (z=0) \label{fz0}
\end{equation}
Similarly, the term $Ra_{SA}=Ra_{th} \alpha _0 T_0 \Delta T_{SA} $ is an approximation for the superadiabatic Rayleigh number, which can be evaluated correctly up to the degree $2$ in $a$ and $\mathcal{D}$. Using the superadiabatic Rayleigh number $Ra_{SA}$, 
the critical disturbance equation can be written:
\begin{eqnarray}
0&=&- (\mathrm{D}^2-k_x^2) \mathrm{D} \left( \mathrm{D}+ \frac{\rho '_b}{\rho _b} \right) f (\mathrm{D}^2 -k_x^2) \widetilde{T} - \frac{k_x^2}{\alpha _0 T_0}  Ra_{SA} \left. \frac{ \partial \rho }{\partial T} \right| _p \widetilde{T} + k_x^2 (\mathrm{D}^2-k_x^2) f (\mathrm{D}^2- k_x^2) \widetilde{T} \nonumber \\
&& + \left. \frac{ \partial \rho }{\partial p} \right| _T \left[ \frac{1}{3} k_x^2 \frac{\rho '_b}{\rho _b} f (\mathrm{D}^2 -k_x^2) - (\mathrm{D}^2-k_x^2) \left( \mathrm{D}+\frac{\rho '_b}{\rho _b} \right) f (\mathrm{D}^2 -k_x^2)  \right] \widetilde{T} . \label{dispersion_full} 
\end{eqnarray}

This equation depends on several functions of $z$, computed along the base profile, namely $f$, ${\rho '_b}/{\rho _b}$, $\left. { \partial \rho }/{\partial T} \right| _p$, $\left. { \partial \rho }/{\partial p} \right| _T$, all depending on the equation of state considered and on the dimensionless governing parameters $a$ and ${\cal{D}}$.  
In the limit of vanishing temperature difference across the convecting layer, $a<< 1$, the temperature becomes
nearly homogeneous $T\simeq T_0$. The variation of density with pressure at $z=0$, $\left. { \partial \rho }/{\partial p} \right| _{T0}=\rho_0/K_{T0}$ ($K_{T0}$ is the isothermal incompressibility at $z=0$) can be expressed on the following form
\begin{equation}
\left. {\partial \rho \over \partial p} \right| _{T0} = \widetilde{\mathcal{D}}\alpha_0 T_0 , 
\end{equation}
using the general Mayer's relation
\begin{equation}
c_p - c_v = \frac{\alpha ^2 K_T T}{\rho} , \label{mayerx}
\end{equation}
where $c_v$ is the heat capacity at constant volume and defining, for the sake of brevity, 
\begin{equation}
\widetilde{\mathcal{D}} = \frac{{\mathcal{D}}}{1-\gamma_0^{-1}}, \mbox{~~~~} \gamma_0=c_{p0}/c_{v0},
\mbox{~~and~~} \hat\alpha=\alpha_0T_0 . \label{notations}
\end{equation}
Note that $\widetilde{\mathcal{D}}$ can also be written
\begin{equation}
\widetilde{\mathcal{D}} = {1\over \hat\alpha }{\rho_0 g L\over K_{T0}},
\end{equation}
$\widetilde{\mathcal{D}}$ is therefore the ratio of compressible to thermal effects. No surprise it will be the central
parameter to discuss the compressible effects in thermal convection.
In the limit of a vanishing compressibility ($\widetilde{\mathcal{D}} << 1$ or ${\mathcal{D}} << 1$), the base density becomes independent of pressure. Therefore, when both $a$ and ${\mathcal{D}}$ are small, the temperature becomes constant, the density independent of pressure and $f=-1$, ${\rho '_b}/{\rho _b}=0$, $\left. { \partial \rho }/{\partial T} \right| _p=-\alpha _0 T_0$, $\left. { \partial \rho }/{\partial p} \right| _T = 0$,
and the critical equation becomes the well-known dispersion relation for Rayleigh-B\'enard stability:
\begin{equation}
 (\mathrm{D}^2 - k_x^2 )^3 \widetilde{T} + k_x^2 Ra_{SA} \widetilde{T} = 0 \label{RBdisp}
\end{equation}
The thermal perturbation $\widetilde{T}$ satisfies $\widetilde{T}=0$ in $z=\pm 1/2$ (fixed temperatures), $\mathrm{D}^2 \widetilde{T} =0$ ($\widetilde{u}_z =0$) and $\mathrm{D}^4 \widetilde{T} =0 $ (no-stress conditions). It has non-zero solutions for a minimal value of $Ra_{SA}=27 \upi ^4 / 4$ and a corresponding wavenumber $k_x=\upi / \sqrt{2} $. The corresponding eigenvector is a cosine function $\cos (\upi z)$.

Now, for a finite temperature gradient $a$ or dissipation number ${\cal{D}}$, the functions $f$ and $\left. { \partial \rho }/{\partial T} \right| _p$, have some $z$-dependence and the functions ${\rho '_b}/{\rho _b}$ and $\left. { \partial \rho }/{\partial p} \right| _T$ are not zero. As a consequence, when an even function of $z$ is initially considered for the temperature eigenvector, there are odd contributions generated in (\ref{dispersion_full}). Hence, the eigenvectors must be a combination of at least an even and an odd contribution. Hence, we decided to expand the eigenmodes as 
\begin{equation}
\widetilde{T} = \cos (\upi z) + \epsilon \sin (2 \upi z ). \label{eigenmode}
\end{equation} 
The motivation for this particular choice $\sin (2 \upi z )$ of odd function of $z$ is that it satisfies the boundary conditions and that it is the second least dissipative harmonic mode after $\cos (\upi z)$. In addition, we have checked on some eigenvectors obtained using Chebyshev expansion that they could be written as the sum of two such modes (\ref{eigenmode}) with negligible residuals (see section \ref{eos} and Fig.~\ref{modepr}). 
We wish to achieve a second order accuracy, in the base temperature gradient $a$ and in the dissipation number $\mathcal{D}$, so that we can evaluate the change in critical Rayleigh number to a similar degree. We thus expand the functions of $z$ related to the base profile $f$, ${\rho '_b}/{\rho _b}$, $\left. { \partial \rho }/{\partial T} \right| _p$ and $\left. { \partial \rho }/{\partial p} \right| _T$ in Taylor expansions of degree two, for instance: 
\begin{equation}
f(z)=f_0+ \left. \frac{\mathrm{d} f}{\mathrm{d} z} \right| _0 z + \frac{1}{2} \left. \frac{\mathrm{d}^2 f}{\mathrm{d} z^2}\right| _0 z^2, \label{expf}
\end{equation}
and similarly for the others. 
The introduction of the expansions of the form (\ref{expf}) and (\ref{eigenmode}) into the critical equation (\ref{dispersion_full}) generates terms which are products between trigonometric functions and powers of $z$. We project these functions back on the two chosen modes $\cos (\upi z)$ and $\sin (2 \upi z )$. The projection is that associated with the $L^2$ functional space on $[-1/2 ; 1/2 ]$
(see Table \ref{tableProj}).
The change in the reference profiles due to the dissipation parameter $\mathcal{D}$ and finite temperature gradient $a$ affects not only $\epsilon$ but also the critical Rayleigh number by a quantity $dRa_{SA}$,
\begin{equation}
Ra_{SA}={27\over 4} \pi^4+dRa_{SA}.
\end{equation}
For any equation of state from which the stable basic state can be computed and Taylor expanded (as in (\ref{expf})), our eigenmodes (\ref{eigenmode}) introduced into the critical equation (\ref{dispersion_full}) lead to two equations (i.e., the terms in factor of
$\cos (\upi z)$ and $\sin (2 \upi z )$)
those solutions are the eigenmode amplitude $\epsilon$ (from the $\sin (2 \upi z )$ part) and the perturbation of the critical Rayleigh number $dRa_{SA}$ (from the $\cos (\upi z)$ part). A close look to the equations indicates that $\epsilon$ depends linearly on the parameters describing the distance of the problem to the classical Boussinesq problem (mainly $a$ and ${\cal{D}}$ the temperature gradient and dissipation number) while $dRa_{SA}$ is only affected by terms of order $2$. Similarly the horizontal wavenumber $k_x$ is also affected by terms of order $2$. Moreover, because the critical Rayleigh number is also such that $\mathrm{d} Ra_{SA} / \mathrm{d} k_x =0 $ (minimal Rayleigh number over wavenumbers), the quadratic disturbance of $k_x$ does not affect the evaluation of the quadratic disturbance of $Ra_{SA}$. It is hence correct to use a constant value $k_x=\upi / \sqrt{2}$ for this analysis.

\begin{table*}
\begin{center}
\begin{tabular}{@{}lcc@{}}
\toprule  & $\cos (\upi z )$ & $\sin (2 \upi z )$  \\
\hline
$\sin (\upi z )$ & 0 & $\frac{8}{3 \upi}$  \\
$z \sin (\upi z )$ & $\frac{1}{2 \pi}$ & 0 \\
$z \cos ( 2 \upi z )$ & 0 & $-\frac{1}{4 \pi}$ \\
$ \cos ( 2 \upi z )$ & $\frac{4}{3 \pi}$ & 0 \\
$ z \cos (\upi z )$ & 0 & \hspace*{3 mm} $\frac{16}{9 \upi ^2}$ \hspace*{3 mm} \\
$ z \sin (2 \upi z )$ & $\frac{16}{9 \upi ^2}$ & 0 \\
$ z^2 \sin (2 \upi z )$ & 0 &  $\frac{2 \upi ^2 -3}{24 \upi ^2}$ \\
$ z^2 \cos (\upi z ) $ \hspace*{5 mm} & \hspace*{3 mm} $\frac{\upi ^2 - 6}{12 \upi ^2}$ \hspace*{3 mm} & 0 \\
$ z^2 \sin (\upi z ) $ & 0 & $ \frac{18 \upi ^2 -112}{27 \upi ^3} $\\
$ z^2 \cos (2 \upi z ) $ & $\frac{9 \upi ^2 -104}{27 \upi ^3}$ & 0 \\
\bottomrule
\end{tabular}
\caption{Projection coefficients of some functions on the modes $\cos (\upi z )$ and $\sin (2 \upi z )$}
\label{tableProj}
\end{center}
\end{table*}

Let us now provide some details on how the equations for $\epsilon$ and $dRa_{SA}$ are derived. We introduce $\widetilde{u} = f (\mathrm{D}^2 - k_x^2) \widetilde{T} $, $a - \mathcal{D}$ times the vertical velocity component $\widetilde{u}_z$, and $\widetilde{v} = \mathrm{D} (\mathrm{D} + {\rho '_b}/{\rho _b} ) \widetilde{u}$, which is $\mathrm{i} (a - \mathcal{D})/k_x$ times the $z$-derivative of the horizontal velocity component (from equation (\ref{cont_eig3})). Using variables $\widetilde{T}$, $\widetilde{u}$ and $\widetilde{v}$, the critical equation (\ref{dispersion_full}) takes the form:
\begin{equation}
0=- (\mathrm{D}^2-k_x^2) \widetilde{v} - \frac{k_x^2}{\alpha _0 T_0 }  Ra_{SA} \left. \frac{ \partial \rho }{\partial T} \right| _p \widetilde{T} + k_x^2 (\mathrm{D}^2-k_x^2)  \widetilde{u} + \left. \frac{ \partial \rho }{\partial p} \right| _T \left[ \frac{4}{3} k_x^2 \frac{\rho '_b}{\rho _b} \widetilde{u} - \mathrm{D} \widetilde{v} + k_x^2  \mathrm{D} \widetilde{u} \right] . \label{dispersion_full2}
\end{equation}
Both $\widetilde{u}$ and $\widetilde{v}$ satisfy the same boundary conditions as $\widetilde{T}$ (zero in $z = \pm 1/2$) so that they are also projected on the same modes defined in (\ref{eigenmode}): 
\begin{eqnarray}
\widetilde{u} &=& U_{c} \cos (\upi z ) + U_s \sin (2 \upi z ) , \label{UcUs}\\
\widetilde{v} &=& V_{c} \cos (\upi z ) + V_s \sin (2 \upi z ) . \label{VcVs}
\end{eqnarray}

From the definition $\widetilde{u} = f (\mathrm{D}^2 - k_x^2) \widetilde{T} $, we have  
\begin{eqnarray}
U_c &=& - \frac{3 \upi ^2}{2} f_0 - 8 \epsilon \left. \frac{\mathrm{d} f}{\mathrm{d} z} \right| _0 + \left( -\frac{\upi ^2}{16} + \frac{3}{8} \right) \left. \frac{\mathrm{d}^2 f}{\mathrm{d} z^2} \right| _0, \label{UZc} \\
U_s &=&  - 9 \frac{\upi ^2}{2} \epsilon f_0 - \frac{8}{3} \left. \frac{\mathrm{d} f}{\mathrm{d} z} \right| _0 , \label{UZs} 
\end{eqnarray}
where the projections determined in table \ref{tableProj} have been used. 
Using Maxima, a software for formal manipulations, we shall obtain the Taylor coefficients for $f$ and other quantities, once an equation of state will be specified. Next, from $\widetilde{v} = \mathrm{D} (\mathrm{D} + {\rho '_b}/{\rho _b} ) \widetilde{u}$, and using again table \ref{tableProj}, we obtain:
\begin{eqnarray}
V_c &=& \left( -\upi ^2 + \frac{1}{2} \left. \frac{\mathrm{d} \frac{\rho '_b}{\rho _b}}{\mathrm{d} z} \right| _0 \right) U_c + \frac{8}{3} \left. \frac{\rho '_b}{\rho _b} \right| _0 U_s , \label{Vc} \\
V_s &=&	- 4 \upi ^2 U_s - \frac{8}{3} \left. \frac{\rho '_b}{\rho _b} \right| _0 U_c 
.   \label{Vs}
\end{eqnarray}
Before we can write equation (\ref{dispersion_full2}) onto our two base functions, we need to express auxiliary variables:
\begin{eqnarray}
\left. { \partial \rho }/{\partial T} \right| _p  \widetilde{T} &=& A_c \cos (\upi z ) + A_s \sin (2 \upi z ), \label{AcAs} \\
\frac{\rho '_b}{\rho _b} \widetilde{u} &=& B_c \cos (\upi z ) + B_s \sin (2 \upi z ), \label{BcBs} 
\end{eqnarray}
with coefficients: 
\begin{eqnarray}
A_c &=& - \hat\alpha + \left( \frac{1}{24} - \frac{1}{4 \upi ^2} \right) \left. \frac{\mathrm{d}^2 \left. { \partial \rho }/{\partial T} \right| _p }{\mathrm{d} z^2} \right| _0 + \frac{16}{9 \upi ^2 } \left. \frac{\mathrm{d} \left. { \partial \rho }/{\partial T} \right| _p }{\mathrm{d} z} \right| _0 \epsilon , \label{Ac} \\
A_s &=& \frac{16}{9 \upi ^2} \left. \frac{\mathrm{d} \left. { \partial \rho }/{\partial T} \right| _p }{\mathrm{d} z} \right| _0  - \hat\alpha \epsilon, \label{As} \\
B_c &=& \left. \frac{\rho '_b}{\rho _b} \right| _0 U_c + \frac{16}{9 \upi ^2}  \left. \frac{\mathrm{d} \frac{\rho '_b}{\rho _b}}{\mathrm{d} z} \right| _0 U_s , \label{Bc} \\
B_s &=& \left. \frac{\rho '_b}{\rho _b} \right| _0 U_s  + \frac{16}{9 \upi ^2}  \left. \frac{\mathrm{d} \frac{\rho '_b}{\rho _b}}{\mathrm{d} z} \right| _0 U_c  , \label{Bs}
\end{eqnarray}
where we have used $\left.  {\partial \rho }/{\partial T} \right| _{p0} = - \hat\alpha $. 
We can now write the projection of equation (\ref{dispersion_full2}) on $\cos (\upi z )$ and $\sin (2 \upi z )$ keeping only the terms of appropriate order: 
\begin{eqnarray}
&& \hspace*{-5 mm} \frac{3}{2} \upi ^2 V_c - \frac{\upi ^2}{2 \hat\alpha } \left( \frac{27 \upi ^4}{4} + dRa_{SA} \right) A_c - \frac{3 \upi ^4}{4} U_c + \left. \frac{ \partial \rho }{\partial p} \right| _{T0} \left[ \frac{2 \upi ^2}{3} B_c - \frac{8}{3} V_s + \frac{4 \upi ^2}{3} U_s \right]  \nonumber \\
&&  \hspace*{-5 mm} 
+ \left. \frac{\mathrm{d} \left. \frac{ \partial \rho }{\partial p} \right| _T }{\mathrm{d} z} \right| _0  \left[ \frac{1}{2} V_c - \frac{\upi ^2}{4} U_c \right] = 0, \label{dispCOS} \\
&&  \hspace*{-5 mm} \frac{9}{2} \upi ^2 V_s - \frac{\upi ^2}{2 \hat\alpha }  \frac{27 \upi ^4}{4}  A_s - \frac{9 \upi ^4}{4} U_s + \left. \frac{ \partial \rho }{\partial p} \right| _{T0} \left[ \frac{8}{3} V_c - \frac{4 \upi ^2}{3} U_c \right] = 0, \label{dispSIN} 
\end{eqnarray}
The second equation (\ref{dispSIN}) is used to determine the coefficient $\epsilon $ (see \ref{eigenmode}). This coefficient $\epsilon$ depends linearly on the parameters describing the distance of the problem to the classical Boussinesq problem, $a$ and ${\mathcal{D}}$ the temperature gradient and dissipation number. The first equation (\ref{dispCOS}) is then solved to obtain $dRa_{SA}$, the change in critical Rayleigh number compared to the classical critical Rayleigh number $27 \upi ^4 /4 $ for no-stress boundary conditions. This change is thus quadratic in $a$ and ${\mathcal{D}}$: the terms of order zero cancel out (Boussinesq limit), the terms of order $1$ are found in the (\ref{dispSIN}) equation used to determine the coefficient $\epsilon$ of the $\sin (2 \upi z )$  mode, and the terms of order $2$ balance $dRa_{SA}$ in (\ref{dispCOS}), with $A_c$ containing a term of order $0$ in $a$ and ${\mathcal{D}}$. 

Equations (\ref{dispSIN}) then (\ref{dispCOS}) are solved  explicitly  in terms of the quantities $f$, $\mathrm{d} f / \mathrm{d} z$, $\mathrm{d}^2 f / \mathrm{d} z^2$, $\rho' _b / \rho _b $, $\mathrm{d} / \mathrm{d} z \left( \rho' _b / \rho _b \right) $, $ \left. \partial \rho / \partial T \right| _p $, $\mathrm{d} / \mathrm{d} z ( \left. \partial \rho / \partial T \right| _p ) $, $\mathrm{d}^2 / \mathrm{d} z^2 ( \left. \partial \rho / \partial T \right| _p ) $, $ \left. \partial \rho / \partial p \right| _T $, $\mathrm{d} / \mathrm{d} z (  \left. \partial \rho / \partial p \right| _T ) $, evaluated at $z=0$. Equation (\ref{dispSIN}) leads to 
\begin{equation}
\epsilon = \frac{8}{117 \pi^2} \left[ 9 \left. \frac{\mathrm{d} f }{ \mathrm{d} z} \right| _0 - \frac {1}{\hat\alpha } \left. \frac{\mathrm{d} }{ \mathrm{d} z} \frac{\partial \rho }{ \partial T} \right| _{p0}  - \left. \frac{\partial \rho }{ \partial p } \right| _{T0} - 3 \left. \frac{ \rho' _b }{ \rho _b } \right| _0 \right]. \label{sol_epsilon} 
\end{equation}
With this value for $\epsilon$, $dRa_{SA}$ is obtained from equation (\ref{dispCOS})
\begin{eqnarray}
\hspace*{-5 mm}dRa_{SA} &=& - \frac{9}{4} \pi^2 \left. \frac{\mathrm{d} }{ \mathrm{d} z}  \frac{ \rho' _b }{ \rho _b } \right| _0  - \left( 36 \upi ^2 \epsilon + \left[2 \upi ^2 + \frac{64}{3} \right] \left. \frac{\partial \rho }{ \partial p} \right| _{p0} - \frac{64}{3}  \left. \frac{\mathrm{d} f }{ \mathrm{d} z} \right| _0 \right) \left. \frac{ \rho' _b }{ \rho _b } \right| _0  \nonumber \\
&& - \frac{27 \upi ^4}{4} \left( \left. f \right| _0 + 1 \right) + 
\left(
\frac{9 \upi ^4}{32  } - \frac{27 \upi ^2}{16 }
\right)
\left(
\frac{1}{ \hat\alpha }
 \left. \frac{\mathrm{d}^2 }{ \mathrm{d} z^2}  \frac{\partial \rho }{ \partial T} \right| _{p0}-  \left. \frac{\mathrm{d}^2 f }{ \mathrm{d} z^2} \right| _0  \right)
 \nonumber \\
&&  - \left( 36 \left. \frac{\mathrm{d} f }{ \mathrm{d} z} \right| _0 - \frac{12}{\hat\alpha } \left. \frac{\mathrm{d} }{ \mathrm{d} z}  \frac{\partial \rho }{ \partial T} \right| _{p0} + 108 \left.\frac{\partial \rho }{ \partial p} \right| _{p0} \right) \upi ^2 \epsilon 
\nonumber \\
&& + \frac{9 \pi^2 }{4} \left.  \frac{\mathrm{d} }{ \mathrm{d} z}  \frac{\partial \rho }{ \partial p} \right| _{T0} + 64 \left. \frac{\partial \rho }{ \partial p } \right| _{T0} \left. \frac{\mathrm{d} f }{ \mathrm{d} z} \right| _0 .  \label{sol_dRaSA} 
\end{eqnarray}

\section{The quasi-Boussinesq and quasi-ALA models}
\label{themodels}

We refer to the stability analysis presented in sections \ref{eigenvalue} and \ref{estimate} as to the exact model for Rayleigh-B\'enard Stability, since it is based on the continuity, Navier-Stokes and entropy equations without any approximation. In the exact model the linearized density perturbation is therefore
\begin{equation}
\rho' = \left. \frac{\partial \rho}{\partial T}\right| _p T' + \left. \frac{\partial \rho}{\partial p} \right| _T p' , \label{eos_exact}
\end{equation}
like in (\ref{eos_rho}). 
 We will now introduce two models, corresponding to changes in the governing equations, with different assumptions on compressibility. For both models, the base solution is kept unchanged, which means that compressible effects are fully taken into account. The assumptions concern the fluctuations. 
The quasi-Boussinesq model consists in neglecting the pressure dependence of the density fluctuations in equation (\ref{eos_exact}) and therefore in using
\begin{equation}
\rho' = \left. \frac{\partial \rho}{\partial T}\right| _p T'. \label{eos_quasibouss}
\end{equation}
The quasi-Boussinesq critical superadiabatic Rayleigh number $Ra_{SA}^B$ is obtained from the same Chebyshev collocation expansion as described in section \ref{eigenvalue}. This model is not called a Boussinesq model, because the base profile takes into account compressibility effects, contrary to the original Boussinesq model.
Similarly, the quasi-ALA model is reminiscent but not identical to the anelastic liquid approximation (ALA) as described in \citep{ajs05} as the base profile is the conduction profile, not the adiabatic profile. Density fluctuations are first expressed in terms of fluctuations of pressure and entropy, instead of pressure and temperature in (\ref{eos_exact}): 
\begin{equation}
 \rho ' = \left. \frac{\partial \rho}{\partial p}\right| _s p' + \left. \frac{\partial \rho}{\partial s} \right| _p s' . \label{eos_rho2}
\end{equation}
Then two assumptions are made: the first term is evaluated as though the base density gradient were close to the adiabat and pressure dependence of entropy fluctuations are neglected compared to their temperature dependence:
\begin{equation}
\rho ' = -\frac{1}{\rho _b} \frac{{\rm d} \rho _b}{{\rm d} z} p'   + \left. \frac{\partial \rho}{\partial T} \right| _p T' .  \label{gradrho} 
\end{equation}
The first assumption on the density gradient does not need to be made in the classical ALA model, as the solutions are indeed expanded from the (hydrostatic) adiabatic profile which is not possible in a stability analysis, as the adiabatic profile is always stable. The quasi-ALA critical Rayleigh number $Ra_{SA}^{ALA}$ is obtained from a similar analysis as described in section \ref{eigenvalue}.
In summary, the terms $ - {\bf \nabla } p' - { \rho ' }  {\bf e}_z$ in equation (\ref{NS_eig}) are changed for $ - {\bf \nabla } p' - \left. {\partial \rho}/{\partial T} \right| _p T' {\bf e}_z$ in the quasi-Boussinesq model and for $- \rho _b {\bf \nabla } \left( p' / \rho _b \right) - \left. {\partial \rho}/{\partial T} \right| _p T' {\bf e}_z$ in the quasi-ALA model. 

For the quasi-Boussinesq and quasi-ALA models, a two-modes approximation analysis is also carried out (see section \ref{estimate}), providing $\epsilon ^B$ and $\epsilon ^{ALA}$ the $\sin (2 \upi z)$ contributions of the eigenmodes of the quasi-Boussinesq and quasi-ALA approximations, as well as $dRa_{SA}^B$ and $dRa_{SA}^{ALA}$ the departures from $27 \upi ^4 / 4$ of the critical Rayleigh numbers for each approximation respectively. Equations (\ref{dispCOS}) and (\ref{dispSIN}) are modified in the following way: for the quasi-Boussinesq approximation, all terms involving $\left. \partial \rho / \partial p \right| _T$ or its derivative with respect to $z$ are removed, while for the quasi-ALA approximation, $\left. \partial \rho / \partial p \right| _T$ is replaced by $- \rho _b' / \rho _b$ and $\mathrm{d} / \mathrm{d} z \left( \left. \partial \rho / \partial p \right| _T \right) $ by $- \mathrm{d} / \mathrm{d} z  \left( \rho _b' / \rho _b \right)$. The same changes are therefore made on the solutions for $\epsilon $ and $dRa_{SA}$ in equations (\ref{sol_epsilon}) and (\ref{sol_dRaSA}). The differences $\delta \epsilon ^B = \epsilon ^B - \epsilon $ and $\delta \epsilon ^{ALA} = \epsilon ^{ALA} - \epsilon $ can then be expressed as
\begin{eqnarray}
\delta \epsilon ^B  &=&  \frac{8}{117 \pi^2} \left. \frac{\partial \rho }{ \partial p } \right| _{T0} , \label{sol_delta_epsilonB} \\
\delta \epsilon ^{ALA} &=&  \frac{8}{117 \pi^2} \left( \left. \frac{\partial \rho }{ \partial p } \right| _{T0} + \left. \frac{ \rho' _b }{ \rho _b } \right| _0 \right) . \label{sol_delta_epsilonALA} 
\end{eqnarray}
The differences of $dRa_{SA}$ induced by the quasi-Boussinesq and quasi-ALA approximations, $\delta Ra_{SA}^B = d Ra_{SA}^B - d Ra_{SA}$ and $\delta Ra_{SA}^{ALA} = d Ra_{SA}^{ALA} - d Ra_{SA}$, take the following form 
\begin{eqnarray}
\delta Ra_{SA}^B & = & - \left( 36 \upi ^2 \delta \epsilon ^B - \left[2 \upi ^2 + \frac{64}{3} \right] \left. \frac{\partial \rho }{ \partial p} \right| _{p0} \right) \left. \frac{ \rho' _b }{ \rho _b } \right| _0 \nonumber \\
&& - \left( 36 \left. \frac{\mathrm{d} f }{ \mathrm{d} z} \right| _0 - \frac{12}{\hat\alpha } \left. \frac{\mathrm{d} }{ \mathrm{d} z}  \frac{\partial \rho }{ \partial T} \right| _{p0} \right) \upi ^2 \delta \epsilon ^B + 108 \left.\frac{\partial \rho }{ \partial p} \right| _{p0} \upi ^2 \epsilon \nonumber \\ 
&&- \frac{9 \pi^2 }{4} \left.  \frac{\mathrm{d} }{ \mathrm{d} z}  \frac{\partial \rho }{ \partial p} \right| _{T0} - 64 \left. \frac{\partial \rho }{ \partial p } \right| _{T0} \left. \frac{\mathrm{d} f }{ \mathrm{d} z} \right| _0 ,   \label{sol_deltaRaB} \\
\delta Ra_{SA}^{ALA} & = & - \left( 36 \upi ^2 \delta \epsilon ^{ALA} - \left[2 \upi ^2 + \frac{64}{3} \right] \left[ \left. \frac{\partial \rho }{ \partial p} \right| _{p0} + \left. \frac{ \rho' _b }{ \rho _b } \right| _0 \right] \right) \left. \frac{ \rho' _b }{ \rho _b } \right| _0 \nonumber \\
&& - \left( 36 \left. \frac{\mathrm{d} f }{ \mathrm{d} z} \right| _0 - \frac{12}{\hat\alpha } \left. \frac{\mathrm{d} }{ \mathrm{d} z}  \frac{\partial \rho }{ \partial T} \right| _{p0} \right) \upi ^2  \delta \epsilon ^{ALA} + 108 \upi ^2 \left( \left.\frac{\partial \rho }{ \partial p} \right| _{p0} \epsilon + \left. \frac{ \rho' _b }{ \rho _b } \right| _{0} \epsilon ^{ALA} \right) \nonumber \\
&& - \frac{9 \pi^2 }{4} \left( \left. \frac{\mathrm{d} }{ \mathrm{d} z}  \frac{\partial \rho }{ \partial p} \right| _{T0} + \frac{\mathrm{d} }{ \mathrm{d} z} \left. \frac{ \rho' _b }{ \rho _b } \right| _{0} \right)  - 64 \left( \left. \frac{\partial \rho }{ \partial p } \right| _{T0} + \left. \frac{ \rho' _b }{ \rho _b } \right| _{0} \right) \left. \frac{\mathrm{d} f }{ \mathrm{d} z} \right| _0 . \label{sol_deltaRaALA}
\end{eqnarray}

\section{Stability results for various equations of state}
\label{eos}

We now consider different equations of state and perform the stability analyses, numerical Chebyshev expansion and two-modes analysis, for the exact, quasi-Boussinesq and quasi-ALA approximations. 

\subsection{Ideal gas EoS}
\label{ideal}

The following dimensional equation of state is considered: 
\begin{equation}
p = \rho R T, \label{idealgas}
\end{equation}
where $R = {\cal{R}}/M$ is the gas constant, while ${\cal{R}}$ and $M$ are the universal gas constant and molar mass of the gas respectively. In addition, ideal gases are characterized by the choice of a constant heat capacity at constant volume $c_v$. It can then be shown that $c_p$ is constant as well and obeys Mayer's relation: $c_p - c_v = R$. The ratio of heat capacities is $\gamma = c_p / c_v$. Using the scales already defined, $\rho_0 g L$ for pressure, $\rho_0$ for density and $T_0$ for temperature, the equation of state takes the following dimensionless form:
\begin{equation}
p = \rho  T \frac{1 - \gamma ^{-1}}{{\cal{D}}} = \frac{\rho  T}{\widetilde{\mathcal{D}}} . \label{idealgasa}
\end{equation}
Finally, for ideal gases, the marginal stability problem depends on four dimensionless numbers: $Ra_{th}$, ${\cal{D}}$, $a$ and $\widetilde{\mathcal{D}}$.  
It can be shown that the product $\alpha T$ is always unity for an ideal gas. The base thermal profile is given by (\ref{basetemperature}).  
The adiabatic profile $T_a (z)$ is also derived, so that the superadiabatic temperature difference can be evaluated later. The dimensionless adiabatic gradient is:
\begin{equation}
\frac{{\rm d} T_a }{{\rm d} z} = - \frac{g L }{c_p T_0} = - \frac{\alpha _0 g L }{c_p} = - {\cal{D}}, \label{adiabgrad}
\end{equation}
and its solution is:
\begin{equation}
T_a=1 - {z} {{\cal{D}}}, \label{adiab_profile}
\end{equation}
and therefore $\Delta T_{SA}$ is exactly $a-{{\mathcal{D}}}$, (see (\ref{deltaTSA})).

Then the dimensionless hydrostatic equation $\mathrm{d} p_b / \mathrm{d} z = - \rho _b$ is used with the equation of state (\ref{idealgasa}) to derive the density and pressure profiles:
\begin{equation}
\frac{{\rm d} p_b}{{\rm d} z }= \frac{1}{{\widetilde{\cal{D}}}}  \left( \frac{ {\rm d} \rho _b }{{\rm d} z}  T_b + \rho _b  \frac{{\rm d} T_b}{{\rm d} z} \right) = - \rho _b , 
\label{baseprof} 
\end{equation}
Having already derived the temperature profile (\ref{basetemperature}), this is a differential equation for $\rho _b$. With $\rho _b = 1$ when $z=0$, imposed by our normalization, the solution is:
\begin{equation}
\rho _b = T_b^{ -1 - \frac{{\widetilde{\mathcal{D}}}}{a}}. \label{solrhob}
\end{equation}
The corresponding pressure profile can then be derived from the equation of state:
\begin{equation}
p_b = \frac{1}{{\widetilde{\mathcal{D}}}} T_b^{ - \frac{\widetilde{\mathcal{D}}}{a}} . \label{solpb}
\end{equation}
Every quantity, related to the base profile and needed in the eigenvalue problem (\ref{cont_eig2}), (\ref{NSx_eig2}), (\ref{NSz_eig2}) and (\ref{entropy_eig2}), is now available and we can solve exactly for the critical Rayleigh number
using a Chebyshev collocation expansion. 

In addition to this exact problem (no approximation was made in the governing equations), two models are considered: quasi-Boussinesq and quasi-ALA, described in section \ref{themodels} and using respectively the approximated density variations (\ref{eos_quasibouss}) and (\ref{gradrho}).
The critical Rayleigh number is expressed through the superadiabatic Rayleigh number (\ref{RaSA}).
The critical (superadiabatic) Rayleigh numbers for the exact, quasi-Boussinesq and quasi-ALA models are denoted $Ra_{SA}^x$, $Ra_{SA}^B$ and $Ra_{SA}^{ALA}$, respectively.

\begin{table*}
\begin{center}
\begin{tabular}{@{}ll@{}}
\toprule expression \hspace*{1 cm} & value   \\
\hline
$\left. f \right| _0 $ &  $$-1$$  \\
$\left. \frac{\mathrm{d} f }{\mathrm{d} z}  \right| _0$  &  $a - \widetilde{\mathcal{D}}$ \\
$\left. \frac{\mathrm{d}^2 f }{\mathrm{d} z^2}  \right| _0$  & $(a  - \widetilde{\mathcal{D}})\widetilde{\mathcal{D}} $ \\
$\left.  \frac{\rho '_b}{\rho _b} \right| _0 $ & $ a - \widetilde{\mathcal{D}}$ \\
$\left. \frac{\mathrm{d} }{\mathrm{d} z} \frac{\rho '_b}{\rho _b} \right| _0 $  & $ (a - \widetilde{\mathcal{D}})a $\\
$\left. \frac{ \partial \rho }{\partial T} \right| _{p0}$ & $ -1 $ \\
$\left. \frac{\mathrm{d} }{\mathrm{d} z}  \frac{ \partial \rho }{\partial T} \right| _{p0}  $  & $-2 a + \widetilde{\mathcal{D}}$\\
$\left. \frac{\mathrm{d}^2 }{\mathrm{d} z^2}  \frac{ \partial \rho }{\partial T} \right| _{p0} $  & $ -6 a^2 +5 a \widetilde{\mathcal{D}} - \widetilde{\mathcal{D}}^2$ \\
$\left. \frac{ \partial \rho }{\partial p} \right| _{T0}$ & $ \widetilde{\mathcal{D}} $ \\
$\left. \frac{\mathrm{d} }{\mathrm{d} z}  \frac{ \partial \rho }{\partial p} \right| _{T0}  $ & $ a\widetilde{\mathcal{D}}$ \\
\bottomrule
\end{tabular}
\caption{Some quantities related to the base flow, needed for the two-modes approximation, for the equation of state of an ideal gas.}
\label{tableidealgas}
\end{center}
\end{table*}

We also apply the analysis based on just two eigenmodes ($\cos ( \upi z )$ and $\sin ( 2 \upi z )$), leading to equations (\ref{dispCOS}) and (\ref{dispSIN}), which are themselves issued from the critical relation (\ref{dispersion_full2}). We need to derive some expressions from the equation of state: they are values of quantities at $z=0$, relative to the base profile $f$, ${\rho '_b}/{\rho _b}$, $\left. { \partial \rho }/{\partial T} \right| _p$,  $\left. { \partial \rho }/{\partial p} \right| _T$ and their derivatives at $z=0$. They are listed in table \ref{tableidealgas} for the case of an ideal gas. 
The expressions for the base profile in table \ref{tableidealgas} are simple enough to be substituted in the two-modes general solutions (\ref{sol_epsilon}) and (\ref{sol_dRaSA}). The $\sin (2 \upi z)$ contributions $\epsilon $, $\epsilon ^B$ and $\epsilon ^{ALA}$ to the exact model, quasi-Boussinesq and quasi-ALA approximations take the form  
\begin{eqnarray}
\epsilon &=& \frac{64}{117 \upi ^2} \left( a - \tilde{\mathcal{D}} \right)  , \label{eps_GPx} \\
\epsilon ^B &=& \frac{64}{117 \upi ^2} \left( a - \frac{7}{8} \widetilde{\mathcal{D}} \right)   , \label{eps_GPB} \\
\epsilon ^{ALA} &=& \frac{64}{117 \upi ^2} \left(\frac{9}{8}  a - \widetilde{\mathcal{D}} \right)   . \label{eps_GPALA}
\end{eqnarray}

\begin{figure}
\begin{center}
\includegraphics[width=12 cm, keepaspectratio]{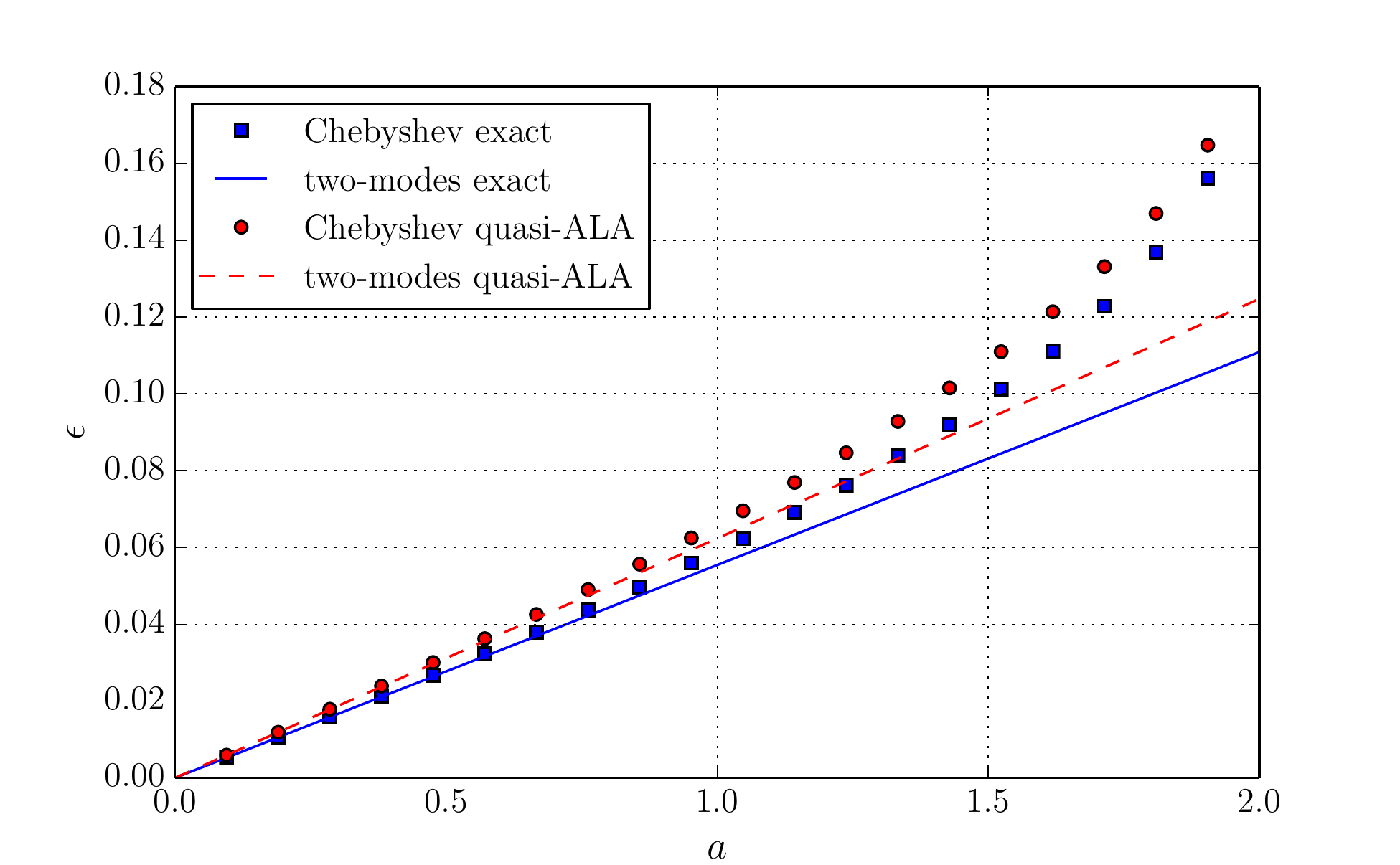}
\caption{Asymmetrical contribution $\epsilon$ of the $\sin (2 \upi z) $ mode to the critical eigenmode, for an ideal gas as a function of the temperature gradient $a$ of the base linear solution, for a negligible ${\cal{D}} = 10^{-8}$ and a ratio of heat capacities $\gamma$ equal to $5/3$. The label 'Chebyshev exact' denotes the numerical solution of the exact model using a Chebyshev expansion (usually 17 polynomials), while the label 'Chebyshev quasi-ALA' corresponds to the solutions of the quasi-ALA model. The labels 'two-modes exact' and 'two-modes quasi-ALA' correspond to the approximate two-modes analytical solutions for the exact and quasi-ALA models. Note that when the dissipation number is negligible, the quasi-Boussinesq model and the exact model coincide.}
\label{GPalineps}
\end{center}
\end{figure}

\begin{figure}
\begin{center}
\includegraphics[width=12 cm, keepaspectratio]{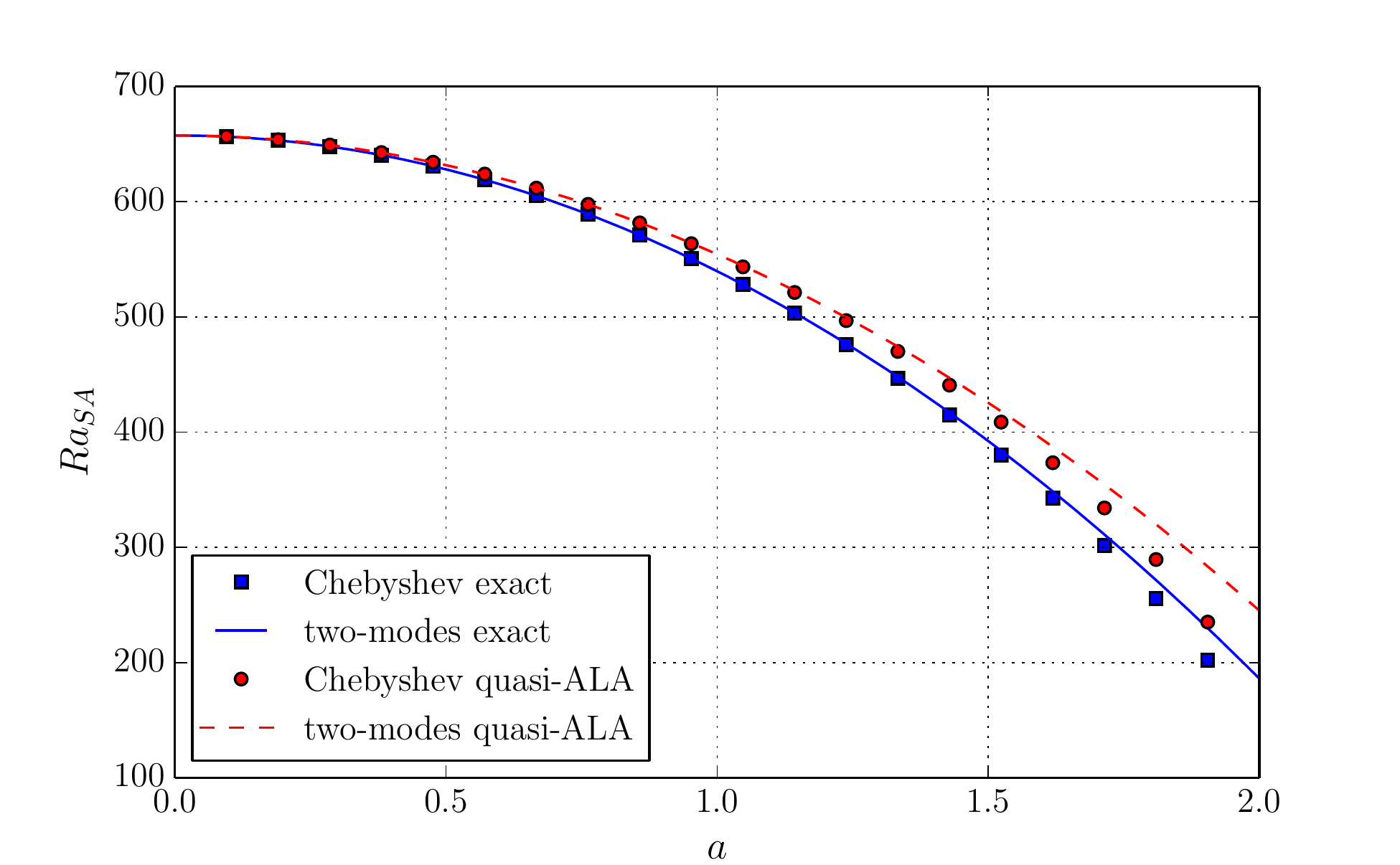}
\caption{Linear stability critical threshold for the Rayleigh number for an ideal gas as a function of the temperature gradient $a$ of the base linear solution, for a negligible ${\cal{D}} = 10^{-8}$ and a ratio of heat capacities $\gamma$ equal to $5/3$. Labels and linestyles correspond to that of figure \ref{GPalineps}. }
\label{GPalin}
\end{center}
\end{figure}

 The corresponding critical superadiabatic Rayleigh number is obtained from (\ref{sol_dRaSA}) as an expansion of degree $2$ in $a$ and ${\cal{D}}$. We also obtain approximate critical Rayleigh numbers in the quasi-Boussinesq and quasi-ALA approximations. The difference between these critical Rayleigh numbers and the classical Boussinesq value $27 \upi ^4 / 4$ are denoted $dRa_{SA}^x$, $dRa_{SA}^B$ and $dRa_{SA}^{ALA}$
\begin{eqnarray}
\hspace*{-1 mm}dRa_{SA}^x &=&  \left[2 \upi ^2 - \frac{320}{39} \right] \widetilde{\mathcal{D}}^2
             + \left[ \frac{9 \upi ^4}{8} - \frac{17 \upi ^2}{4} + \frac{512}{13} \right] a \widetilde{\mathcal{D}} 
- \left[ \frac{27 \upi ^4}{16} - \frac{63 \upi ^2}{8} + \frac{1216}{39} \right] a^2 , \hspace*{6 mm} \label{dRaGPx}\\ 
&& \simeq 11.53 \widetilde{\mathcal{D}}^2 + 107.02 a \widetilde{\mathcal{D}} - 117.83 a^2 , \nonumber \\[3mm]
\hspace*{-1 mm}dRa_{SA}^B &=& - \frac{736}{39} {\widetilde{\mathcal{D}}}^2 
             + \left[ \frac{9 \upi ^4}{8} - \frac{9 \upi ^2}{2} + \frac{640}{13}  \right] a \widetilde{\mathcal{D}}  
             - \left[ \frac{27 \upi ^4}{16} - \frac{63 \upi ^2}{8} + \frac{1216}{39} \right] a^2, \label{dRaGPB} \\
&& \simeq - 18.87 \widetilde{\mathcal{D}}^2 + 114.40 a \widetilde{\mathcal{D}} - 117.83 a^2, \nonumber \\[3mm] 
\hspace*{-1 mm}dRa_{SA}^{ALA} &=& \left[2 \upi ^2 - \frac{320}{39} \right] {\widetilde{\mathcal{D}}}^2 + \left[ \frac{9 \upi ^4}{8} - \frac{25 \upi ^2}{4} + \frac{64}{3} \right]  a \widetilde{\mathcal{D}}  
              - \left[ \frac{27 \upi ^4}{16} - \frac{61 \upi ^2}{8} + \frac{544}{39} \right] a^2, \label{dRaGPALA} \\
&& \simeq 11.53 \widetilde{\mathcal{D}}^2 + 69.23 a \widetilde{\mathcal{D}} - 103.07 a^2. \nonumber 
\end{eqnarray} 

The eigenmode odd contribution $\epsilon$ obtained from the Chebyshev analysis is compared to that obtained from the two-modes analysis on Fig.~\ref{GPalineps} and for an ideal gas. As experimentally, it is much easier to impose a large temperature gradient than large compressible effects, we first consider the case of a negligible dissipation number ($\mathcal{D} = 10^{-8}$). Exact and approximate eigenmode odd contributions are very similar throughout the whole range of $a$ (between $0$ and $2$).
Figure \ref{GPalin} shows how the critical Rayleigh number depends on the temperature ratio, $r$, imposed between the bottom and the top. The Boussinesq value $27 \upi ^4 / 4$ is obtained in the limit $a=0$ (corresponding to a unity temperature ratio $r=1$). Increasing $r$ causes a decrease in the value of the superadiabatic critical Rayleigh number $Ra_{SA}^x$. The approximate analysis (\ref{dRaGPx}) with two eigenmodes ($\cos (\upi z)$ and $\sin (2 \upi z )$) fits the numerical solution very well up to $a=1.5$ (corresponding to $r=7$). With a negligible $\mathcal{D}$, the quasi-Boussinesq approximation is identical to the exact analysis. The quasi-ALA approximation results are also plotted on Fig.~\ref{GPalin}, although this approximation is clearly not best at small $\mathcal{D}$. Again, the quadratic two-modes approximation is very good for small values of $a$. The results on Fig.~\ref{GPalin} are independent of the ratio of heat capacities $\gamma$ as can be seen also on the two-modes approximations (\ref{dRaGPx}), (\ref{dRaGPB}) and (\ref{dRaGPALA}). 

\begin{figure}
\begin{center}
\includegraphics[width=12 cm, keepaspectratio]{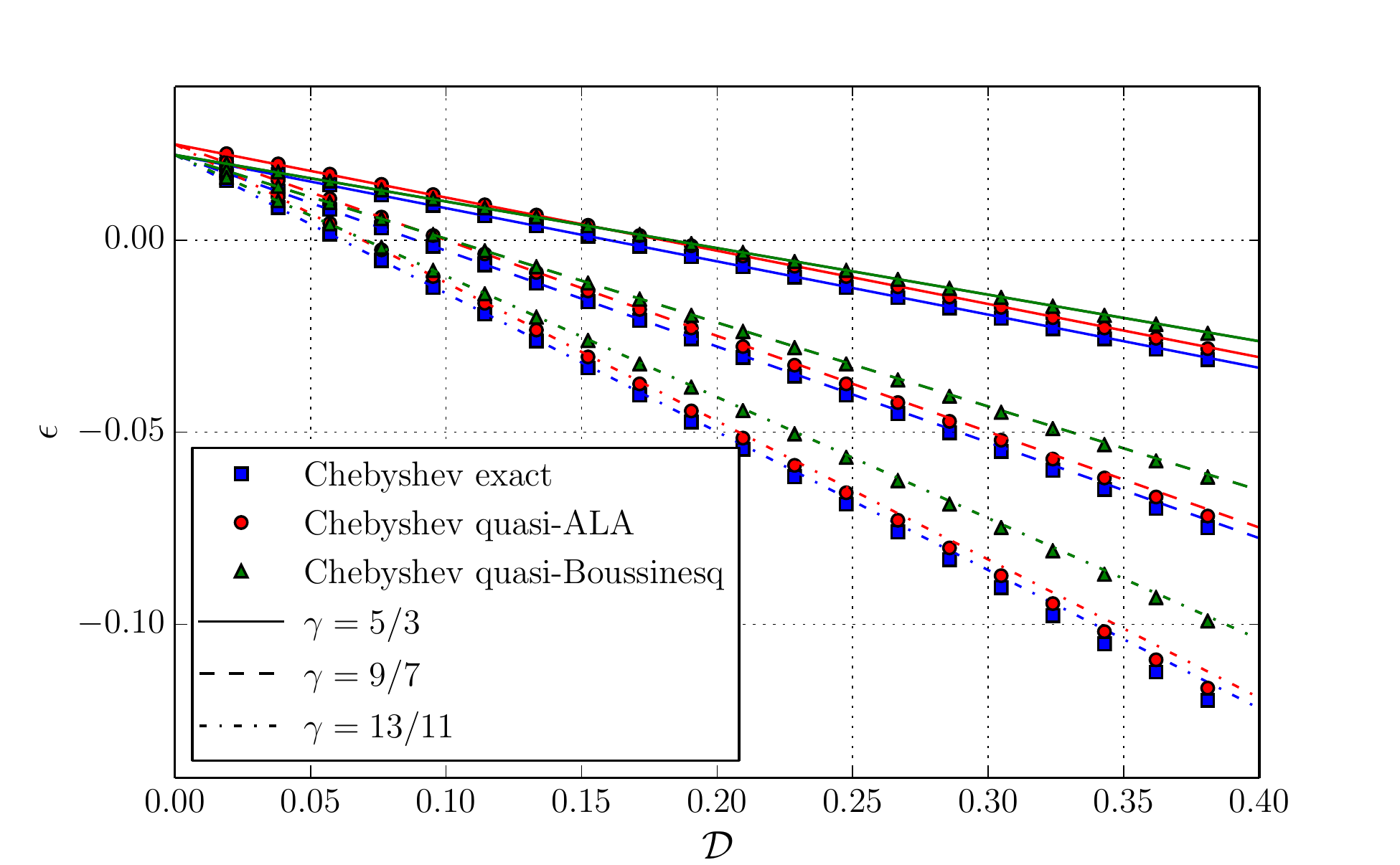}
\caption{Asymmetrical contribution $\epsilon$ of the $\sin (2 \upi z) $ mode to the critical eigenmode, for an ideal gas as a function of the dissipation number $\mathcal{D}$, for a fixed temperature gradient $a=0.4$ (corresponding to a temperature ratio $r=1.5$). The labels Chebyshev exact, quasi-ALA and quasi-Boussinesq correspond to numerical solutions obtained using the Chebyshev collocation eigenvalue calculations described in section \ref{eigenvalue}, for the exact equations, quasi-ALA and Boussinesq models respectively. The lines are the approximate two-modes analytical solutions described in section \ref{estimate}. Solid, dashed and dash-dot lines correspond to three different values of the heat capacity ratio $\gamma = 5/3$, $9/7$ and  $13/11$ respectively.}
\label{GPDlineps}
\end{center}
\end{figure}

\begin{figure}
\begin{center}
\includegraphics[width=12 cm, keepaspectratio]{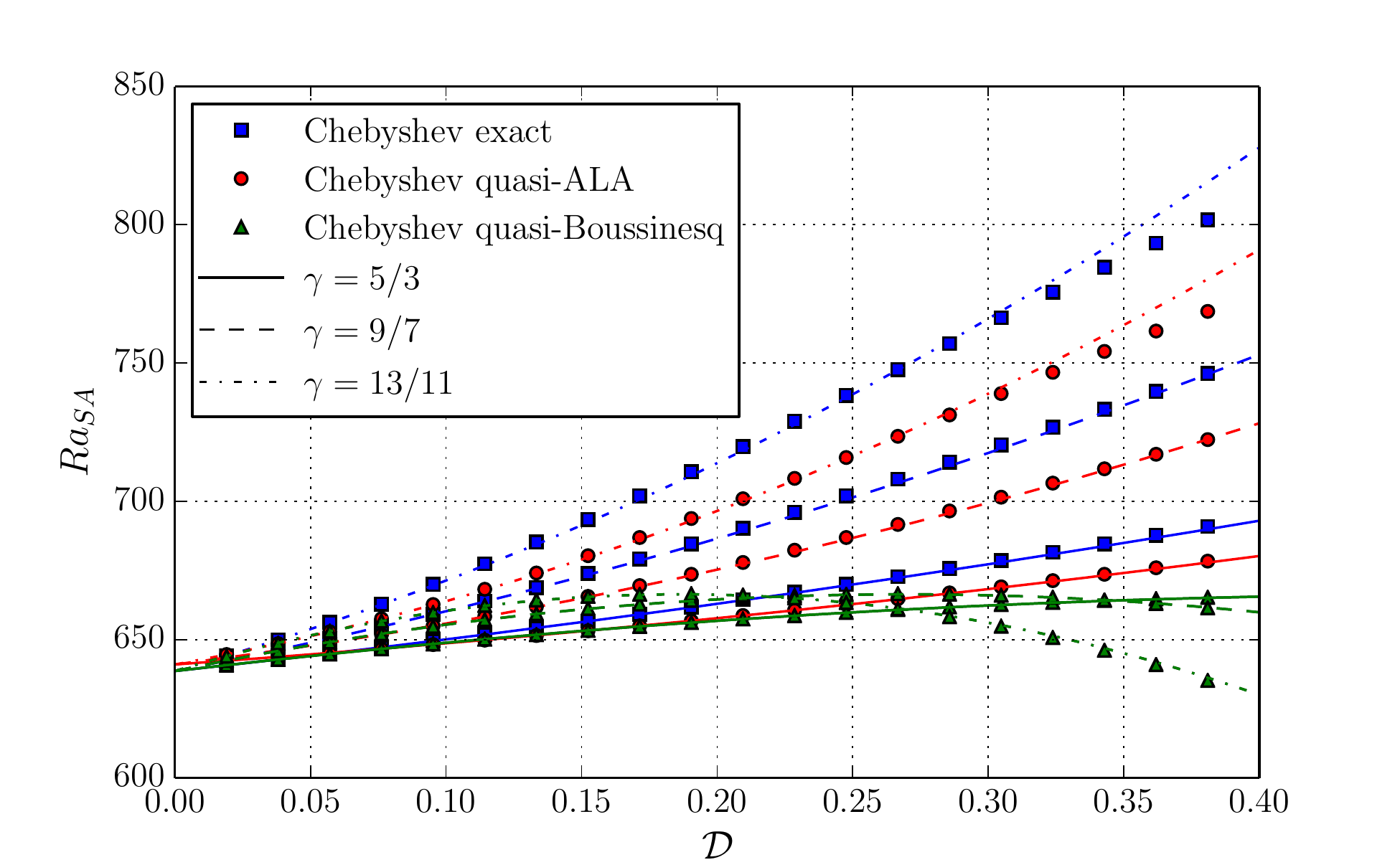}
\caption{Linear stability critical threshold for the Rayleigh number for an ideal gas as a function of the dissipation number $\mathcal{D}$, for a fixed temperature gradient $a=0.4$ (corresponding to a temperature ratio $r=1.5$).  The labels and linestyles correspond to that of figure \ref{GPDlineps}.}
\label{GPDlin}
\end{center}
\end{figure}

Figures \ref{GPDlineps} and \ref{GPDlin} show how the asymmetrical contribution $\epsilon$ and the critical Rayleigh number depend on the dissipation number $\mathcal{D}$ for a fixed value of $a=0.4$. The maximum value for $\mathcal{D}$ is $0.4$ so that superadiabaticity is ensured: for an ideal gas EoS, this happens exactly when  $\mathcal{D} < a$, since the adiabatic gradient is uniform $\mathrm{d} T_a / \mathrm{d} z = - \mathcal{D}$. At small $\mathcal{D}$, the critical Rayleigh numbers increase with $\mathcal{D}$ and that tendency is enhanced as $\gamma$ becomes closer to unity. We can see on Fig.~\ref{GPDlin} that the two-modes results (\ref{dRaGPx}), (\ref{dRaGPB}) and (\ref{dRaGPALA}) are in excellent agreement with the Chebyshev calculations except for the largest values of $\mathcal{D}$.

\begin{figure}
\begin{center}
\includegraphics[width=12 cm, keepaspectratio]{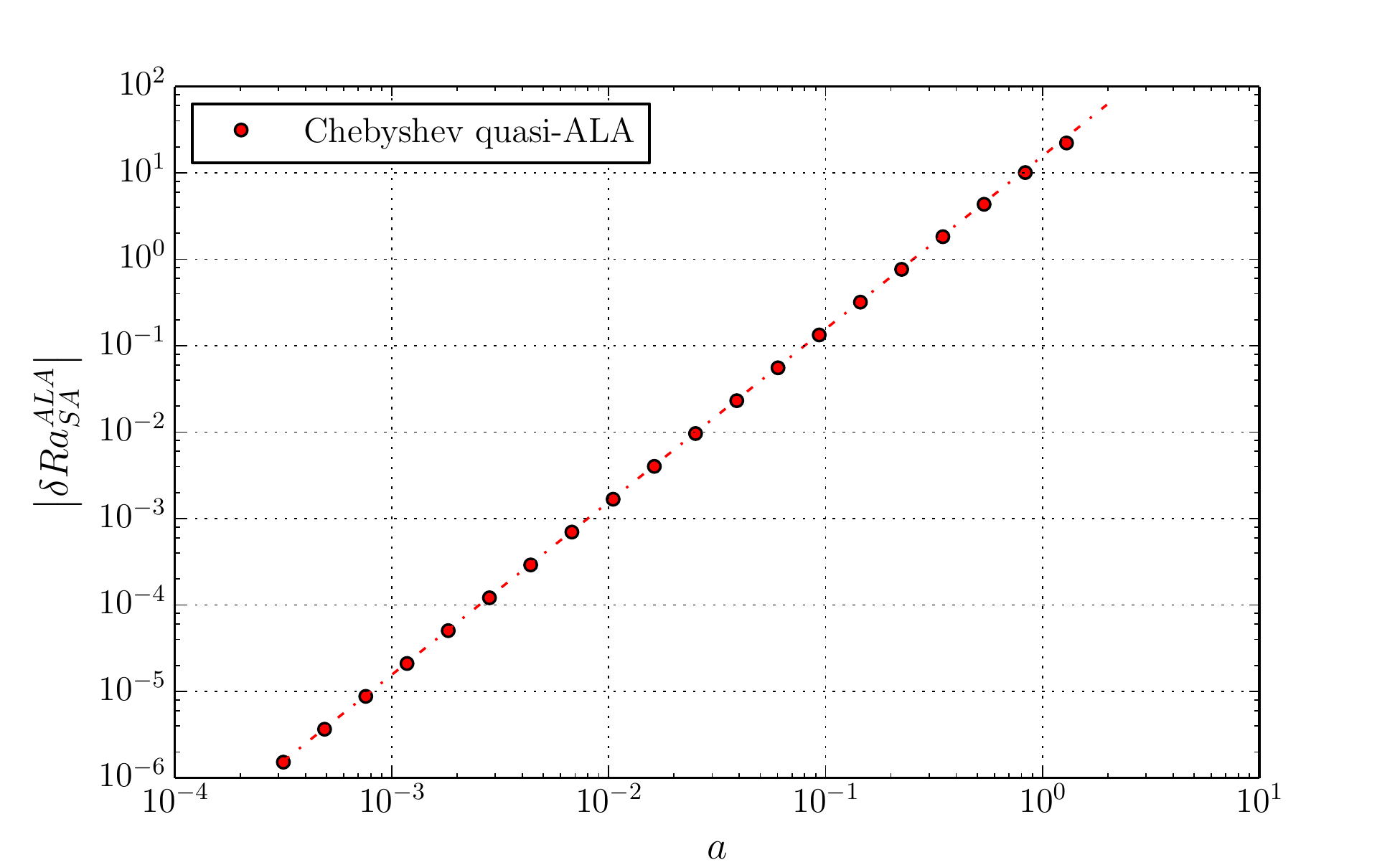}
\caption{Absolute difference between the ALA approximation critical Rayleigh number and the exact critical Rayleigh number, for an ideal gas as a function of $a$, for $\mathcal{D} = 10^{-8}$ and three values of the ratio of heat capacities, $\gamma = 5/3$, $9/7$ and $13/11$, The results using these
three values are undistinguishable as expected from the approximated solutions (\ref{deltaRaGPB}) and (\ref{deltaRaGPALA}). }
\label{GPalog}
\end{center}
\end{figure}

\begin{figure}
\begin{center}
\includegraphics[width=12 cm, keepaspectratio]{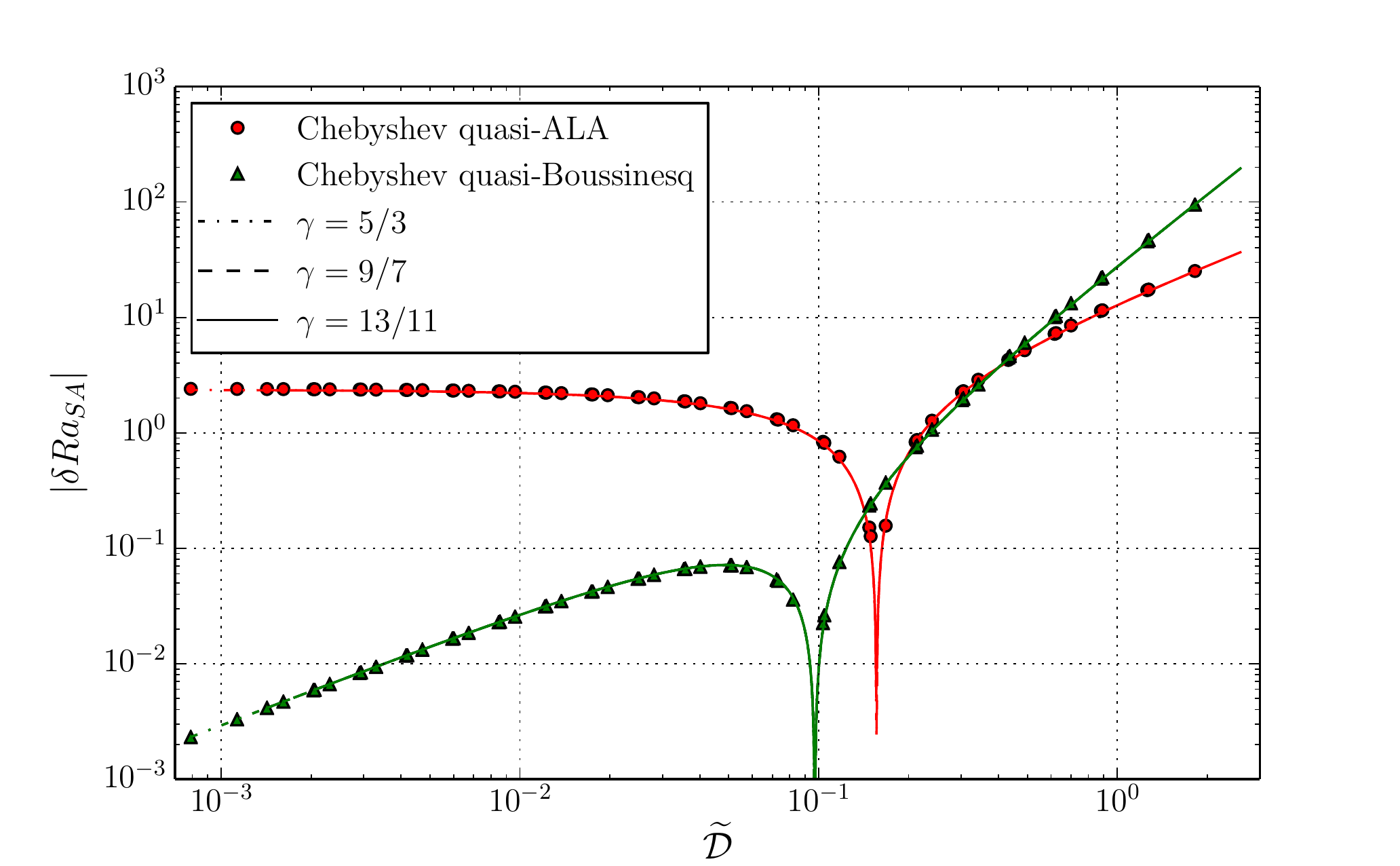}
\caption{Absolute difference between the 'Boussinesq' approximation critical Rayleigh number and the exact critical Rayleigh number and absolute difference between the ALA and exact Rayleigh numbers, for an ideal gas as a function of ${\widetilde{\mathcal{D}}}$, for $a = 0.4$ and three values of the ratio of heat capacities, $\gamma = 5/3$, $9/7$ and $13/11$. }
\label{GPDloga0v4}
\end{center}
\end{figure}

\begin{figure}
\begin{center}
\includegraphics[width=12 cm, keepaspectratio]{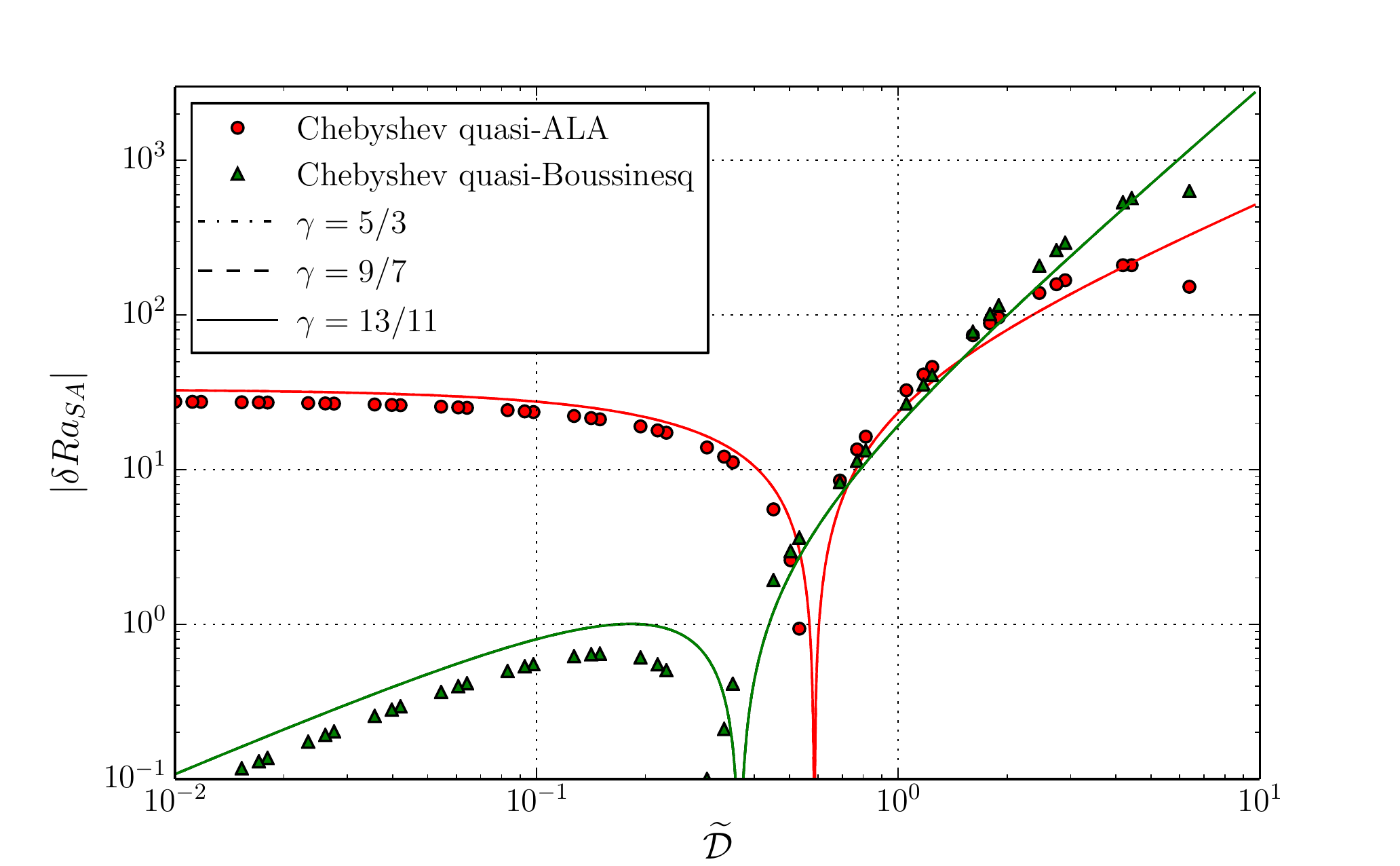}
\caption{Similar to Fig.~\ref{GPDloga0v4}, but with a temperature ratio $r=7$ ($a=1.5$) instead of $r=1.5$ ($a=0.4$). }
\label{GPDloga1v5}
\end{center}
\end{figure}

We shall now consider the results from a different point of view: instead of looking at the Rayleigh numbers dependence, we shall plot the differences between the critical Rayleigh numbers of the quasi-Boussinesq and exact models and between the quasi-ALA and exact models: $\delta Ra_{SA}^B = Ra_{SA}^B - Ra_{SA}^x$ and $\delta Ra_{SA}^{ALA} = Ra_{SA}^{ALA} - Ra_{SA}^x$. From (\ref{dRaGPx}), (\ref{dRaGPB}) and (\ref{dRaGPALA}), we can extract the two-modes approximations for these differences:
\begin{eqnarray}
\delta Ra_{SA}^B &=& - \frac{6 \upi ^2 + 32}{3} \widetilde{\mathcal{D}}^2 - \frac{13 \upi ^2 - 512}{52} a \widetilde{\mathcal{D}} \simeq - 30.41 \widetilde{\mathcal{D}}^2 + 7.38 a \widetilde{\mathcal{D}}  , \label{deltaRaGPB} \\
\delta Ra_{SA}^{ALA} &=& - \frac{78 \upi ^2 + 704}{39} a \widetilde{\mathcal{D}} - \frac{13 \upi ^2 - 896}{52} a^2 \simeq - 37.79 a \widetilde{\mathcal{D}} + 14.76 a^2 . \label{deltaRaGPALA}
\end{eqnarray}
Plotting these differences provides an assessment of the quasi-Boussinesq and quasi-ALA models. Moreover, as we are interested in evaluating small departures from the exact model, we decide to plot the absolute value of these differences in logarithmic coordinates. Figure \ref{GPalog} shows the difference between the quasi-ALA approximation and the exact models, for $\mathcal{D}=10^{-8}$, as a function of $a$. This difference is quadratic in $a$, in agreement with (\ref{deltaRaGPALA}). 

\begin{figure}
\begin{center}
\includegraphics[width=12 cm, keepaspectratio]{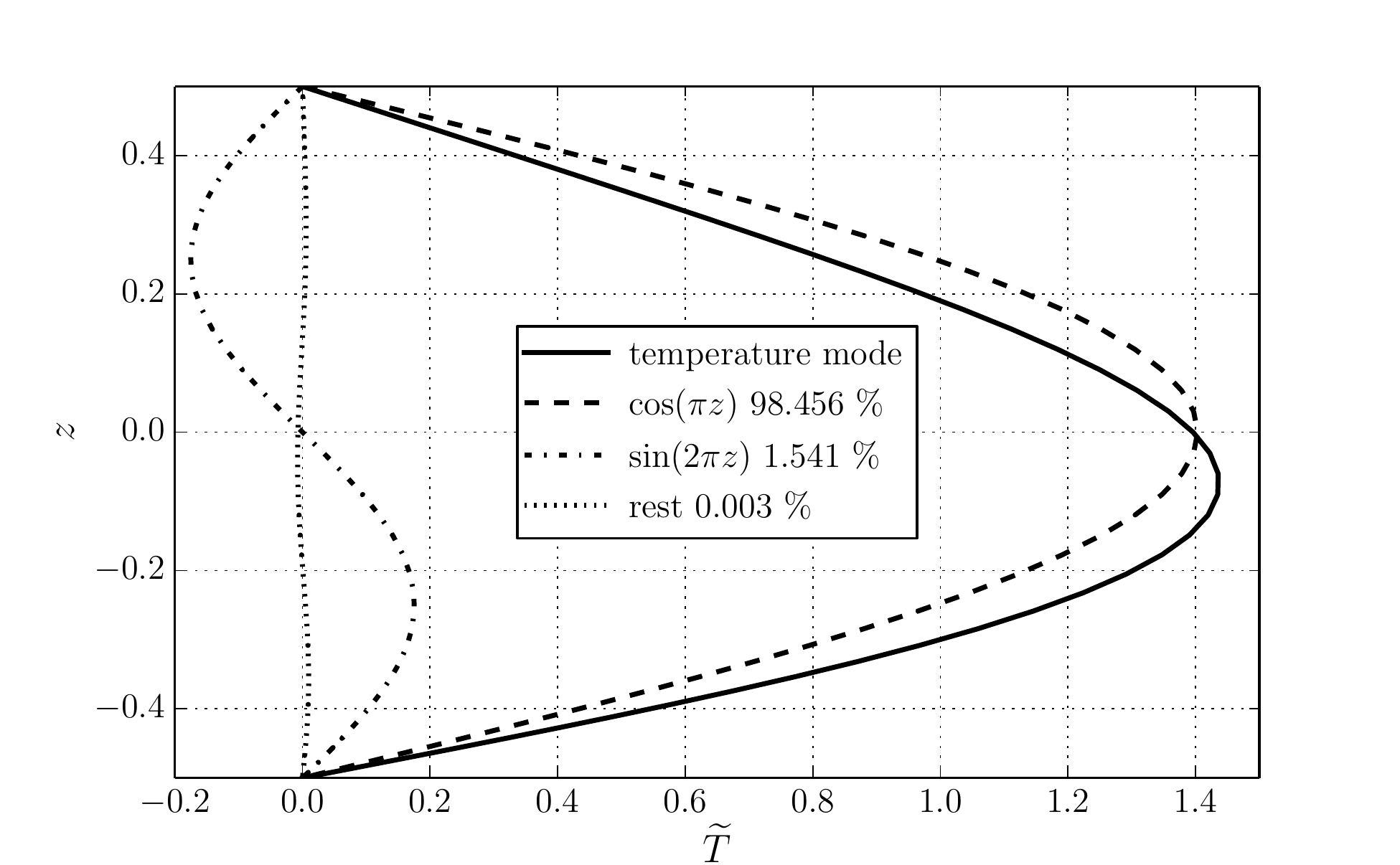}
\caption{Temperature eigenmode, at the critical threshold for an ideal gas of $\gamma = 5/3$, $a=1.5$ (equivalently $r=7$), $\mathcal{D}=1.3$. Its $\cos (\upi z) $ and $\sin (2 \upi z)$ parts represent 98.456~\% and 1.541~\% of its $\mathrm{L}^2$ norm. The rest (0.003~\%) can hardly be distinguished from zero. }
\label{modepr}
\end{center}
\end{figure}

On Fig.~\ref{GPDloga0v4}, we plot the differences between the quasi-ALA and exact models and between the quasi-Boussinesq and exact models, for a constant value of $a=0.4$, as a function of $\widetilde{\mathcal{D}}$. Plotting these differences in terms of $\widetilde{\mathcal{D}}$ instead of $\mathcal{D}$ removes the dependency in $\gamma$ that was observed on Fig.~\ref{GPDlin}. All points collapse on a single curve (for each model quasi-Boussinesq and quasi-ALA) and the two-modes approximations (\ref{deltaRaGPB}) and (\ref{deltaRaGPALA}) are in very good agreement with those obtained through the collocation Chebyshev eigenvalue solutions. These differences are quadratic in $a$ and $\widetilde{\mathcal{D}}$, hence our plot for a constant $a$ and varying $\widetilde{\mathcal{D}}$ can exhibit constant values ($a^2$ contribution), linear regimes ($a \widetilde{\mathcal{D}}$ contribution) or quadratic regimes ($\widetilde{\mathcal{D}}^2$ contributions). Indeed the quasi-Boussinesq model differs first linearly from the exact model at small $\widetilde{\mathcal{D}}$, then quadratically when $\widetilde{\mathcal{D}}$ exceed $a=0.4$. The quasi-ALA model is different from the exact model at $\widetilde{\mathcal{D}}=0$, so that $\delta Ra_{SA}^{ALA}$ is first constant as a function of $\widetilde{\mathcal{D}}$, and is then a linear function of $\widetilde{\mathcal{D}}$ because it has no quadratic contribution (see (\ref{deltaRaGPALA}). A cusp between different regimes indicates simply a change of sign, as we plot the absolute value of the differences: use (\ref{deltaRaGPB}) and (\ref{deltaRaGPALA}) to determine the sign. Figure \ref{GPDloga0v4} shows that the quasi-Boussinesq model is better at small $\widetilde{\mathcal{D}}$ and the quasi-ALA model is better at larger values. For a given value of the dissipation parameter $\mathcal{D}$, decreasing the heat capacity ratio $\gamma$ towards unity has the effect of increasing $\widetilde{\mathcal{D}}$, so that the quasi-ALA model may be better than the quasi-Boussinesq model even for a relatively small dissipation parameter, provided $\gamma$ is close enough to unity. Figure \ref{GPDloga1v5} corresponds to a larger temperature ratio of $r=7$ ($a=1.5$), for which the quadratic two-modes approximation is less good, although still acceptable.  

Figure \ref{modepr} shows an eigenmode, for temperature, corresponding to the critical threshold, obtained for a temperature ratio equal to $7$ and a dissipation number equal to $1.3$. The value of the ratio of heat capacities is $\gamma = 5/3$. The eigenmode is projected on $\cos ( \upi z)$ and $\sin (2 \upi z)$ using the standard $\mathrm{L}^2$ inner product on the interval $-0.5 < z < 0.5$. The $\mathrm{L}^2$ norm contributions of the $\cos 
( \upi z)$ and $\sin (2 \upi z)$ modes are 98.456 \% and 1.541 \% respectively, while the rest is 0.003 \% only. This example is chosen so that the $\sin (2 \upi z)$ contribution can be seen easily, {\it i. e.} with large values of the temperature gradient $a=1.5$ (corresponding to a temperature ratio of $r=7$) and $\mathcal{D}=1.3$. For small values of $a$ and $\mathcal{D}$, suitable for our expansion near $a=0$ and $\mathcal{D}=0$, the modes are closer to a pure $\cos (\upi z )$ function and the  $\sin (2 \upi z)$ function captures even better the difference between the mode and its cosine part. The example on Fig.~\ref{modepr} shows that the choice of the two functions  $\cos 
( \upi z)$ and $\sin (2 \upi z)$ is a good choice for an approximate representation for the eigenmodes. 

\subsection{Murnaghan's EoS}
\label{murnaghan}

Let us now consider an equation of state suitable for condensed matter, liquid or solid, proposed by \citet{murnaghan} 
with a temperature dependence appropriate for models of solid state planetary interiors \citep{ricardTOG}. This equation of state can be written as
\begin{equation}
 \left( \frac{\rho }{\rho _0} \right) ^n = 1 + \frac{n p}{K_0} - n \alpha _0 ( T - T_0 ) , \label{eos_murnaghan} 
\end{equation}
with $n=3$ or $n=4$ for most solid materials and $K_0$ and $\alpha_0$ are constants. The reference density $\rho _0$ is obtained for the reference temperature $T_0$ and pressure $p=0$ (the reference pressure is irrelevant as only pressure gradients play a role in the dynamical equations).
This equation reproduces the observations that, for liquids and solids, the isothermal incompressibility $K_T = \rho \left. \partial p / \partial \rho \right| _T $ increases with compression
\begin{equation}
K_T=K_0\left(\frac{\rho}{\rho_0}\right)^n, \label{incomp} 
\end{equation}
and that the coefficient of thermal expansion diminishes with compression
\begin{equation}
\alpha=\alpha_0\left(\frac{\rho_0}{\rho}\right)^n. \label{alph} 
\end{equation}

We also need to derive the heat capacity from the equation of state. 
The thermodynamic relation $\left. \partial c_v / \partial v \right| _T = T\left. \partial^2 p / \partial T^2 \right| _\nu$ (where $\nu$ is the specific volume $1/\rho$) indicates that
for a solid following the equation (\ref{eos_murnaghan}), $c_v$
is not a function of $\rho$ as the pressure is linear in $T$ for a given density. So $c_v$ can only be a function of temperature $T$: any choice is valid in principle. We make the choice of a constant $c_{v0}$ which is in agreement with the Dulong and Petit rule for condensed matter. It follows then from Mayer's relation (\ref{mayerx}) that
\begin{equation}
c_{p} = c_{v0} + \alpha_0 K_0 T\frac{\rho_0^n}{\rho^{n+1}}. \label{cpMurnx}
\end{equation}

Notation (\ref{notations}) is still in use 
Using our dimensional scales, Murnaghan's EoS takes therefore the following dimensionless form
\begin{equation}
 {\rho}^n = 1 + \hat{\alpha} \widetilde{\mathcal{D}} n p - n \hat{\alpha} ( T - 1 ). \label{adim_murnaghan} 
\end{equation}
The base profile is determined as follows. The temperature base profile is independent of the EoS, hence equation (\ref{basetemperature}) is still valid. The derivative of (\ref{adim_murnaghan}) and the hydrostatic equation $\mathrm{d} p_b / \mathrm{d} z = - \rho _b$ lead to a differential equation for the base density profile $\rho _b$
\begin{equation}
\frac{\mathrm{d} \rho _b}{\mathrm{d} z} = - \hat{\alpha} \widetilde{\mathcal{D}} \rho _b^{2-n} + \hat{\alpha} a \rho _b^{1-n} .  \label{baseMurnrho}
\end{equation}
This equation is integrated numerically, under the condition that $\rho _b = 1$ at $z=0$ in accordance with our choice for the dimensional reference density $\rho _0$.
The base pressure profile $p_b$ is then obtained from the equation of state (\ref{adim_murnaghan}).

In the resolution of the eigenvalue problem (\ref{cont_eig2}), (\ref{NSx_eig2}), (\ref{NSz_eig2}) and (\ref{entropy_eig2}), we also need to determine the base profile for the dimensionless specific heat capacity $c_{pb}$ and expansivity $\alpha _b $. After nondimensionalisation (\ref{alph}) writes
\begin{equation}
\alpha _b = \rho _b^{-n}, \label{alphabTb}
\end{equation}
and (\ref{cpMurnx}),
\begin{equation}
c_{pb} = \frac{1}{\gamma _0} + \frac{\gamma _0 -1 }{\gamma _0} T_b \rho _b^{-1-n}. \label{cpMurn}
\end{equation}

We also need to compute some quantities for the two-modes analysis. The third derivative of the adiabatic temperature profile, at $z=0$, is obtained from the expression of the adiabatic gradient, $\mathrm{d} T_a / \mathrm{d} z = - \mathcal{D} \alpha_a T_a / c_{pa} $ (see (\ref{adiabgrad})), the equation of state (\ref{adim_murnaghan}) and the expression for $c_p$ above, by successive derivatives 
\begin{eqnarray}
\frac{\mathrm{d}^3 T_a}{\mathrm{d} z^3} &=&  \left[ 2 \gamma _0  - 3 - 3 \hat{\alpha}^2 \right] \frac{{\mathcal{D}}^3}{\gamma _0^2} - \left[ (3 \gamma _0 -6 ) n - \gamma _0^2 + 7 \gamma _0 - 6 \right] \frac{\hat{\alpha} \widehat{\mathcal{D}} {\mathcal{D}}^2}{\gamma _0^3} \nonumber \\
&& + \left[ (\gamma _0 - 3 ) n^2 + (7 \gamma _0 - 6 ) n \right] \frac{\hat{\alpha}^2 {\widehat{\mathcal{D}}}^2 \mathcal{D}}{\gamma _0^4}  . \label{dTadztrois}
\end{eqnarray}
The function $f(z)$ and its derivatives at $z=0$ can then be determined using (\ref{dTadztrois}),  (\ref{alphabTb}) and (\ref{cpMurn}) up to degree $2$ in $a$ and $\mathcal{D}$, like other quanities.

\begin{figure}
\begin{center}
\includegraphics[width=12 cm, keepaspectratio]{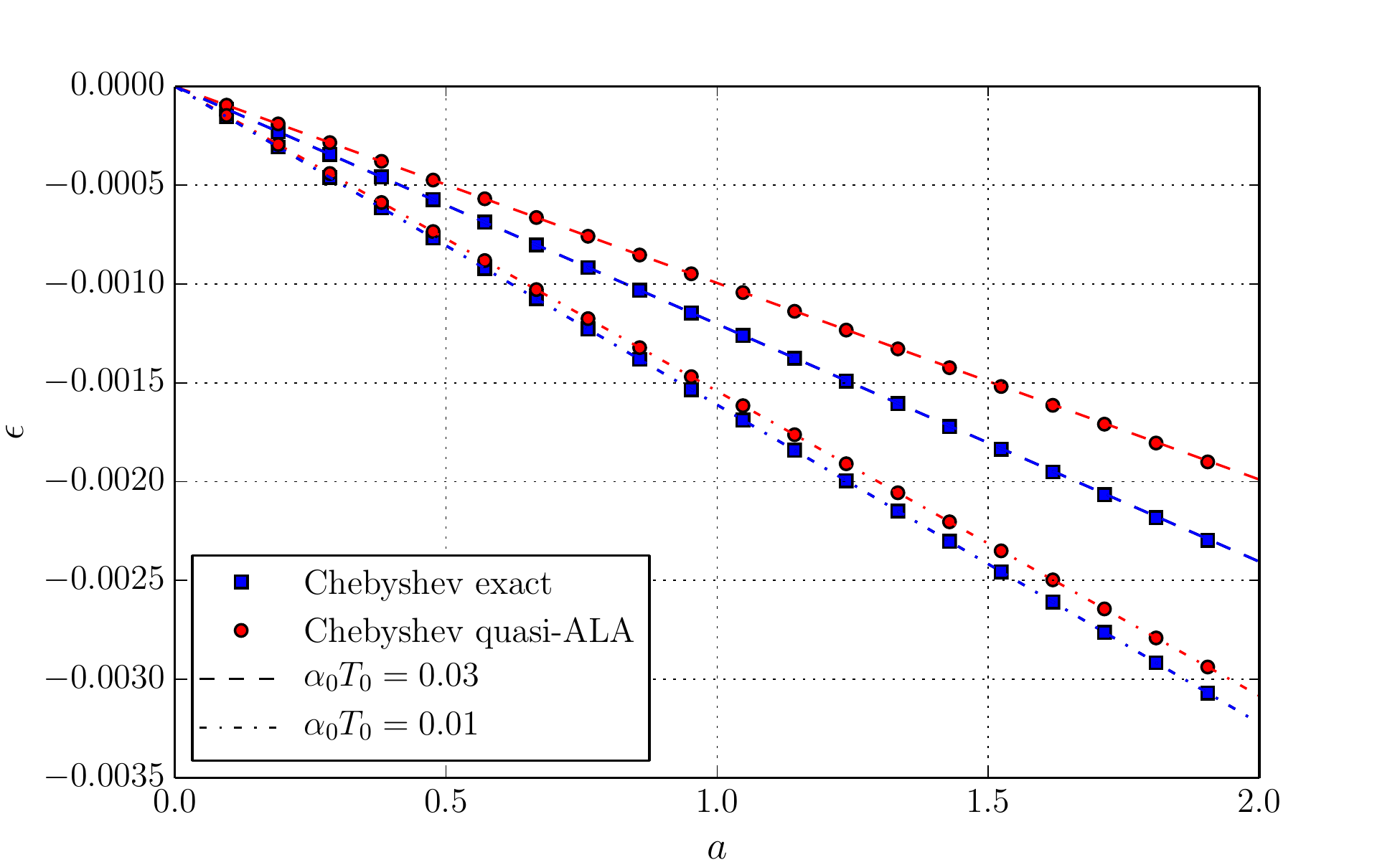}
\caption{Asymmetrical contribution of the $\sin ( 2 \upi z)$ mode to the critical eigenmode, for a Murnaghan EoS as a function of the temperature gradient $a$ of the base linear solution, for a negligible ${\cal{D}} = 10^{-8}$. The label 'Chebyshev exact' denotes the numerical solution of the exact model using a Chebyshev expansion (usually 17 polynomials), while the label 'Chebyshev quasi-ALA' corresponds to the solutions of the quasi-ALA model. The dashed and dash-dot lines correspond to the approximate two-modes analytical solutions for the exact and quasi-ALA models, at $\hat{\alpha} = \alpha _0 T_0 = 0.03$ and $0.01$ respectively. The ratio of heat capacities and integer $n$ in the equation of state (\ref{adim_murnaghan}) are kept constant $\gamma _0 =1.03$ and $n=3$. Note that when the dissipation number is negligible, the quasi-Boussinesq model and the exact model coincide.}
\label{Murnalineps}
\end{center}
\end{figure}

\begin{figure}
\begin{center}
\includegraphics[width=12 cm, keepaspectratio]{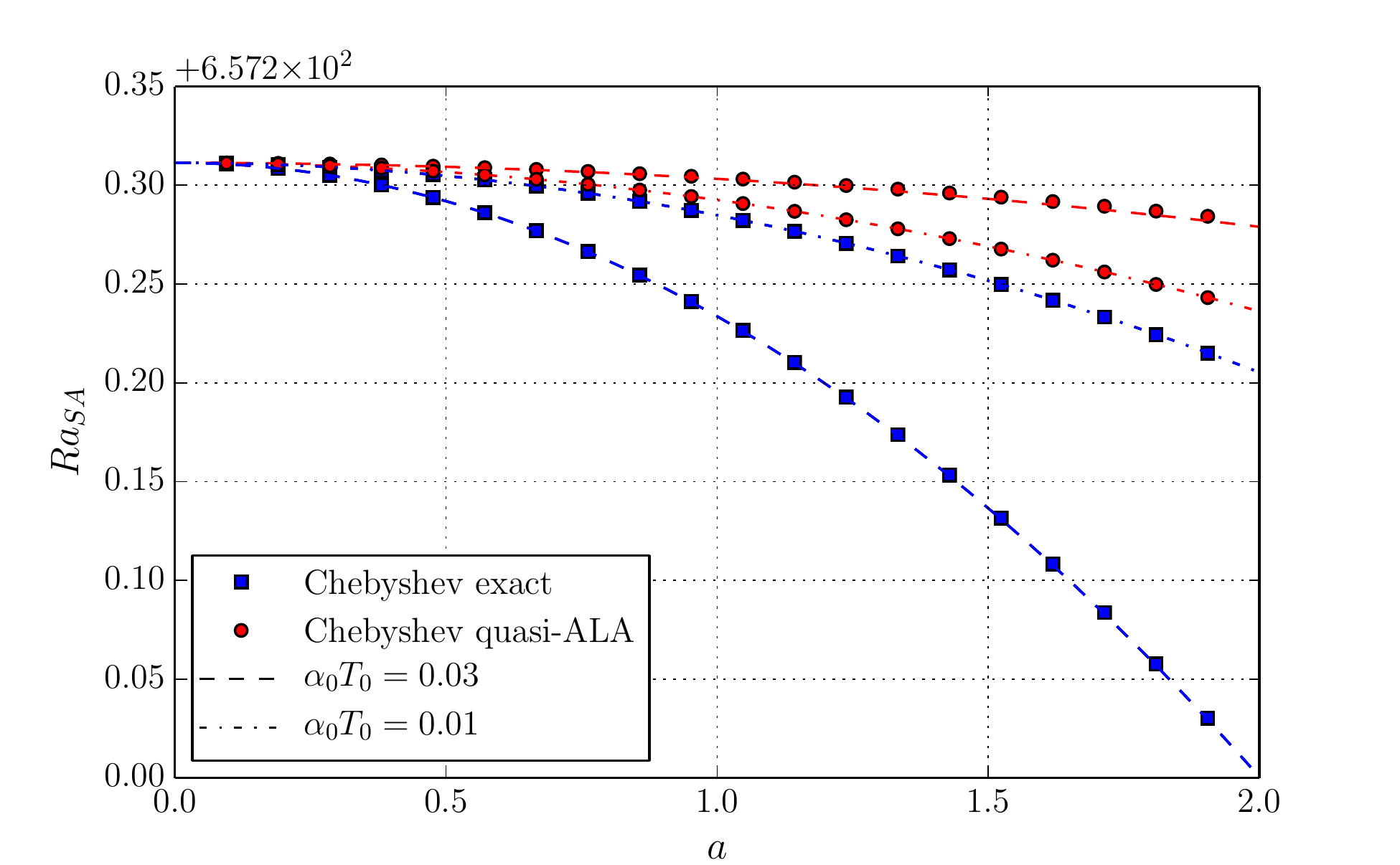}
\caption{Linear stability critical threshold for the Rayleigh number for a Murnaghan EoS as a function of the temperature gradient $a$ of the base linear solution, for a negligible ${\cal{D}} = 10^{-8}$. The labels are similar to those of figure \ref{Murnalineps}.} 
\label{Murnalin}
\end{center}
\end{figure}

\begin{figure}
\begin{center}
\includegraphics[width=12 cm, keepaspectratio]{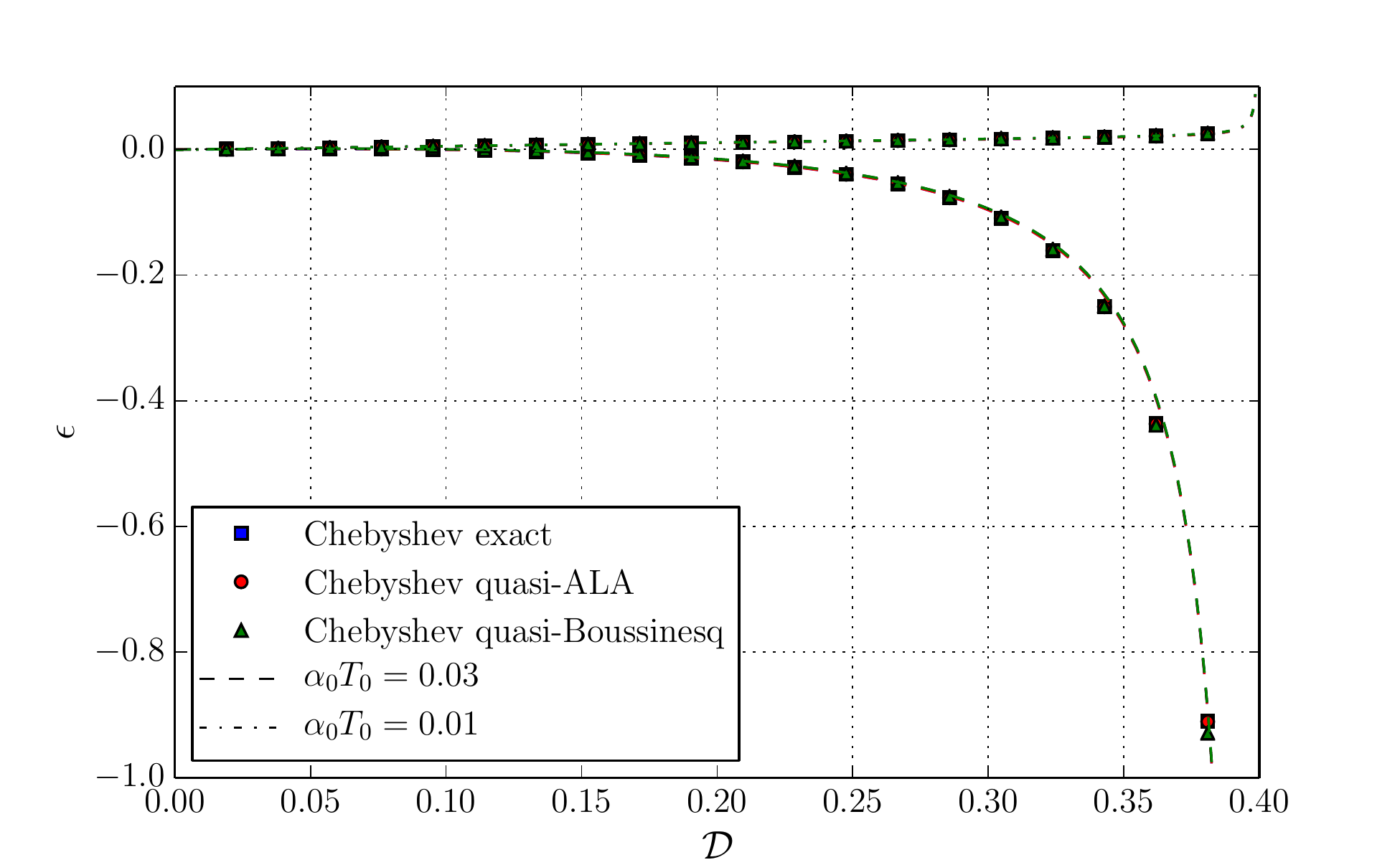}
\caption{Asymmetrical contribution of the $\sin ( 2 \upi z)$ mode to the critical eigenmode, for the Rayleigh number for a Murnaghan EoS as a function of the dissipation number $\mathcal{D}$, for a fixed temperature gradient $a=0.4$ (corresponding to a temperature ratio $r=1.5$). The labels Chebyshev exact, quasi-ALA and quasi-Boussinesq correspond to numerical solutions obtained using the Chebyshev collocation eigenvalue calculations described in section \ref{eigenvalue}, for the exact equations, quasi-ALA and Boussinesq approximations respectively. The lines are the analytical two-modes solutions described in section \ref{estimate}. Dashed and dash-dot lines correspond to two different values for the product of the expansion coefficient and temperature at $z=0$, $\hat{\alpha} = \alpha _0 T_0 = 0.03$ and $0.01$, while the heat capacity ratio at $z=0$ is kept constant $\gamma _0 = 1.03$ and $n=3$.}
\label{MurnDlineps}
\end{center}
\end{figure}

\begin{figure}
\begin{center}
\includegraphics[width=12 cm, keepaspectratio]{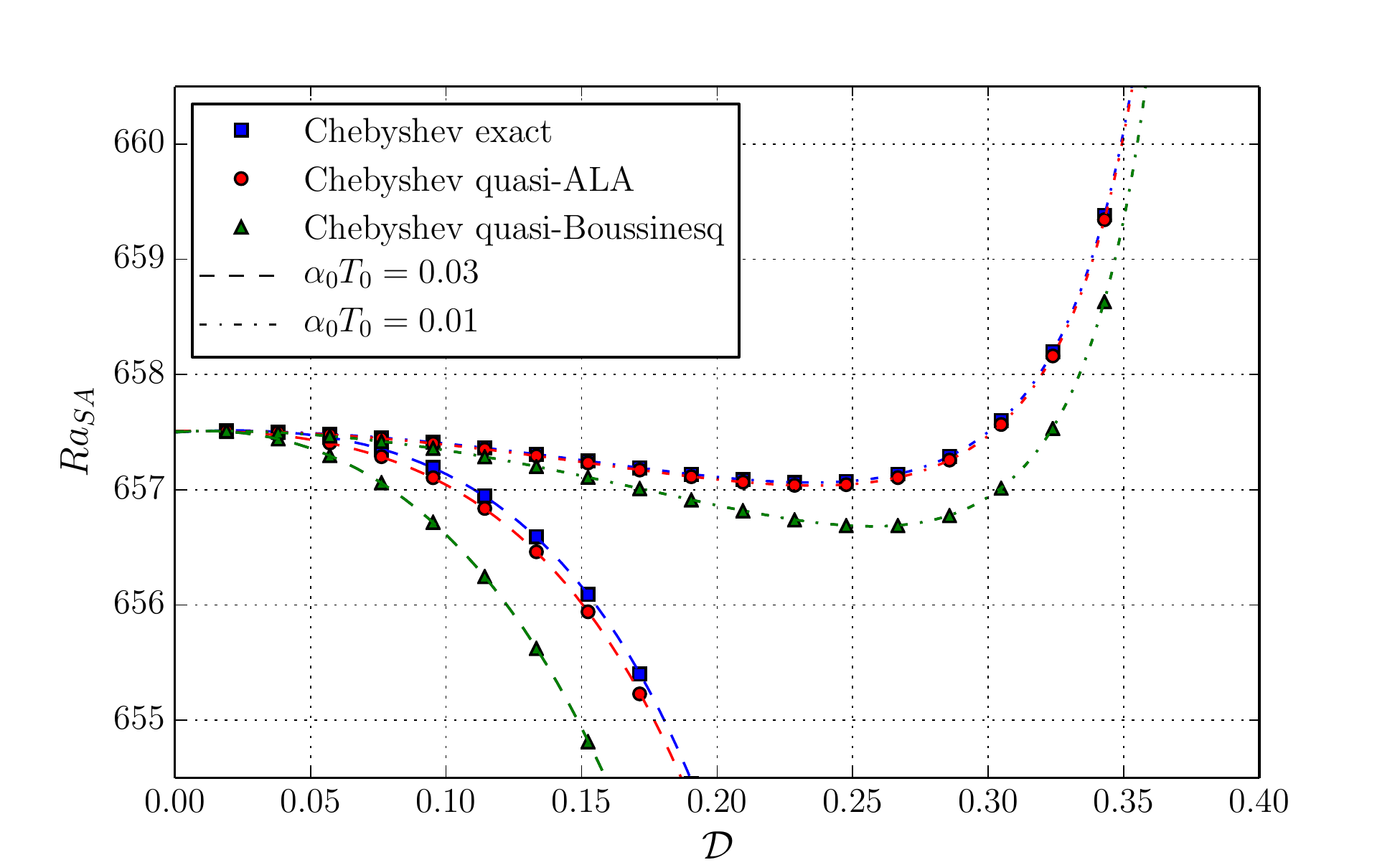}
\caption{Linear stability critical threshold for the Rayleigh number for a Murnaghan EoS as a function of the dissipation number $\mathcal{D}$, for a fixed temperature gradient $a=0.4$ (corresponding to a temperature ratio $r=1.5$). The labels are defined on Fig.~\ref{MurnDlineps}.}
\label{MurnDlin}
\end{center}
\end{figure}

\begin{figure}
\begin{center}
\includegraphics[width=12 cm, keepaspectratio]{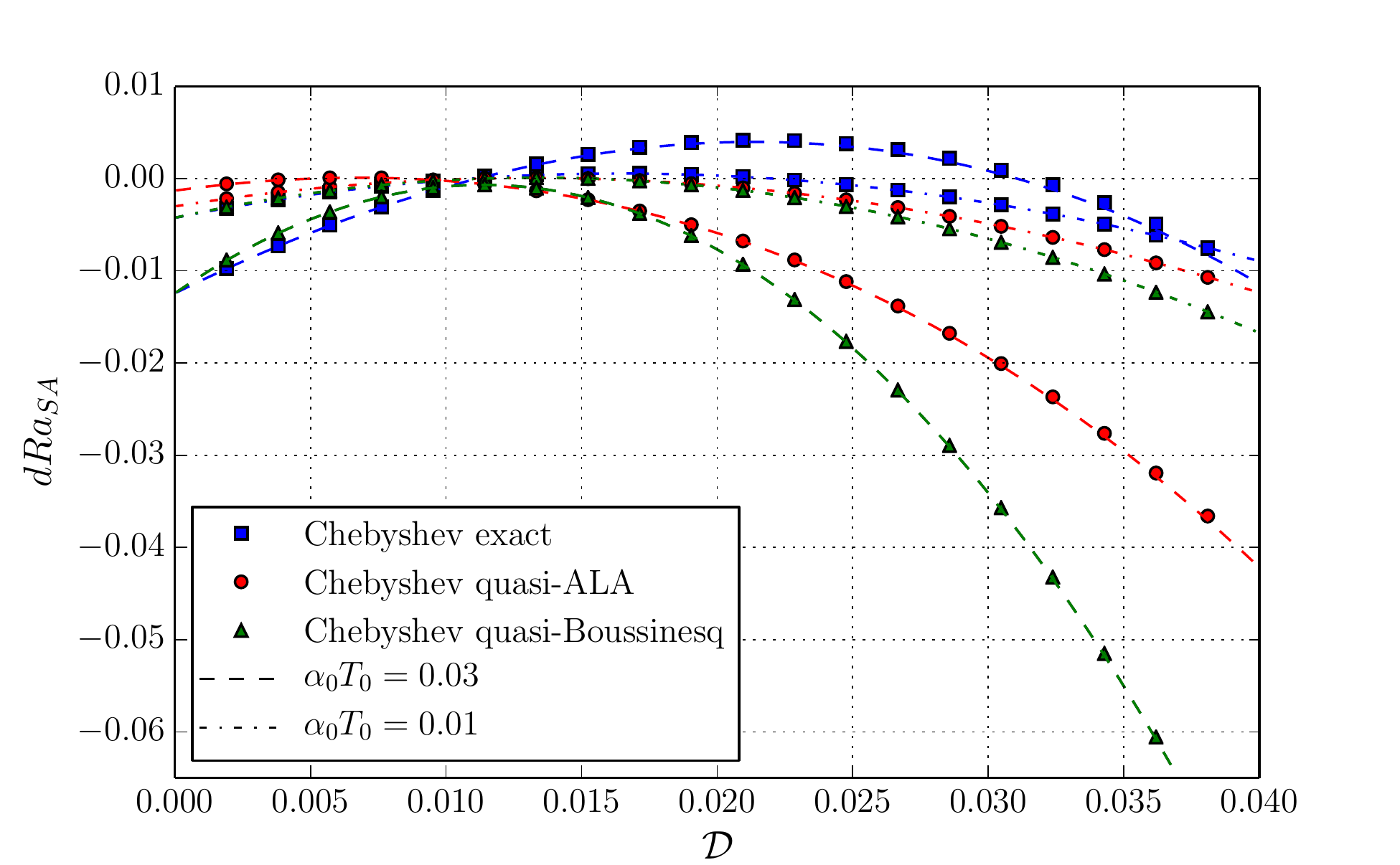}
\caption{Same as Fig.~\ref{MurnDlin} with a close-up around small values of $\mathcal{D}$, between $0$ and $0.04$. The difference $dRa_{SA}=Ra_{SA}-27 \upi ^4 / 4$ is plotted instead of $Ra_{SA}$. }
\label{MurnDlin_ter}
\end{center}
\end{figure}

On table \ref{tableMurnaghan}, we show the expression of all quantities needed for the approximate two-mode analysis. From these expressions, using equations (\ref{dispCOS}) and (\ref{dispSIN}), we obtain the approximate expressions for the critical Rayleigh numbers with and without the effect of compressibility for the small disturbances. 

\begin{table*}
\begin{center}
\begin{tabular}{@{}ll@{}}
\toprule expression \hspace*{1 cm} & value   \\
\hline
$\left. f \right| _0 $ &  $-1 - \left[ 2 \gamma _0  - 3 - 3 \hat{\alpha}^2 \right] \frac{{\mathcal{D}}^3}{24 \gamma _0^2 (a-\mathcal{D})} + \left[ (3 \gamma _0 -6 ) n - \gamma _0^2 + 7 \gamma _0 - 6 \right] \frac{\hat{\alpha} \widehat{\mathcal{D}} {\mathcal{D}}^2}{24 \gamma _0^3 (a-\mathcal{D})} $ \\ 
& $ - \left[ (\gamma _0 - 3 ) n^2 + (7 \gamma _0 - 6 ) n \right] \frac{\hat{\alpha}^2 {\widehat{\mathcal{D}}}^2 \mathcal{D}}{24 \gamma _0^4 (a-\mathcal{D})} $ \\ 
$\left. \frac{\mathrm{d} f }{\mathrm{d} z}  \right| _0$  &  
$\frac{a-\widetilde{\mathcal{D}}}{a-\mathcal{D} } \left(\hat{\alpha} (n-1){\mathcal{D}} -((n\hat{\alpha}+1)(1-\frac{1}{\gamma _0})-\frac{\hat\alpha}{\gamma _0})a\right) $ \\
$\left. \frac{\mathrm{d}^2 f }{\mathrm{d} z^2}  \right| _0$  & 
{$ \frac{a(a- \widetilde{\mathcal{D}})}{\gamma _0^2 ( \mathcal{D}-a)^2} 
\Big[ \Big(\hat\alpha n(3(2+\hat\alpha)\gamma_0-4(\hat\alpha+1))-2(\gamma_0-1)^2+(\gamma_0-2)\hat\alpha^2+4\hat\alpha(\gamma_0-1)\Big)a^2$}\\
&{$+\Big(\hat\alpha n(-2(5+\hat\alpha)\gamma_0+6+3\hat\alpha)
+(\gamma_0-1)^2(2-\hat\alpha^2)+2\gamma_0\hat\alpha (1-\gamma_0)\Big)a\widetilde{\mathcal{D}}$}\\
&{$+\Big(\hat\alpha n((4-\hat\alpha)\gamma_0+\hat\alpha-2) + \hat\alpha (\hat\alpha+2)(\gamma_0-1)^2\Big)\widetilde{\mathcal{D}}^2+2\hat\alpha n(\hat\alpha n(\gamma_0-1)-\gamma_0^2)(a-\widetilde{\mathcal{D}})^2\Big]$}\\
$\left.  \frac{\rho '_b}{\rho _b} \right| _0 $ & $ \hat{\alpha} (a - \widetilde{\mathcal{D}} ) $ \\
$\left. \frac{\mathrm{d} }{\mathrm{d} z} \frac{\rho '_b}{\rho _b} \right| _0 $  & 
$\hat{\alpha}^2 (a - \widetilde{\mathcal{D}} ) \left(( n  - 1 ) \widetilde{\mathcal{D}} -na \right)$ \\
$\left. \frac{ \partial \rho }{\partial T} \right| _{p0}$ & $ - \hat{\alpha}  $ \\
$\left. \frac{\mathrm{d} }{\mathrm{d} z}  \frac{ \partial \rho }{\partial T} \right| _{p0}  $  & $ (n-1) \hat{\alpha}^2  (a - \widetilde{\mathcal{D}} ) $ \\
$\left. \frac{\mathrm{d}^2 }{\mathrm{d} z^2}  \frac{ \partial \rho }{\partial T} \right| _{p0} $  & 
$(n-1)\hat{\alpha}^3 (a - \widetilde{\mathcal{D}} ) \left(2( n  - 1 ) \widetilde{\mathcal{D}} -(2n-1)a \right) $\\
$\left. \frac{ \partial \rho }{\partial p} \right| _{T0}$ & $ \hat{\alpha} \widetilde{\mathcal{D}} $ \\
$\left. \frac{\mathrm{d} }{\mathrm{d} z}  \frac{ \partial \rho }{\partial p} \right| _{T0}  $ & $ - (n - 1) \hat{\alpha}^2 \widetilde{\mathcal{D}} (a -   \widetilde{\mathcal{D}}  )  $ \\ 
\bottomrule
\end{tabular}
\caption{Coefficients of Taylor expansion of some quantities related to the base flow, for Murnaghan's equation of state, for arbitrary values of $\gamma _0 = c_{p0} / c_{v0}$ and $\hat{\alpha} = \alpha _0 T_0$.} 
\label{tableMurnaghan}
\end{center}
\end{table*}

With table \ref{tableMurnaghan} and the general solutions (\ref{sol_epsilon}) and (\ref{sol_dRaSA}), we have the quadratic departure of the superadiabatic critical Rayleigh number in terms of the parameters $a$ and $\mathcal{D}$. It would actually be too long to display $d Ra_{SA}$ once the quantities in table \ref{tableMurnaghan} are substituted in those general equations. However, it is possible to do so for the $\sin (2 \upi z)$ contributions:  
the coefficients $\epsilon$ (coefficient of $\sin (2 \upi z )$) obtained by the two-modes analysis (\ref{sol_epsilon}), (\ref{sol_delta_epsilonB} and (\ref{sol_delta_epsilonALA}) are the followings, for the exact model, quasi-Boussinesq and quasi-ALA approximations
\begin{eqnarray}
\hspace*{-10 mm}\epsilon ^x &=& \frac{8 \hat{\alpha} \left[  a - \widetilde{\mathcal{D}} \right]}{117 \upi ^2} \left[9 \frac{ (n-1) \mathcal{D} - \left[ ( 1 - \gamma _0^{-1} ) (n+ \hat{\alpha}^{-1} ) - \gamma _0^{-1} \right] a }{ a - \mathcal{D} }  - n - 2) \right]  - \frac{8 \hat{\alpha} \widetilde{\mathcal{D}} }{117 \upi ^2} , \label{epsMurnx} \\
\hspace*{-10 mm}\epsilon ^B &=& \frac{8 \hat{\alpha} \left[ a - \widetilde{\mathcal{D}} \right]}{117 \upi ^2} \left[9 \frac{ (n-1) \mathcal{D} - \left[ ( 1 - \gamma _0^{-1} ) (n+ \hat{\alpha}^{-1} ) - \gamma _0^{-1} \right] a }{a - \mathcal{D} }  - n -2 \right] , \label{epsMurnB} \\
\hspace*{-10 mm}\epsilon ^{ALA} &=& \frac{8 \hat{\alpha} \left[ a - \widetilde{\mathcal{D}} \right]}{117 \upi ^2} \left[9 \frac{ (n-1) \mathcal{D} - \left[ ( 1 - \gamma _0^{-1} ) (n+ \hat{\alpha}^{-1} ) - \gamma _0^{-1} \right] a }{a - \mathcal{D} }  - n - 1  \right] . \label{epsMurnALA}
\end{eqnarray}
Similarly, the differences between critical superadiabatic Rayleigh numbers obtained from the quasi-Boussinesq or quasi-ALA approximations and the exact model (\ref{sol_deltaRaB}) and (\ref{sol_deltaRaALA}) are also short enough to be shown explicitly
\begin{eqnarray}
\hspace*{-5 mm}\delta Ra_{SA}^B &=& \left[ \frac{9 n -1}{4} \upi ^2 - \frac{256 n - 128}{39} \right] \hat{\alpha}^2 a \widetilde{\mathcal{D}} - \left[  \frac{9 n -1}{4} \upi ^2 - \frac{256 n - 416}{39} \right] \hat{\alpha}^2 \widetilde{\mathcal{D}}^2 , \label{deltaRaMurnB} \\
\hspace*{-5 mm}\delta Ra_{SA}^{ALA} &=& \left[ \frac{9 n + 8}{4} \upi ^2 - \frac{256 n - 416}{39}\right] \hat{\alpha}^2 a^2 - \left[ \frac{9 n + 8}{4} \upi ^2 - \frac{256 n - 704}{39} \right] \hat{\alpha}^2 a \widetilde{\mathcal{D}} . \label{deltaRaMurnALA} 
\end{eqnarray}

Figure \ref{Murnalin} shows the dependence of the critical Rayleigh numbers for the exact model and quasi-ALA approximation on the base temperature gradient $a$ for a negligible dissipation parameter $\mathcal{D} = 10^{-8}$ and a constant $\gamma _0 = 1.03$ and $n=3$. These critical Rayleigh numbers are also obtained with the two-modes analysis with an excellent accuracy. Three different values of $\hat{\alpha}$ are considered and we can see that the departure of the critical Rayleigh numbers from $27 \i ^4 / 4$ gets smaller as $\hat{\alpha}$ diminishes. Figure~\ref{MurnDlin} shows the dependence of the critical Rayleigh numbers (exact, quasi-Boussinesq and quasi-ALA) on $\mathcal{D}$ for a fixed value of the base temperature gradient $a=0.4$. The ratio of heat capacities is kept constant $\gamma _0 = 1.03 $ and two values of $\hat{\alpha} = \alpha _0 T_0 = 0.03$ and $0.01$ are considered. The two-modes analysis provides a good fit throughout the whole range of $\mathcal{D}$. A close-up around small values of $\mathcal{D}$ is shown on Fig.~\ref{MurnDlin_ter}.

\begin{figure}
\begin{center}
\includegraphics[width=12 cm, keepaspectratio]{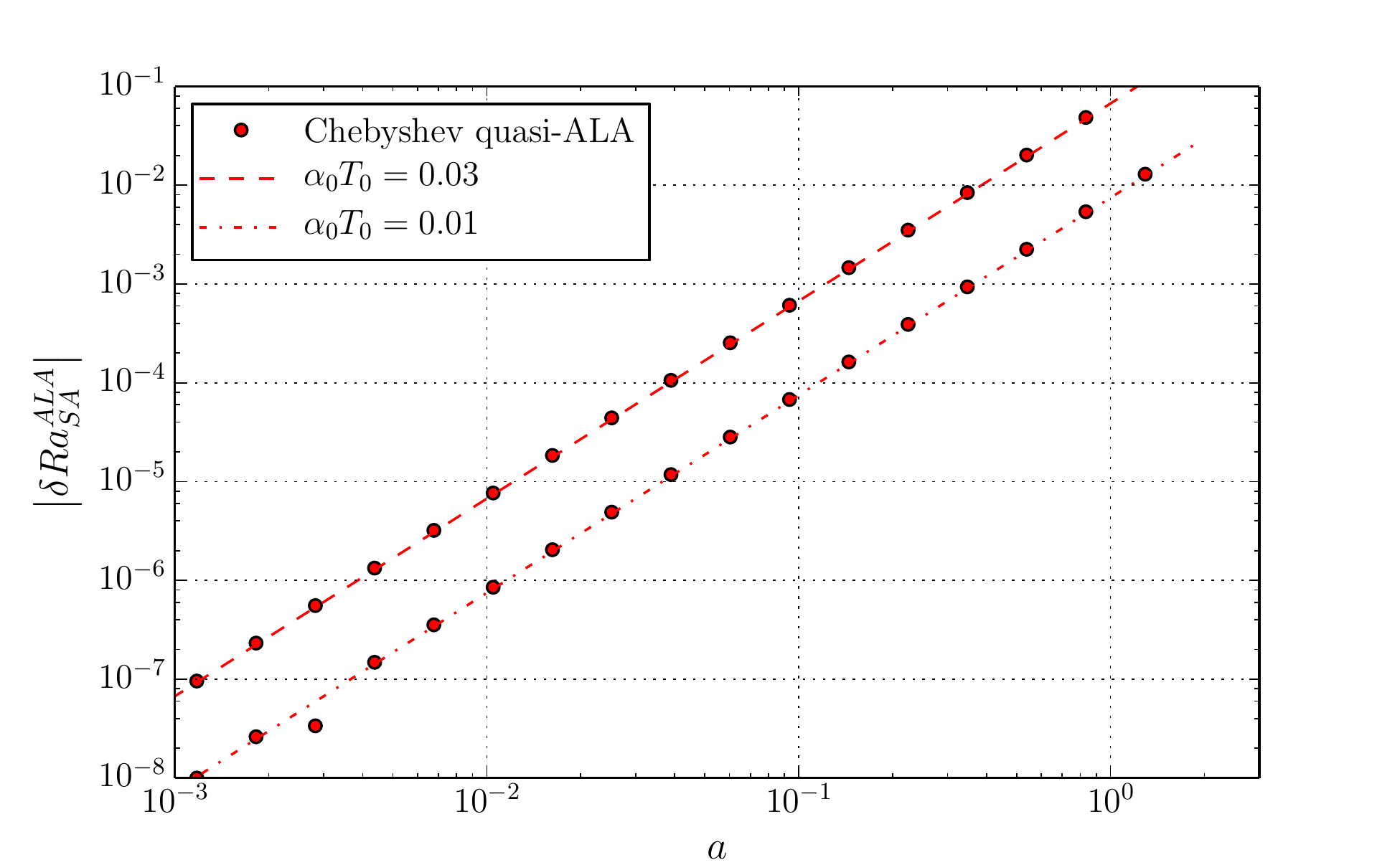}
\caption{Absolute difference between the critical Rayleigh number of the quasi-ALA and exact model, with Murnaghan's equation of state and for a negligible dissipation number equal to $\mathcal{D}=10^{-8}$. The difference is plotted as a function of $a$, for two values of $\alpha _0 T_ 0 = \hat\alpha$ ($0.03$ and $0.01$) and $\gamma _0 = 1.03$. The parameter $n$ in Murnaghan's EoS is equal to $3$ in all cases.}
\label{Murnalog}
\end{center}
\end{figure}

\begin{figure}
\begin{center}
\includegraphics[width=12 cm, keepaspectratio]{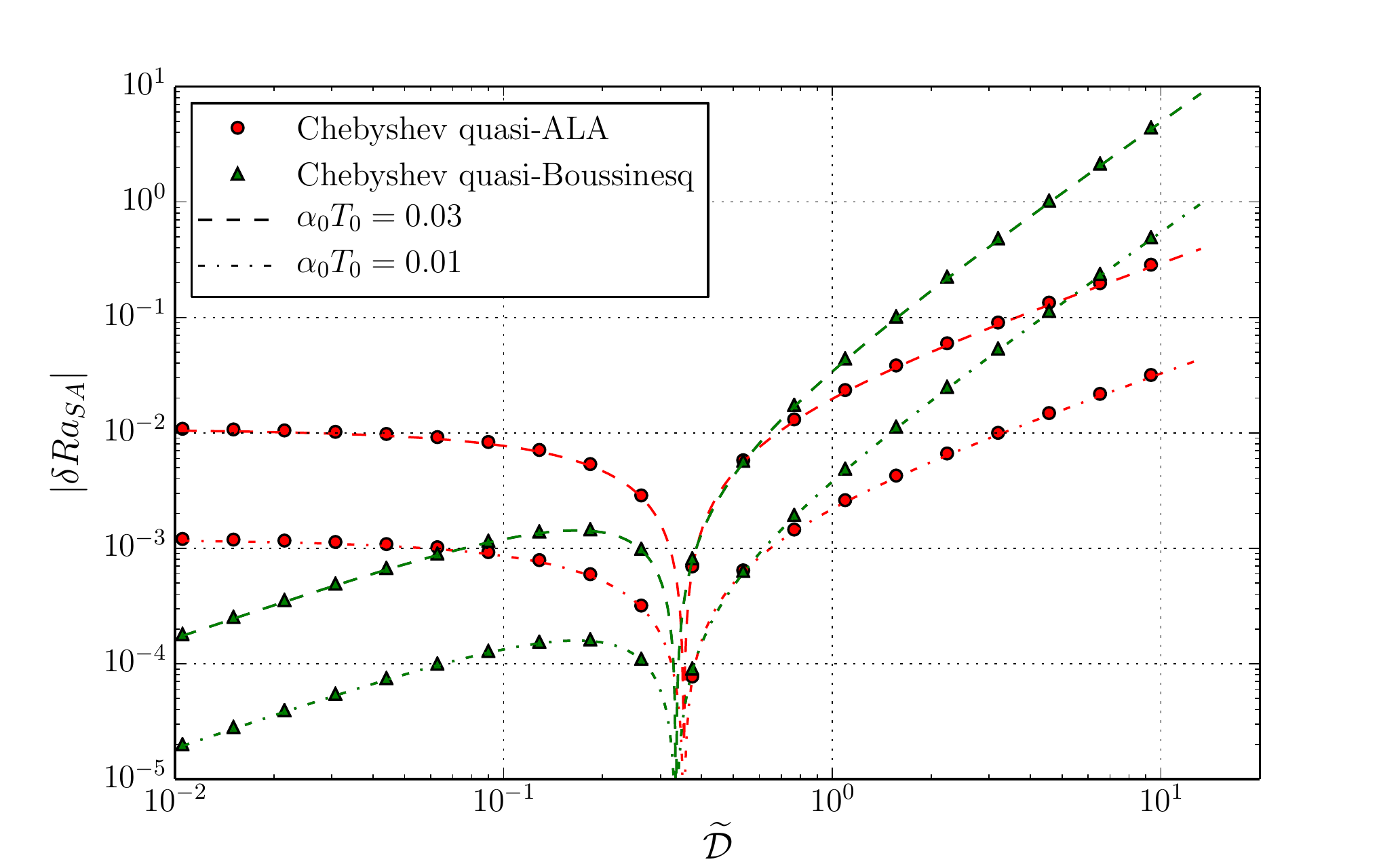}
\caption{Similar to Fig.~\ref{MurnDlin} but in logarithmic coordinates and for the absolute difference of the critical Rayleigh numbers between the approximations and exact model, for two values of $\alpha _0 T_ 0 = \hat\alpha$ ($0.03$ and $0.01$) and $\gamma _0 = 1.03$.}
\label{M_a_log}
\end{center}
\end{figure}

\begin{figure}
\begin{center}
\includegraphics[width=12 cm, keepaspectratio]{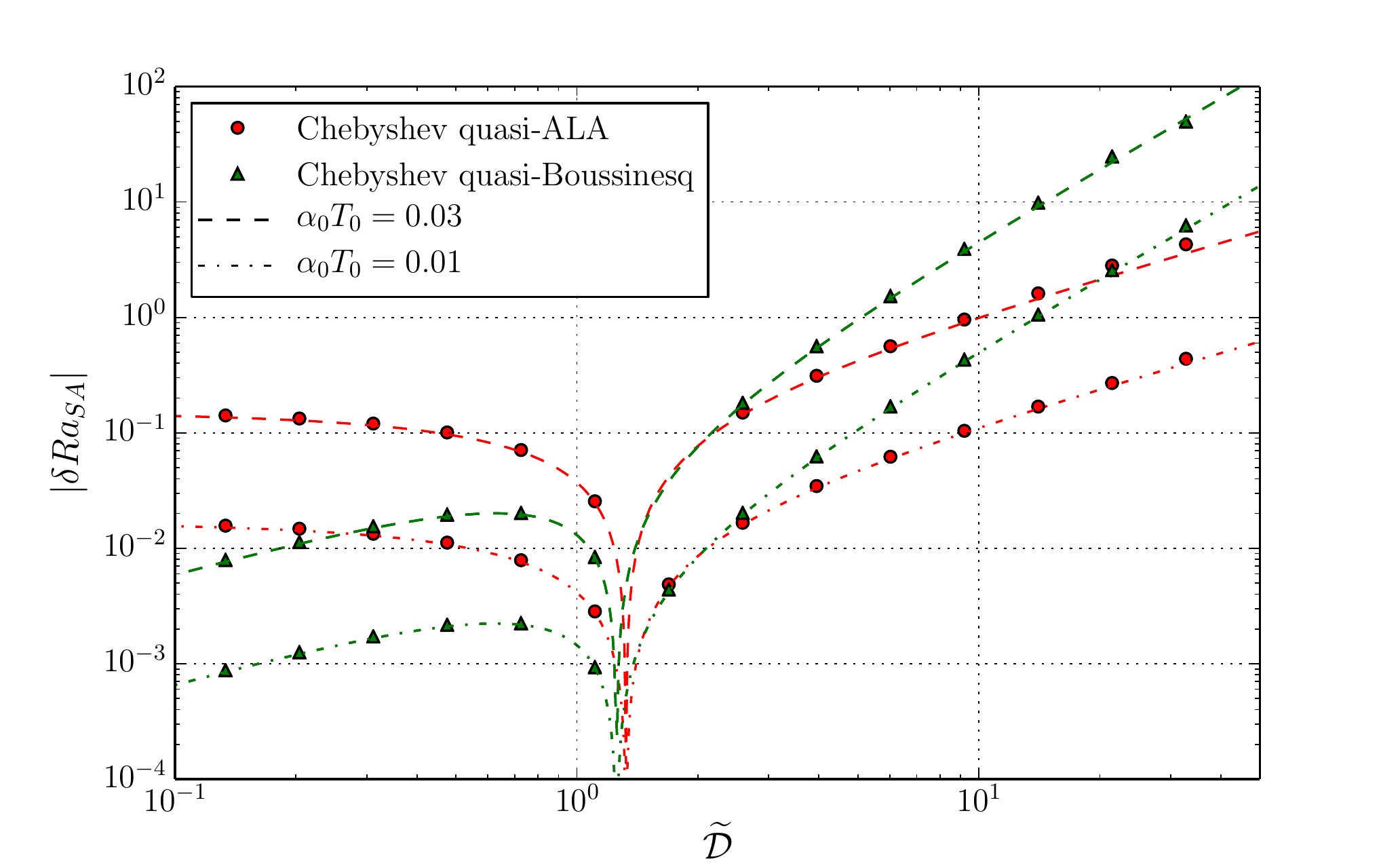}
\caption{Similar to Fig.~\ref{M_a_log} but the temperature ratio is $r=7$ ($a=1.5$) instead of $r=1.5$ ($a=0.4$).}
\label{M_a_log2}
\end{center}
\end{figure}

On Fig.~\ref{Murnalog}, we plot the absolute difference of the quasi-ALA and exact critical Rayleigh numbers, for a negligible dissipation parameter and varying temperature gradient, which can be seen to be very well approximated by the two-modes analysis. This is also the case, for a constant temperature gradient $a=0.4$ and varying dissipation parameter, shown on Fig.~\ref{M_a_log}. With a larger temperature gradient $a=1.5$, and for the largest value of the dissipation parameter, we can detect a small deviation from the two-modes analysis (see Fig.~\ref{M_a_log2}). These results, shown on Fig.~\ref{M_a_log} and \ref{M_a_log2}, confirm that the quasi-ALA approximation is much better than the quasi-Boussinesq approximation when $\widetilde{\mathcal{D}}$ is larger than $a$. Obviously, that condition is most easily fulfilled when $\gamma _0$ is very close to~$1$. 

\subsection{A generic EoS}
\label{generic}

An examination of the previous results -- for instance (\ref{deltaRaMurnB}) or (\ref{deltaRaMurnALA}) -- reveals that the first derivatives of density with respect to temperature or pressure (related to $\hat{\alpha}$ and $\mathcal{D}$ respectively) are not the only parameters affecting the critical Rayleigh numbers: the parameter $n$ is not related to the first derivatives and yet affects the critical Rayleigh numbers. The predictions of our two mode semi analytic model
is based from a set of quantities (whose listed in Tables 2 and 3, for the ideal gaz and the Murnaghan fluid). These
quantities involve up to the third degree of the EoS in the terms 
$\left. {\mathrm{d}^2 }/{\mathrm{d} z^2}  ({ \partial \rho }/{\partial T}) \right| _{p0} $  and $\left. {\mathrm{d}^2 f }/{\mathrm{d} z^2}  \right| _0$, because any derivation along $z$ is a combination of derivatives with respect to temperature and pressure, and because $c_p$ (in $f$) is itself already based on a derivative of the EoS. A generic equation of state that would determine completely the Rayleigh number with a second order precision should extend 
to the degree $3$ in $p$ and~$T$.

In fact, it turns out that it is mathematically more convenient to expand the specific volume $\nu = 1/ \rho$ with respect to temperature and pressure, rather than density. So a dimensionless EoS can be written:
\begin{equation}
\begin{split}
\nu = \frac{1}{\rho } = 1 + \hat{\alpha} (T -1) - \hat{\alpha} \widetilde{\mathcal{D}} p + \hat{\alpha}^2 E (T-1)^2 + \hat{\alpha}^2 \widetilde{\mathcal{D}} F p (T-1) + \hat{\alpha}^2 \widetilde{\mathcal{D}}^2 G p^2\\  \label{genericEoSadim}
+ \hat{\alpha}^3 J (T-1)^3 + \hat{\alpha}^3 \widetilde{\mathcal{D}} K p (T-1)^2 + \hat{\alpha}^3 \widetilde{\mathcal{D}}^2 L p^2 (T-1) 
+ \hat{\alpha}^3 \widetilde{\mathcal{D}}^3 M p^3
\end{split}
\end{equation}
The expression for the dimensionless derivative of $\nu$ with respect to $p$ is found to be $-\hat{\alpha} \widetilde{\mathcal{D}}$ and the coefficients $E$, $F$, $G$, $J$, $K$, $L$ and $M$ are dimensionless parameters proportional to the second and third derivatives of the specific volume. We have chosen to make these coefficients independent of gravity $g$ by multiplying systematically any occurrence of the dimensionless pressure $p$ by the dissipation parameter $\widetilde{\mathcal{D}}$.

\begin{table*}
\begin{center}
\begin{tabular}{@{}ll@{}}
\toprule expression \hspace*{1 cm} & value   \\
\hline
$\left. f \right| _0 $ &  $-1 - \left( -2 L + 2 G - 3 \frac{F}{\gamma _0} - \frac{3}{\gamma _0^2}\right) \frac{\hat{\alpha}^2}{24 (a - \mathcal{D} )} \widetilde{\mathcal{D}}^2 \mathcal{D} - \left( 4 A - 18 E - \frac{1}{\hat{\alpha}} + 1 \right) \frac{\hat{\alpha}}{24 (a - \mathcal{D} )} \mathcal{D}^3 $ \\
& $- 3 \left( -2 K - 2 E F + A F + 2 E + F - \frac{F}{\hat{\alpha}} + \frac{1}{\gamma _0} \left[ - 6 E + A - \frac{1}{\hat{\alpha}} \right] \right) \frac{\hat{\alpha}^2 }{24 (a - \mathcal{D} )}  \widetilde{\mathcal{D}} \mathcal{D}^2 $ \\
& $ - \left(-18 J - 24 E^2 + 18 A E + 2 B -3 A^2 - A \right) \frac{\hat{\alpha}^2}{24 (a - \mathcal{D} )} \mathcal{D}^3$  \\
$\left. \frac{\mathrm{d} f }{\mathrm{d} z}  \right| _0$  & $ \frac{\hat{\alpha}}{a-\mathcal{D}} \left( (1-A) a^2 + \left[ 4 E - 2 + \frac{1}{\hat{\alpha}} \right] a \mathcal{D} - a \widetilde{\mathcal{D}} + (2 + F) \mathcal{D} \widetilde{\mathcal{D}}  \right)  $\\
$\left. \frac{\mathrm{d}^2 f }{\mathrm{d} z^2}  \right| _0$  & \hspace*{-1.2 cm}\begin{minipage}[t]{12 cm}{\vspace*{-5 mm} \begin{dmath*} \frac{\hat{\alpha}^2}{(a-\mathcal{D})^2} \left( 2 (A+B-A^2-E) a^4 + \left[ (-1 -2 F) \widetilde{\mathcal{D}} + \left( 4 - \frac{8}{\hat{\alpha}} \right) E \mathcal{D} - 18 J \mathcal{D} + A \left( \frac{4 \mathcal{D}}{\hat{\alpha}} +16 E \mathcal{D} -6 \mathcal{D} -2 \widetilde{\mathcal{D}}  \right) -2 B \mathcal{D} \right] a^3
+ \left[ \left( 2 E + \left( 7 - \frac{2}{\hat{\alpha}} \right) F + 3 - 6 K +2 A (3 + 2 F) \right) \mathcal{D} \widetilde{\mathcal{D}} + (1 - 2 G)  \widetilde{\mathcal{D}}^2 + \left( -32 E^2 + \left( 14 - \frac{8}{\hat{\alpha}} \right) E - 2 \left( \frac{1}{\hat{\alpha}} -1 \right) ^2 + 18 J \right) \mathcal{D}^2 \right] a^2
 +\left[ - ( 3 + F - 6 G + 2 L) \mathcal{D} \widetilde{\mathcal{D}}^2 - \left( 18 E + \frac{4}{\hat{\alpha}} -2 + 16 E F + \left( 1 + \frac{2}{\hat{\alpha}} \right) F -6 K \right) \mathcal{D}^2 \widetilde{\mathcal{D}}  \right]  a - \left[ 4 G + 2 F^2 + 3 F - 2 L \right] \mathcal{D}^2 \widetilde{\mathcal{D}}^2 \right) \nonumber \end{dmath*}  }\end{minipage}    \\
$\left.  \frac{\rho '_b}{\rho _b} \right| _0 $ & $ \hat{\alpha} (a - \widetilde{\mathcal{D}})$ \\
$\left. \frac{\mathrm{d} }{\mathrm{d} z} \frac{\rho '_b}{\rho _b} \right| _0 $  &
$ \hat{\alpha}^2 \left[ (-2 E + 1 ) a^2 - ( 2 F + 3) \widetilde{\mathcal{D}} a + 2 ( 1 - G ) \widetilde{\mathcal{D}}^2 \right] $ \\
$\left. \frac{ \partial \rho }{\partial T} \right| _{p0}$ & $ - \hat{\alpha}   $ \\
$\left. \frac{\mathrm{d} }{\mathrm{d} z}  \frac{ \partial \rho }{\partial T} \right| _{p0}  $  & $  \hat{\alpha}^2  \left[ 2 (E - 1 ) a + (2 + F) \widetilde{\mathcal{D}} \right]  $\\
$\left. \frac{\mathrm{d}^2 }{\mathrm{d} z^2}  \frac{ \partial \rho }{\partial T} \right| _{p0} $  & $ \hat{\alpha}^3 \left[ (12 E - 6 - 6 J ) a^2 + ( 9 F -8 E + 14 - 4 K ) a \widetilde{\mathcal{D}} + ( 4 G - 5 F - 8 - 2 L ) \widetilde{\mathcal{D}}^2 \right] $ \\
$\left. \frac{ \partial \rho }{\partial p} \right| _{T0}$ & $\hat{\alpha} \widetilde{\mathcal{D}} $ \\
$\left. \frac{\mathrm{d} }{\mathrm{d} z}  \frac{ \partial \rho }{\partial p} \right| _{T0}  $ & $ \hat{\alpha}^2 \left[ ( F + 2 ) a \widetilde{\mathcal{D}} + 2 ( G - 1 ) \widetilde{\mathcal{D}}^2 \right] $ \\
\bottomrule
\end{tabular}
\caption{Coefficients of Taylor expansion of some quantities related to the base flow, for a generic equation of state (\ref{genericEoSadim}).}
\label{tableGeneric}
\end{center}
\end{table*}

We apply the same procedure for this generic equation of state as for the equations of state considered previously. In order to obtain an expression for $c_p$, we integrate the relation
\begin{equation}
\left. {\partial c_p \over \partial p}\right| _T=  -\left. {{\mathcal{D}}\over \hat{\alpha}} T{\partial^2 v\over \partial T^2}\right| _p, \label{cpgeneric0}
\end{equation}
which leads to
\begin{equation}
c_p = 1+ \hat{\alpha} A(T-1)+ \hat{\alpha}^2 B(T-1)^2 - \hat{\alpha} \mathcal{D} T (  2E p+6\hat{\alpha}Jp(T-1)+\hat{\alpha}\widetilde{\mathcal{D}} Kp^2), \label{cpgeneric1}
\end{equation}
where the $p$-independent integration term has been expressed up to degree 2 by introducing two extra coefficients $A$ and $B$.

The reference temperature is still
\begin{equation}
T_b(z)=1-az , 
\end{equation}
with a uniform gradient and we need to compute the third derivative of the adiabatic temperature profile, at $z=0$. This is obtained by derivating the adiabatic gradient twice, using the equation of state (\ref{genericEoSadim}) and the expression for $c_p$ (\ref{cpgeneric1})
\begin{eqnarray}
\left. \frac{\mathrm{d}^3 T_a}{\mathrm{d} z^3} \right| _0 &=& \left( -3F+2G-2 L - 3 \right) \hat{\alpha}^2 \widetilde{\mathcal{D}}^2 \mathcal{D} + \Big[3 (2F+2-2K-4E-2EF+AF+A)\hat{\alpha}   \nonumber \\
&-&3(1+F)\Big] \hat{\alpha} \mathcal{D}^2\widetilde{\mathcal{D}}+
\Big[(-4A-3A^2-24E^2+18E+18AE-18J-3+2B)\hat\alpha^2\nonumber\\
&+&(4+4A-18E)\hat\alpha-1\Big]\mathcal{D}^3
\label{d3Tageneric}
\end{eqnarray}

All quantities needed in the approximate analysis have been determined and listed in Table \ref{tableGeneric}.
With table \ref{tableGeneric} and the general solutions obtained in sections \ref{estimate} and \ref{themodels}, the analytic expression for $\epsilon$ and $dRa_{SA}$ (and corresponding results for the quasi-Boussinesq and quasi-ALA approximations) are explicitly determined. Some results, like $dRa_{SA}$ would take a page to display when the substitution is made. Others are shorter. For instance, 
the relative amplitude of the $\sin (2 \upi z )$ component relative to the $\cos ( \upi z)$ component can be entirely written in terms of the elementary governing coefficients:
\begin{eqnarray}
\hspace*{-14 mm}\epsilon ^x &=&  \frac{8 \hat{\alpha} }{ \upi ^2} \left[   
  \frac{ (1-A) a^2 - a \widetilde{\mathcal{D}} + \left[ 4 E + \frac{1}{\hat{\alpha}} -2 \right] a \mathcal{D} +(F+2) \widetilde{\mathcal{D}} \mathcal{D} }{13 (a - \mathcal{D}) } -\frac{(1+2 E) a+ F \widetilde{\mathcal{D}}}{117}    \right]  ,  \label{epsxGene}\\
\hspace*{-14 mm}\epsilon ^B &=&  \frac{8 \hat{\alpha} }{ \upi ^2} \left[   
  \frac{ (1-A) a^2 - a \widetilde{\mathcal{D}} + \left[ 4 E + \frac{1}{\hat{\alpha}} -2 \right] a \mathcal{D} +(F+2) \widetilde{\mathcal{D}} \mathcal{D} }{13 (a - \mathcal{D}) } -\frac{(1+2 E) a+ (F-1) \widetilde{\mathcal{D}}}{117}    \right]  ,  \label{epsBGene}\\
\hspace*{-14 mm}\epsilon ^{ALA} &=&  \frac{8 \hat{\alpha} }{ \upi ^2} \left[   
  \frac{ (1-A) a^2 - a \widetilde{\mathcal{D}} + \left[ 4 E + \frac{1}{\hat{\alpha}} -2 \right] a \mathcal{D} +(F+2) \widetilde{\mathcal{D}} \mathcal{D} }{13 (a - \mathcal{D}) } -\frac{2 E a+ F \widetilde{\mathcal{D}}}{117}    \right]  .  \label{epsALAGene}
\end{eqnarray}

We can also expand the two-modes approximations (\ref{sol_deltaRaB}) and (\ref{sol_deltaRaALA}), using table \ref{tableGeneric}, for the difference between the quasi-Boussinesq approximation and the exact model $\delta Ra_{SA}^{B}$, and between the quasi-ALA approximation and exact model $\delta Ra_{SA}^{ALA}$: 
\begin{eqnarray}
\delta Ra_{SA}^{B} &=&
 - \left[ - \frac{5}{2} \upi ^2 + \frac{224}{13} + \frac{9}{2} \upi ^2 G  + \frac{256}{39} F \right] \hat{\alpha}^2  \widetilde{\mathcal{D}}^2     \nonumber \\
&& - \left[ \frac{5}{2} \upi ^2 - \frac{128}{13} + \frac{9}{4} \upi ^2 F + \frac{512}{39} E \right] \hat{\alpha}^2 a  \widetilde{\mathcal{D}} ,\label{deltaRaBgeneric} \\ 
\delta Ra_{SA}^{ALA} &=&  \left[ \frac{1}{4} \upi ^2 - \frac{320}{13} + \left( \frac{9}{4} \upi ^2 - \frac{256}{39} \right) F \right] \hat{\alpha}^2 a \widetilde{\mathcal{D}} \nonumber \\
&& + \left[ \frac{1}{4} \upi ^2 + \frac{224}{13} + \left( \frac{9}{2} \upi ^2 - \frac{512}{39}  \right) E \right] \hat{\alpha}^2 a^2 . \label{deltaRaALAgeneric}
\end{eqnarray}

The two-mode analysis and table \ref{tableGeneric} indicate that the quadratic departure of the suparadiabatic threshold from the Boussinesq limit (\ref{sol_dRaSA}) depends on all coefficients of the cubic expansion of the generic equation of state (\ref{genericEoSadim}) and on the extra free coefficients $A$ and $B$ in the expression for the heat capacity (\ref{cpgeneric1}). Only $M$ (related to $\partial ^3 \nu / \partial p^3$) has no influence, as expected, because that particular third derivative is not involved in the relevant coefficients $\left. \frac{\mathrm{d}^2 }{\mathrm{d} z^2}  \frac{ \partial \rho }{\partial T} \right| _{p0} $  and $\left. \frac{\mathrm{d}^2 f }{\mathrm{d} z^2}  \right| _0$. The two-mode analyses of the quasi-Boussinesq and quasi-ALA models show that the difference of the critical superadiabatic Rayleigh numbers depends entirely on the second order expansion of the equation of state: $J$, $K$, $L$, $M$ do not affect the differences (\ref{deltaRaBgeneric}) and (\ref{deltaRaALAgeneric}), neither do $A$ and $B$.

\begin{figure}
\begin{center}
\includegraphics[width=12 cm, keepaspectratio]{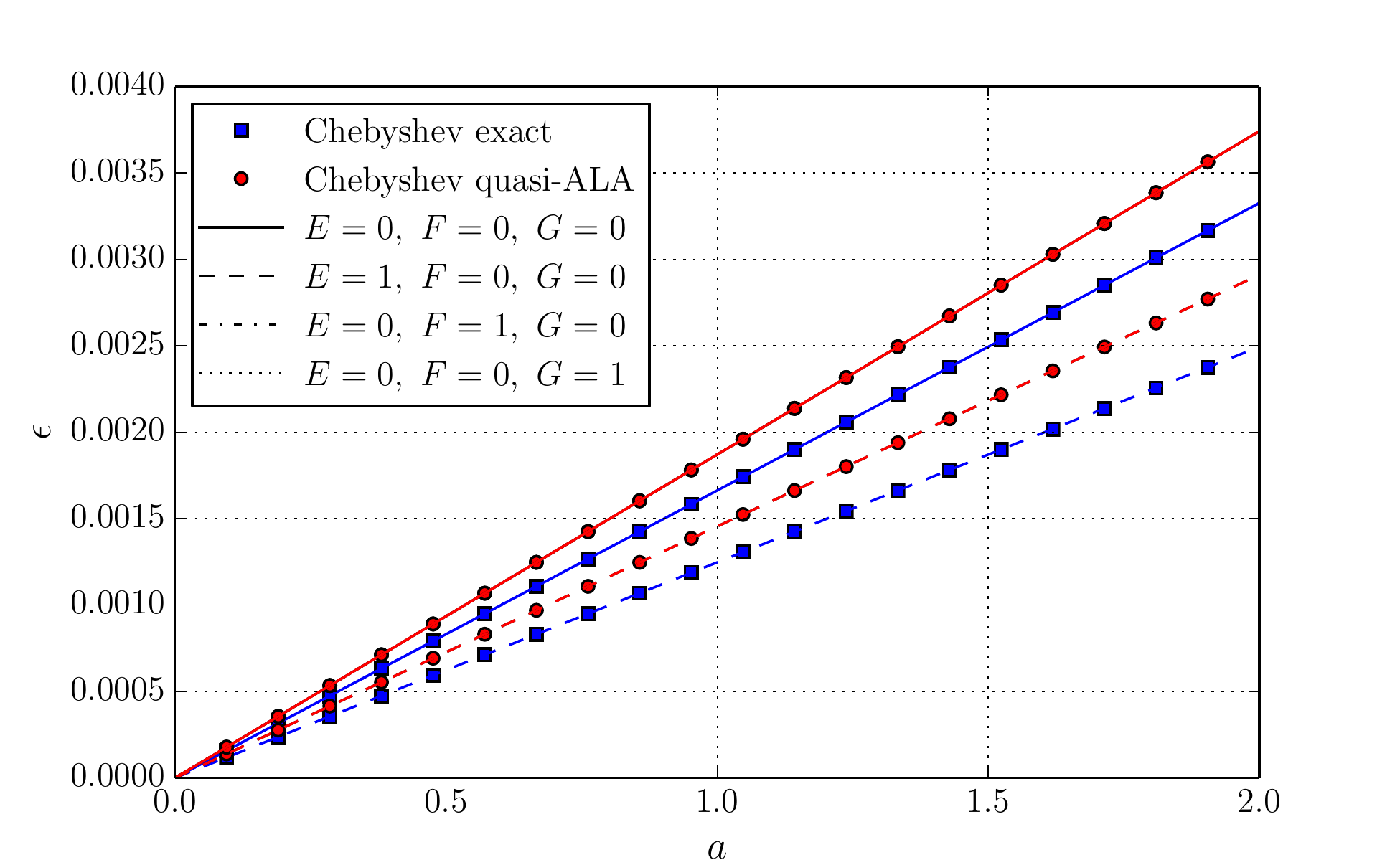}
\caption{Asymmetrical contribution of the $\sin ( 2 \upi z)$ mode to the critical eigenmode, for a generic EoS (\ref{genericEoSadim}) as a function of the temperature gradient $a$ of the base linear solution, for a negligible ${\cal{D}} = 10^{-8}$. The label 'Chebyshev exact' denotes the numerical solution of the exact model using a Chebyshev expansion (usually 17 polynomials), while the label 'Chebyshev quasi-ALA' corresponds to the solutions of the quasi-ALA model. The solid, dashed, dash-dot and dotted lines correspond to the approximate two-modes analytical solutions for the exact and quasi-ALA models, for different choices of the dimensionless parameters $E$, $F$ and $G$ respectively. The ratio of heat capacities and the product of temperature and the thermal expansion coefficient are kept constant $\gamma _0 =1.03$ and $\hat{\alpha} = \alpha _0 T_0 =0.03$. When the dissipation number is negligible, the quasi-Boussinesq model and the exact model coincide.}
\label{Genalineps}
\end{center}
\end{figure}

\begin{figure}
\begin{center}
\includegraphics[width=12 cm, keepaspectratio]{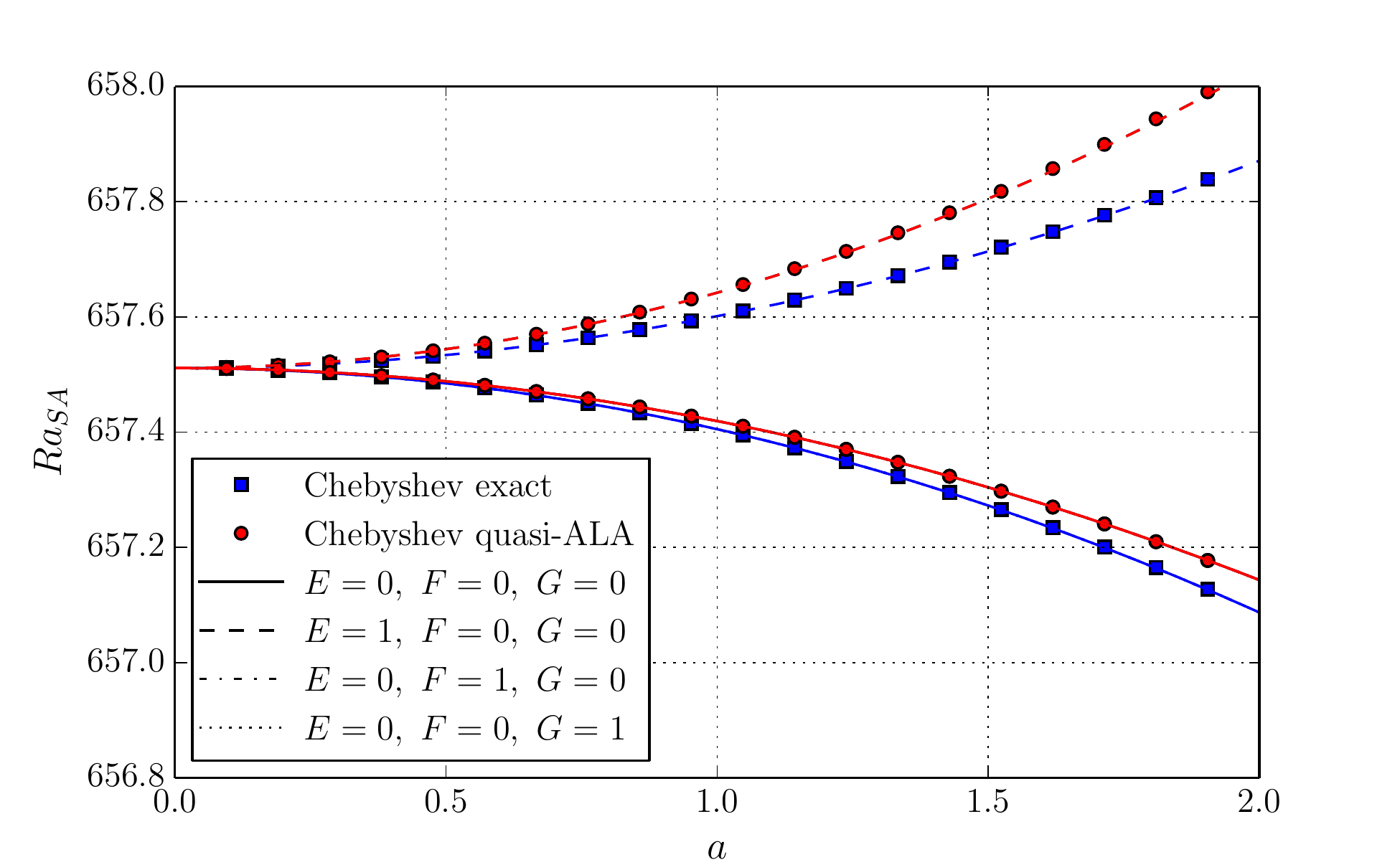}
\caption{Linear stability critical threshold for the Rayleigh number for a generic EoS (\ref{genericEoSadim}) as a function of the temperature gradient $a$ of the base linear solution, for a negligible ${\cal{D}} = 10^{-8}$. The labels are identical to those defined in figure \ref{Genalineps}.}
\label{Genalin}
\end{center}
\end{figure}

With so many parameters, (9 parameters without counting $\hat\alpha$ and ${\widetilde{\mathcal{D}}}$) it is impossible to show and explore all the possible cases. Similarly to what we have computed for the ideal gas and the Murnaghan EoS, we start by depicting a few cases where the compressible effects are small ${\cal{D}}=10^{-8}$ but
the temperature difference large which are conditions that could be easily reproduced experimentally. In Figures
\ref{Genalineps} and \ref{Genalin}, we plot the asymmetrical contributions of the critical eigenmode, $\epsilon$ and the
corresponding changes in the critical Ra number when only the second order coefficients of the generic EoS (i.e., $E$, $F$ and $G$) are changed. In agreement with (\ref{epsxGene}) or (\ref{epsALAGene}), when ${\cal{D}}<<1$ and $A$ constant, $\epsilon$ is only a function of $E$ which corresponds very precisely to the numerical estimates (see Figure \ref{Genalineps}). The Rayleigh numbers of the exact and ALA cases are also only functions of $E$, (see Figure \ref{Genalineps}).

\begin{figure}
\begin{center}
\includegraphics[width=12 cm, keepaspectratio]{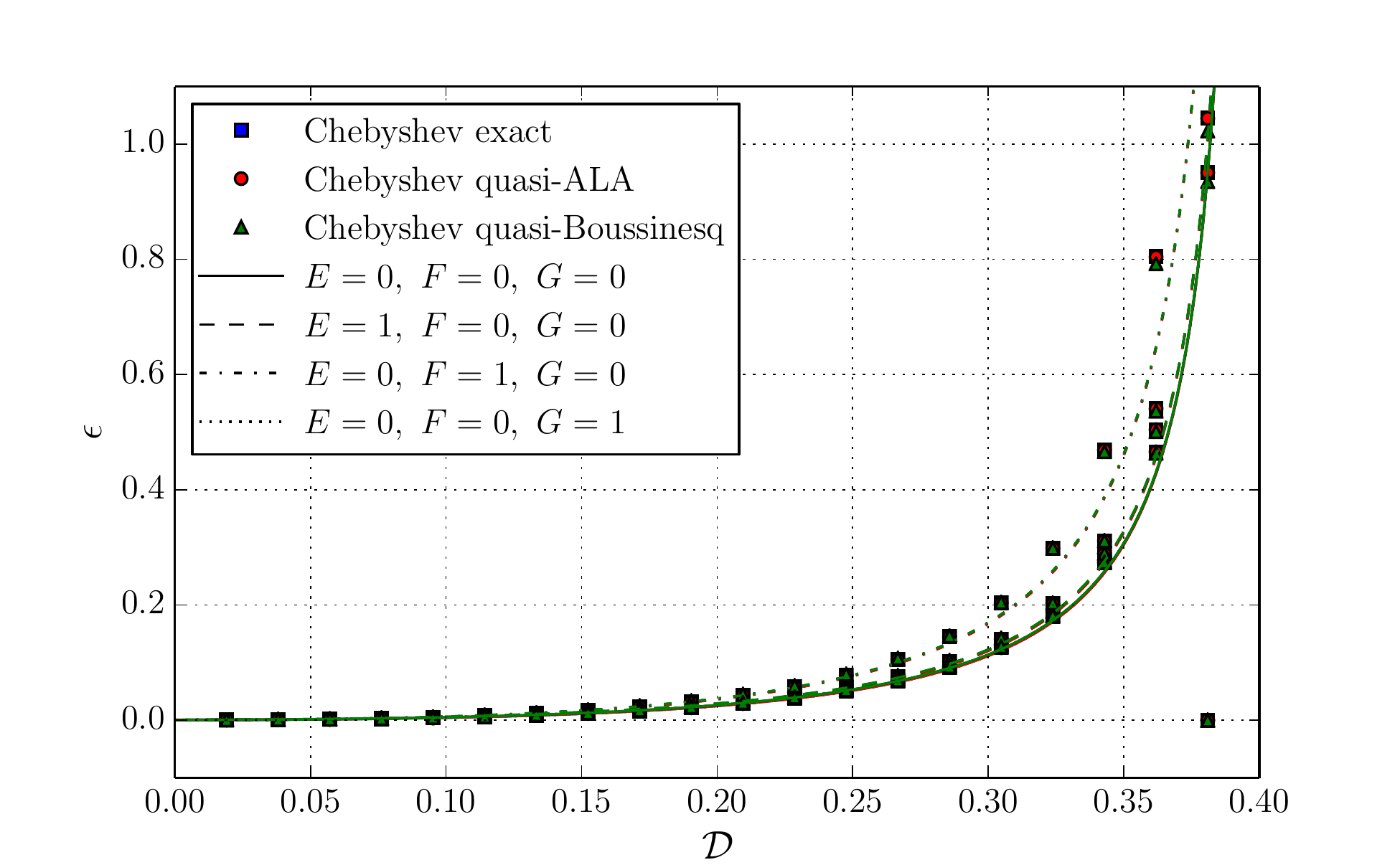}
\caption{Asymmetrical contribution of the $\sin ( 2 \upi z)$ mode to the critical eigenmode, for the Rayleigh number for a generic EoS (\ref{genericEoSadim}) as a function of the dissipation number $\mathcal{D}$, for a fixed temperature gradient $a=0.4$ (corresponding to a temperature ratio $r=1.5$). The labels Chebyshev exact, quasi-ALA and quasi-Boussinesq correspond to numerical solutions obtained using the Chebyshev collocation eigenvalue calculations described in section \ref{eigenvalue}, for the exact equations, quasi-ALA and Boussinesq approximations respectively. The lines are the analytical two-modes solutions described in section \ref{estimate}. Solid, dashed, dash-dot and dotted lines correspond to different selections of the parameters $E$, $F$ and $G$, while the heat capacity ratio and $\alpha T$ at $z=0$ are kept constant $\gamma _0 = 1.03$ and $\hat{\alpha} = \alpha _0 T_0 =0.03$, and the other parameters of the generic EoS are set to zero: $J=K=L=M=A=B=0$.}
\label{GenDlineps}
\end{center}
\end{figure}

\begin{figure}
\begin{center}
\includegraphics[width=12 cm, keepaspectratio]{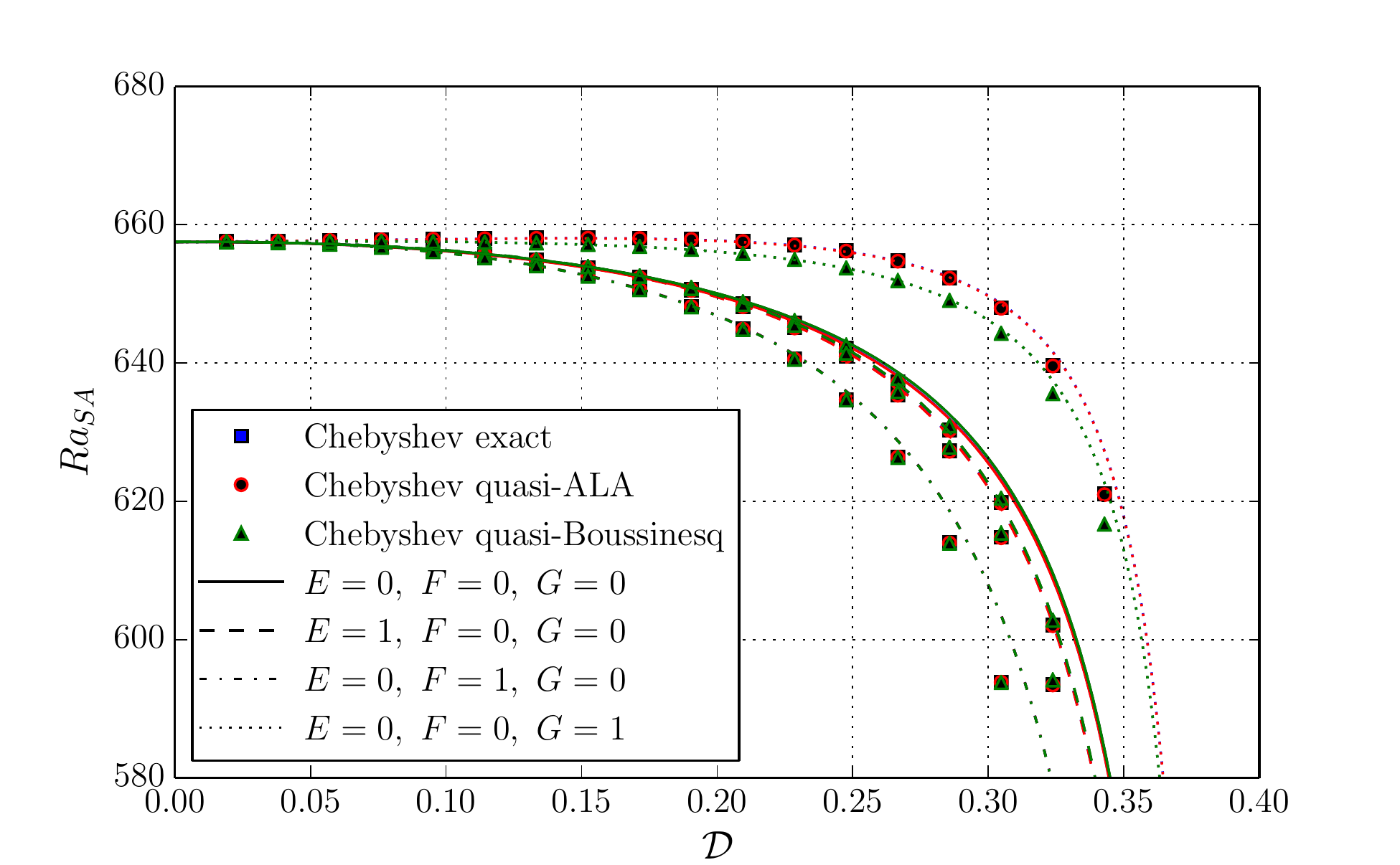}
\caption{Linear stability critical threshold for the Rayleigh number for a generic EoS (\ref{genericEoSadim}) as a function of the dissipation number $\mathcal{D}$, for a fixed temperature gradient $a=0.4$ (corresponding to a temperature ratio $r=1.5$). The labels are similar to those in Fig.~\ref{GenDlineps}.}
\label{GenDlin}
\end{center}
\end{figure}

\begin{figure}
\begin{center}
\includegraphics[width=12 cm, keepaspectratio]{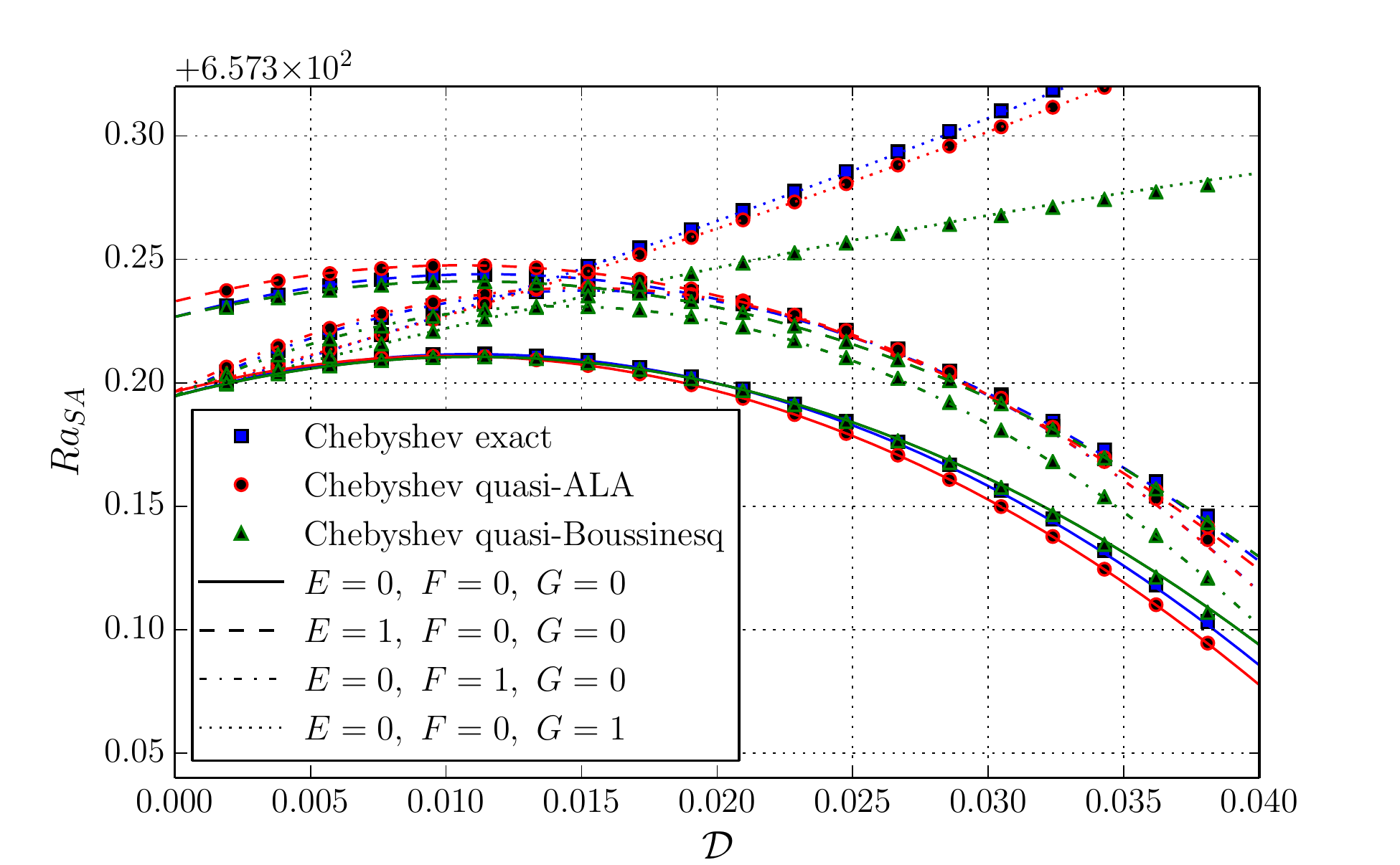}
\caption{Same as Fig.~\ref{GenDlin} with a close-up around small values of $\mathcal{D}$, between $0$ and $0.04$. }
\label{GenDlin_bis}
\end{center}
\end{figure}

We then compute a few cases with a fixed temperature interval $a=0.4$ but for varying the compressible effects. Like for the cases illustrated in the two previous figures, we only vary the second order coefficients of the EoS. The asymmetrical contributions of the critical eigenmode, $\epsilon$ and the corresponding change in the critical Ra number are depicted In Figures \ref{GenDlineps} and \ref{GenDlin}. In the two figures, an asymptote is present at $\mathcal{D}=0.4$ because of the singular term in $a-\mathcal{D}$ the various analytical expressions. Here again, the two-mode expansion captures reasonably accurately the numerical results. However, the two-mode expansion is even better for small values of $\mathcal{D}$ since it is a Taylor expansion of degree two. On Fig.~\ref{GenDlin_bis}, we show a close-up of Fig.~\ref{GenDlin} at small $\mathcal{D}$ (between $0$ and $0.04$) and it is apparent that each coefficient $E$, $F$ and $G$ has a specific influence of the critical Rayleigh number, which is very accurately modelled by the two-mode analysis. 

\begin{figure}
\begin{center}
\includegraphics[width=12 cm, keepaspectratio]{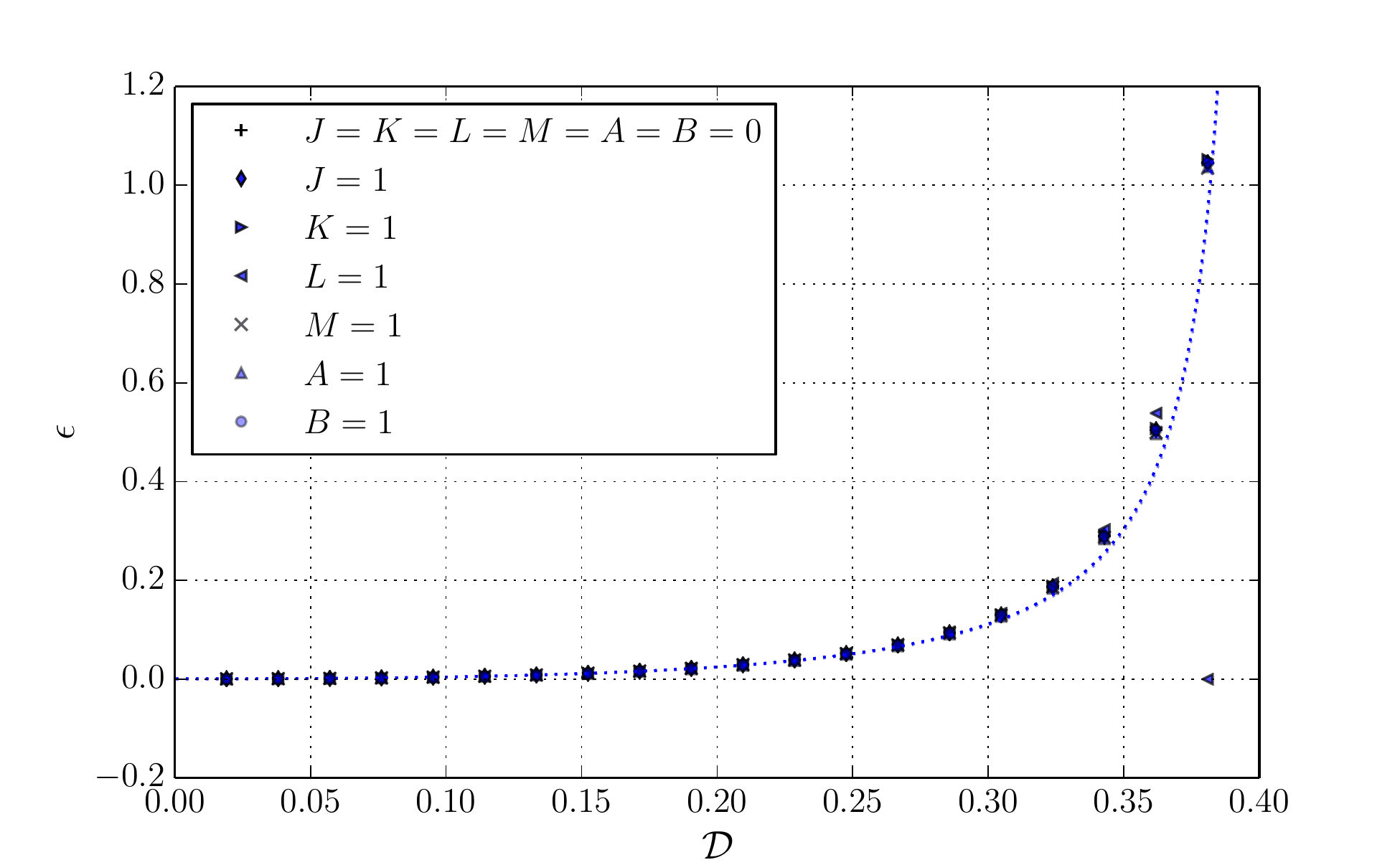}
\caption{Same as Fig.~\ref{GenDlineps}, but only the results from the exact governing equations are shown, along with its two-modes approximation. Now the parameters $E$, $F$ and $G$ of the equation of state (\ref{genericEoSadim}) are set to zero, while each of the other parameters ($J$, $K$, $L$, $M$, $A$, and $B$) is set to $1$ in turn. }
\label{GenDlin_deg3biseps}
\end{center}
\end{figure}

\begin{figure}
\begin{center}
\includegraphics[width=12 cm, keepaspectratio]{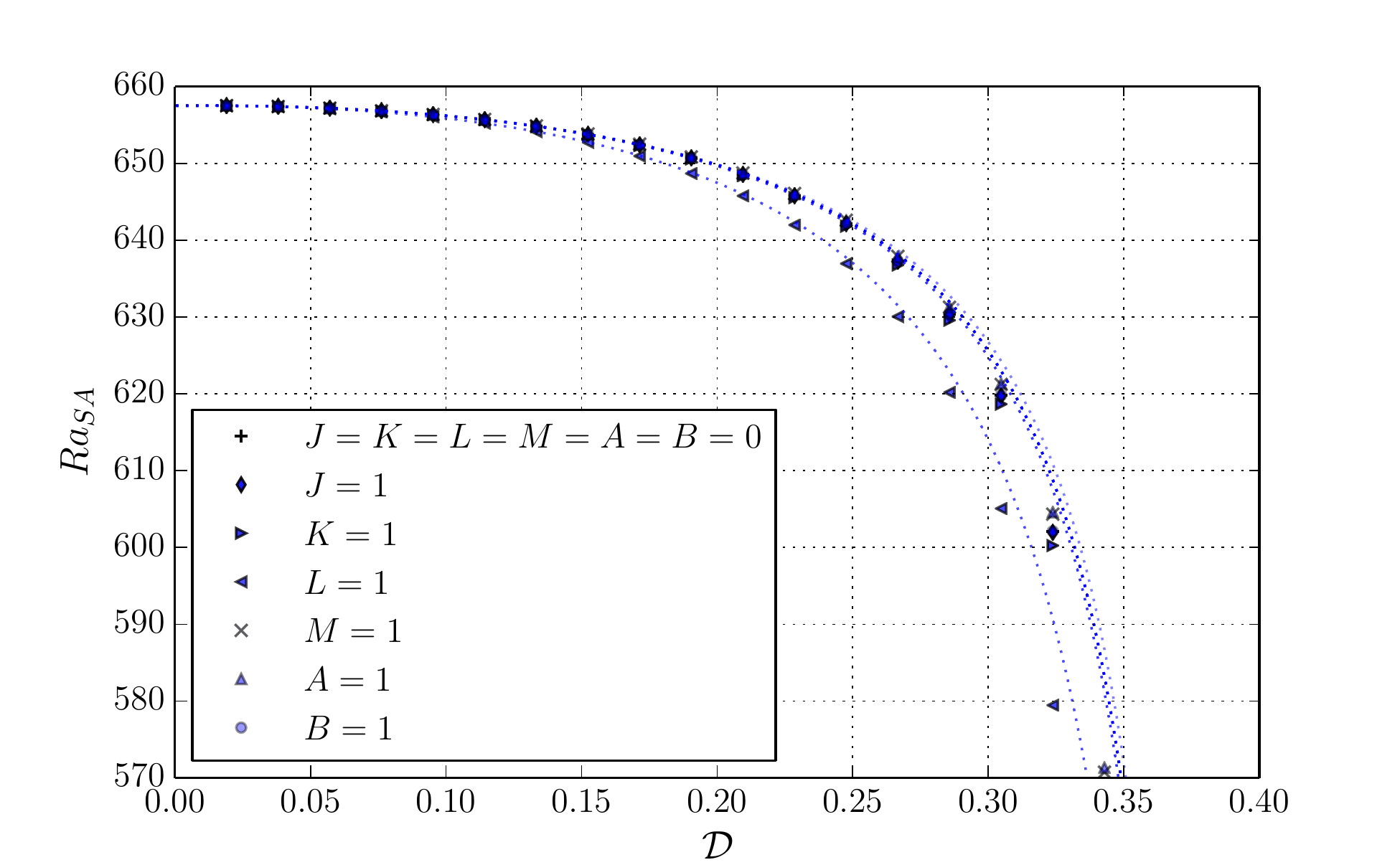}
\caption{Same as Fig.~\ref{GenDlin}, but only the results from the exact governing equations are shown, along with its two-modes approximation. Now the parameters $E$, $F$ and $G$ of the equation of state (\ref{genericEoSadim}) are set to zero, while each of the other parameters ($J$, $K$, $L$, $M$, $A$, and $B$) is set to $1$ in turn. }
\label{GenDlin_deg3bis}
\end{center}
\end{figure}

\begin{figure}
\begin{center}
\includegraphics[width=12 cm, keepaspectratio]{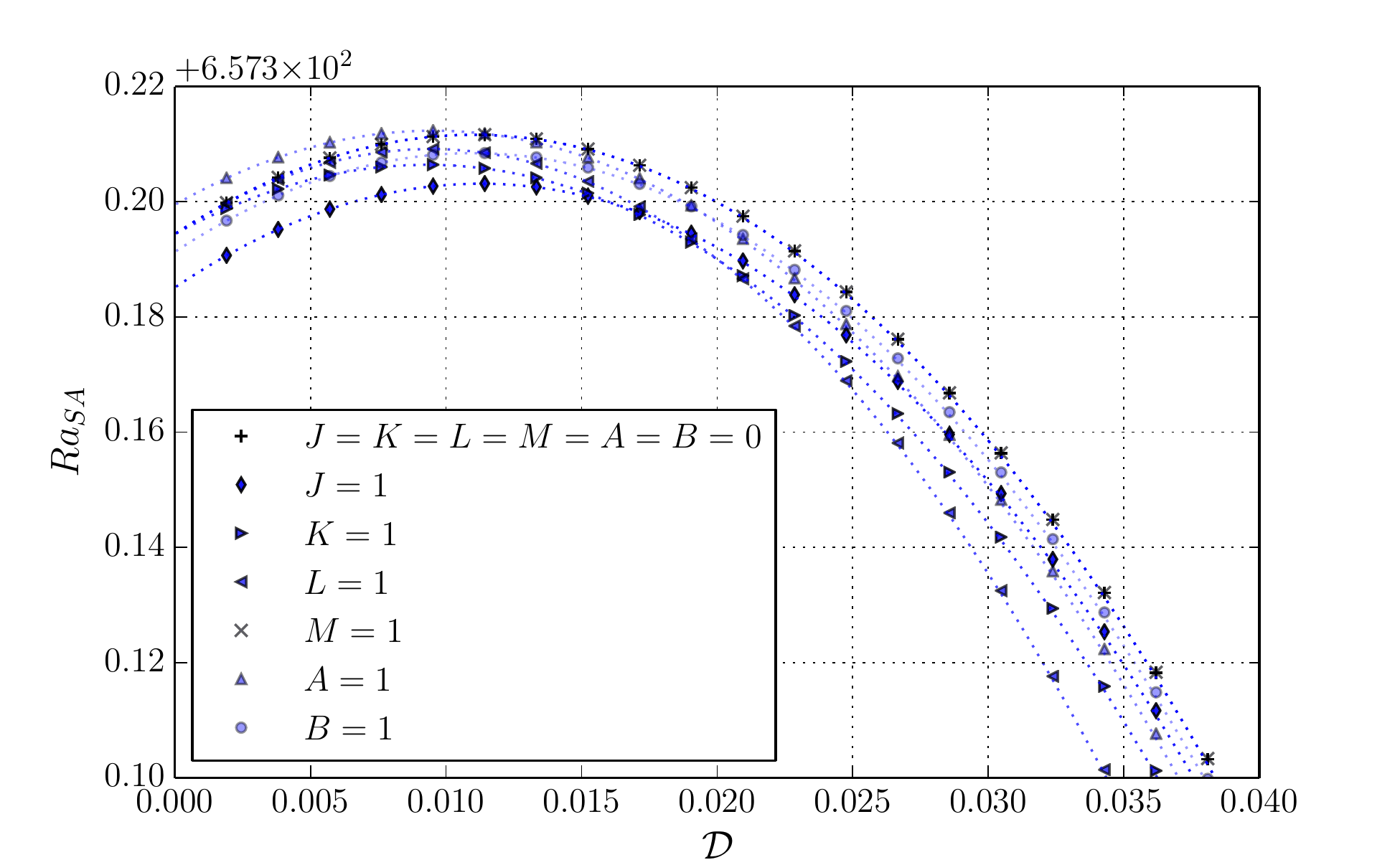}
\caption{Same as Fig.~\ref{GenDlin_deg3bis} with a close-up around small values of $\mathcal{D}$, between $0$ and $0.04$. It is clearly visible that all parameters have an impact on the critical Rayleigh number, except $M$ ($+$ and $\times$ symbols superimpose), as predicted by the approximate analysis. }
\label{GenDlin_deg3}
\end{center}
\end{figure}

We now test the effects of the third order terms (i.e., $J$, $K$, $L$ and $M$, see (\ref{genericEoSadim})) as well as of the two terms controlling the heat capacity at reference pressure ($A$ and $B$, see (\ref{cpgeneric1})) in
Figures \ref{GenDlin_deg3biseps}, \ref{GenDlin_deg3bis} and \ref{GenDlin_deg3}. In all these simulations the temperature gradient is fixed to $a=0.4$. We only compare  the solutions of the exact equations solved numerically (symbols) or using analytical two-mode approximations (dotted lines). In agreement with
with (\ref{epsxGene}), the analytical approximations for $\epsilon$ are independent of all these parameters.
The fit to the numerical solutions is very good, although we notice a slight difference between the numerical solutions when the parameters are varied, likely due to the contributions of higher degrees above our second order approximation. In agreement with (\ref{sol_dRaSA}) and table \ref{tableGeneric}, the exact values of the critical Rayleigh numbers are affected by each of these coefficients, except for $M$ (see Figure \ref{GenDlin_deg3bis}). This is more obvious on the close-up Fig.~\ref{GenDlin_deg3}, for small values of $\mathcal{D}$, as the second-order two-modes analysis provides accurate estimates for the critical Rayleigh number: changing $M$ from $0$ to $1$ does not affect the critical Rayleigh number, while changing any of the other third-order coefficients $J$, $K$, $L$, $A$ and $B$ produces a change in $Ra_{SA}$. 

\begin{figure}
\begin{center}
\includegraphics[width=12 cm, keepaspectratio]{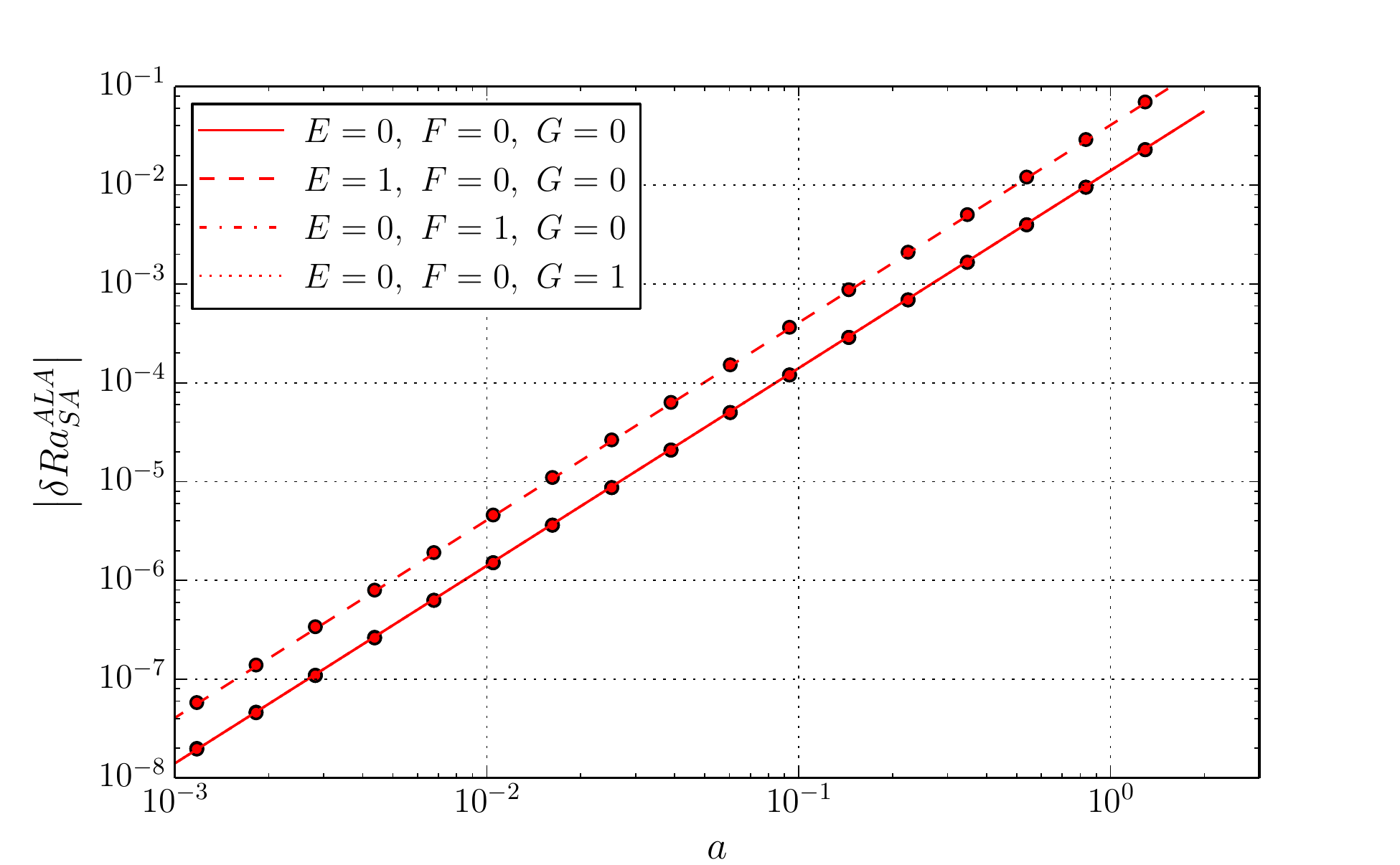}
\caption{Difference between the critical Rayleigh number of the quasi-ALA and exact model, with the generic equation of state (\ref{genericEoSadim}) and for a negligible dissipation number equal to $\mathcal{D}=10^{-8}$. The difference is plotted as a function of $a$, for different selections of the parameters $E$, $F$ and $G$. The heat capacity ratio and $\alpha T$ at $z=0$ are kept constant $\gamma _0 = 1.03$, $\hat{\alpha} = \alpha _0 T_0 =0.03$ and $J=K=L=M=A=B=0$.}
\label{Genalog}
\end{center}
\end{figure}

\begin{figure}
\begin{center}
\includegraphics[width=12 cm, keepaspectratio]{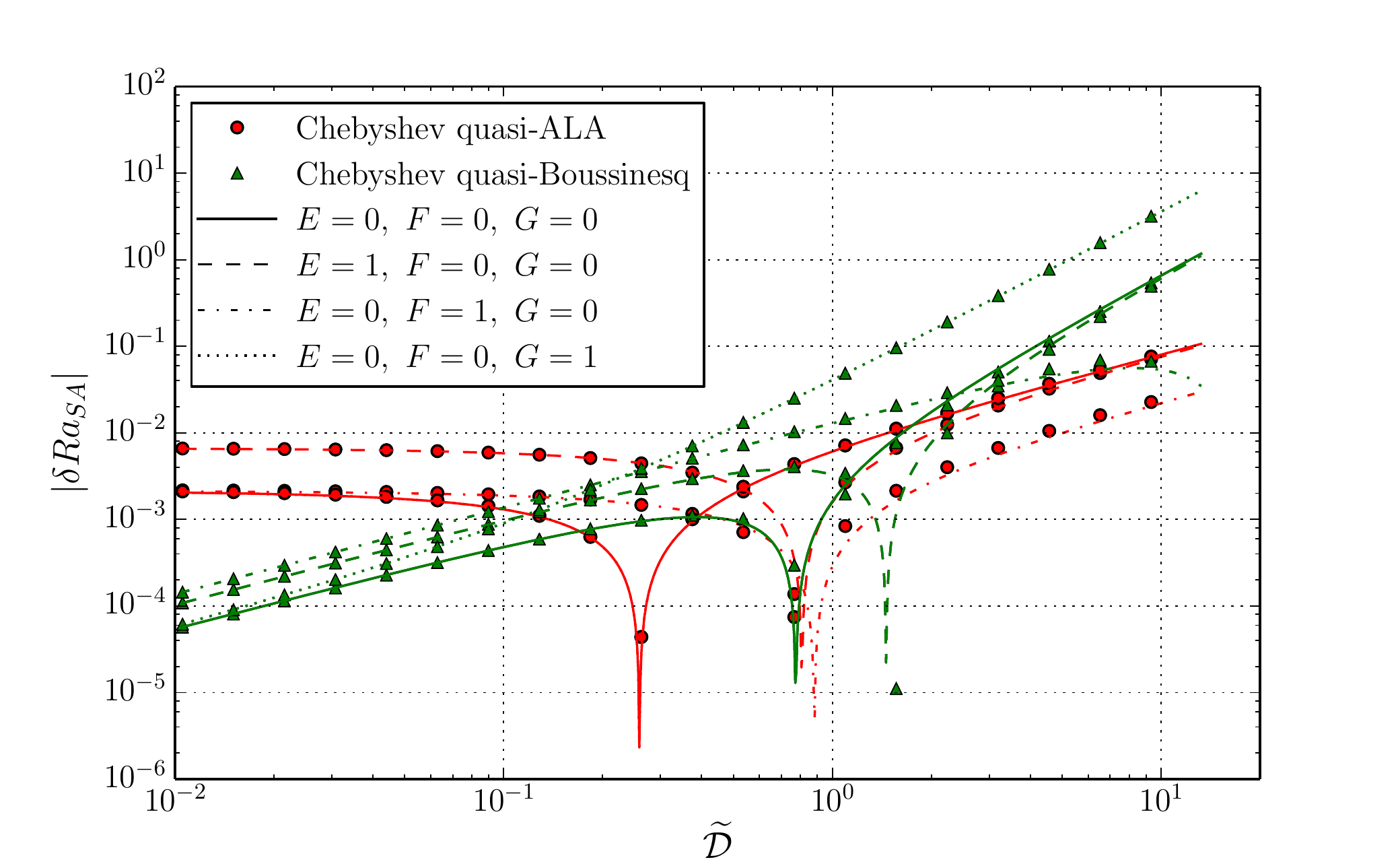}
\caption{Difference between the critical Rayleigh number of the quasi-Boussinesq and exact model, and between the quasi-ALA and exact model, with the generic equation of state (\ref{genericEoSadim}), for a constant temperature gradient $a=0.4$, as a function of the modified dissipation number $\widetilde{\mathcal{D}}=\mathcal{D}/(1-\gamma _0^{-1})$. The heat capacity ratio and $\alpha T$ at $z=0$ are kept constant $\gamma _0 = 1.03$, $\hat{\alpha} = \alpha _0 T_0 =0.03$ and $J=K=L=M=A=B=0$.}
\label{GenDlog}
\end{center}
\end{figure}

\vspace{3 cm}

\begin{figure}
\begin{center}
\includegraphics[width=12 cm, keepaspectratio]{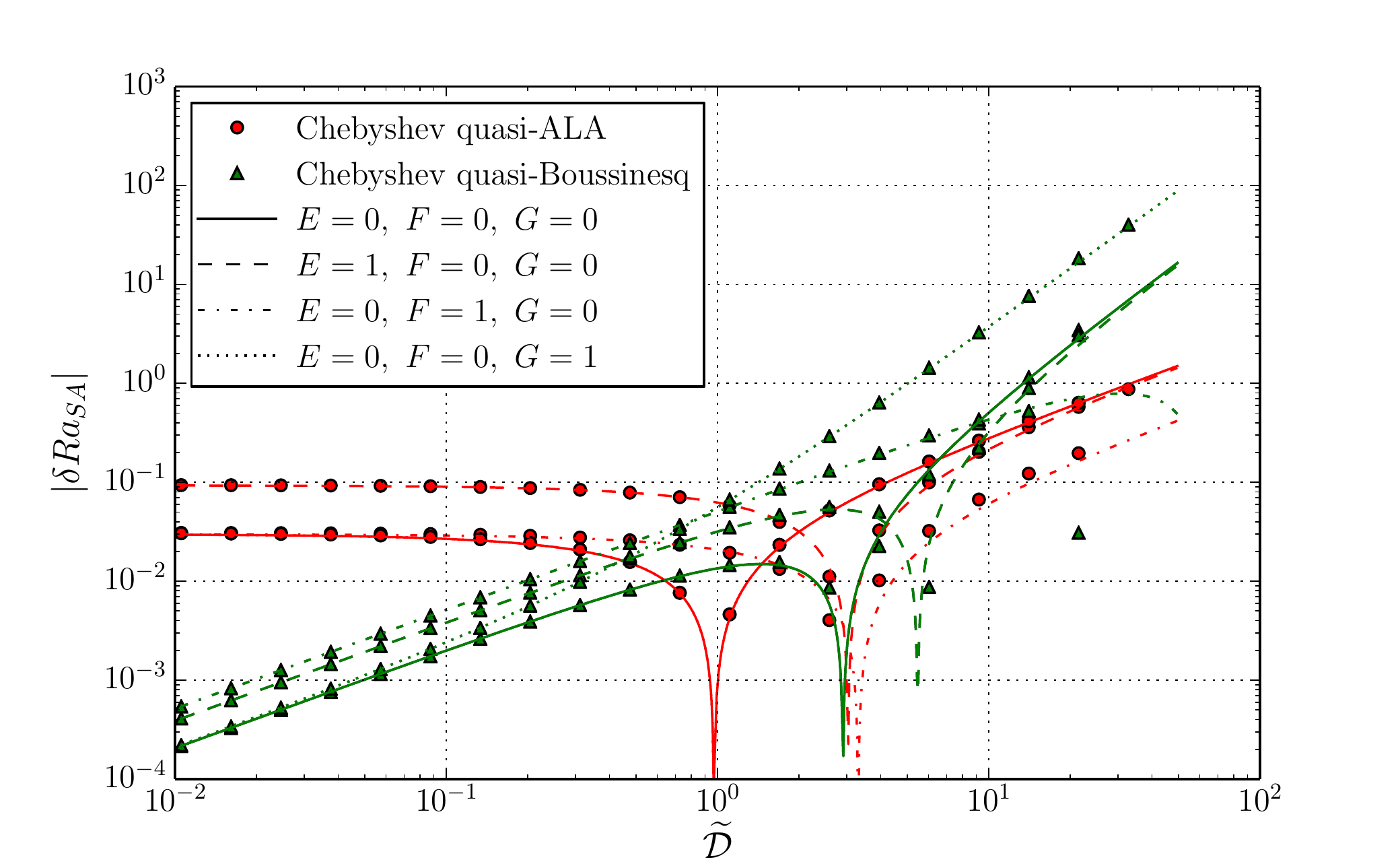}
\caption{Similar to Fig.~\ref{GenDlog}, but with a temperature ratio $r=7$ ($a=1.5$) instead of $r=1.5$ ($a=0.4$).}
\label{GenDlog2}
\end{center}
\end{figure}

Finally we compute the departures between the exact, Boussinesq and ALA approximations, solved numerically (symbols) or
analytically (lines). In Figure \ref{Genalog}, we only vary the second order coefficients keeping $\mathcal{D}\approx 0$. In agreement with the analytical results (\ref{deltaRaBgeneric}), the exact and Boussinesq models coincide. The difference between the ALA and exact solution (\ref{deltaRaALAgeneric}) is a function of $E$ only, {\it i.e.} independent of $F$ and $G$.
To prove the quality of the analytical model, in the Figure \ref{GenDlog2} we maintain a rather large temperature
gradient $a=1.5$
across the layer (i.e. a temperature ratio $r=7$), and we vary the dissipation number and the second order coefficients of the EoS.
In this Figure, like in all the previous figures using the generic EoS, the two-mode approximation gives an accurate fit to the numerical computations. We performed a number of other simulations that we do not show here, varying rather systematically all the parameters. All these simulations confirmed the quality of the two-mode approximation.

\subsection{Universality of the generic EoS}
\label{universality}

The generic EoS (\ref{genericEoSadim}) is meant to represent any equation of state, as an expansion up to degree three in temperature and pressure: the quadratic departure in $a$ and $\mathcal{D}$ from $27 \upi ^4 / 4$ is recovered exactly. We test here its applicability, or universality, when compared to the ideal gas (\ref{idealgasa}) and Murnaghan's (\ref{adim_murnaghan}) equations. For the ideal gas equation, the $c_p$ is constant and we expand $\nu = 1/ \rho$ around $T=1$ and $p=1/ \widetilde{\mathcal{D}}$ (the pressure for the base profile at $z=0$) and identify the coefficients of equation (\ref{genericEoSadim}). We obtain:
\begin{equation}
A=B=E=J=K=0, \hspace*{.5 cm}F=M=-1, \hspace*{.5 cm}G=L=1. \label{coeffidealgas}
\end{equation}
When these values are substituted in the expressions of Table \ref{tableGeneric} we obtain exactly the results obtained for the ideal gas, Table \ref{tableMurnaghan}.

For Murnaghan's EoS (\ref{adim_murnaghan}), the second order expansion leads to identify:
\begin{equation}
E = G=\frac{n+1}{2}, \hspace*{.5 cm}F=-(n+1), \hspace*{.5 cm}L=-K=3J=-3M=\frac{(n+1)(2n+1)}{2}, \label{coeffMurn}
\end{equation}
and the expansion of $c_p$ implies that:
\begin{equation}
A={{\mathcal{D}}\over \hat{\alpha}\widetilde{\mathcal{D}}} (\hat{\alpha}+\hat{\alpha}n+1), \hspace*{.5 cm}B={{\mathcal{D}}\over 2\hat{\alpha}\widetilde{\mathcal{D}}}\left(\hat{\alpha}(2 n+1)+2\right)(n+1).
\end{equation}
Again, when substituted in the expressions of Table \ref{tableGeneric} we obtain exactly the results obtained for the Murnaghan fluid, Table \ref{tableMurnaghan}. Hence all expressions for the superadiabatic Rayleigh number are retrieved: (\ref{deltaRaMurnB}) from (\ref{deltaRaBgeneric}) and (\ref{deltaRaMurnALA}) from (\ref{deltaRaALAgeneric}).

\subsection{On the singularity at $\mathcal{D} = a$}
\label{singularity}

Singularities at $\mathcal{D}=a$ appear in the coefficients obtained for the Murnaghan and generic equations of state (see tables \ref{tableMurnaghan} and \ref{tableGeneric}). They lead to a divergence of the $\sin (2 \upi z)$ coefficient and of the Rayleigh departure $d Ra_{SA}$. The physical interpretation of this singular limit is related to the curvature of the adiabatic profile. The conduction profile has no curvature because we have imposed a uniform thermal conductivity. However, the adiabatic profile has a non-zero curvature in general, the ideal gas case being an exception. So the difference between the conduction and adiabatic profiles has a non-zero curvature. The case $\mathcal{D}=a$ corresponds roughly to a vanishing superadiabatic temperature difference between the bottom and top of the cavity, but the finite curvature implies that half of the layer is stably stratified and the other half is unstably stratified hence subjected to instability. When an instability is obtained for a vanishing superadiabatic temperature difference, the (total) superadiabatic critical Rayleigh number vanishes, hence the departure $d Ra_{SA}$ diverges.

\section{Discussion of the stability analysis}
\label{discussion}

Let us first analyze the departure $d Ra_{SA}^x$ of the critical superadiabatic Rayleigh number from the Boussinesq limit $27 \upi ^4 / 4$. 
The numerical (Chebyshev) results are very well retrieved by the two-modes analytical results, when $\mathcal{D}$ and $a$ are very small and still reasonably well retrieved over the whole range of $a$ and $\mathcal{D}$.   
From the two-modes analysis result (\ref{sol_dRaSA}), we can see that those departures are quadratic in $a$ and $\widetilde{\mathcal{D}}$. A striking point is that $\widetilde{\mathcal{D}}$ may reach much larger values than $a$: although $\mathcal{D}$ is restricted to be less than $a$ so that the configuration is superadiabatic -- hence prone to convective instability -- the ratio of specific heat capacities may be very close to one which makes $\widetilde{\mathcal{D}}$ much larger than $\mathcal{D}$ and potentially much larger than $a$. A consequence is that pressure effects are significantly larger than temperature effects on the departure from the Boussinesq stability threshold. The quadratic non-Boussinesq departure depends on the structure of the equation of state: the expansion of density in terms of pressure and temperature has to be made up to the degree 3 (see equation (\ref{genericEoSadim})). The fact that the higher degrees play no role is confirmed by the excellent comparison between numerical Chebyshev results and the two-modes analytical results. 

The difference of critical threshold between the approximation models and the exact model are of special interest because we use them as a proxy for the validity of the Boussinesq and ALA approximations. 
The corresponding two-modes analytical differences, (\ref{deltaRaGPB}) and (\ref{deltaRaGPALA}) for ideal gases, (\ref{deltaRaMurnB}) and (\ref{deltaRaMurnALA}) for a Murnaghan equation of state, (\ref{deltaRaBgeneric}) and (\ref{deltaRaALAgeneric}) for a generic equation of state, have a simple analytical expression. They are quadratic in $a$ and $\widetilde{\mathcal{D}}$, but the $a^2$ contribution is zero for the difference between the quasi-Boussinesq and exact models, while the $\widetilde{\mathcal{D}}^2$ contribution is absent in the difference between quasi-ALA and exact models. Both differences contain a cross-product contribution $a \widetilde{\mathcal{D}}$. As expected, the quasi-Boussinesq approximation is better than the quasi-ALA when $\widetilde{\mathcal{D}} < \mathcal{O} (a)$, and conversely for large $\widetilde{\mathcal{D}} > \mathcal{O} (a)$. Also, we observe that all analytical threshold differences are proportional to $(\alpha _0 T_0 )^2 = \hat{\alpha}^2$. This seems to indicate that the approximations should always be much better for condensed matter than for gases, but that conclusion must include a discussion on the Gr\"uneisen number. 

We have not mentioned the Gr\"uneisen number so far in this paper. This parameter is a dimensionless number associated to any equation of state, is often denoted $\gamma$, sometimes $\Gamma$, and we choose the latter to avoid any confusion with the ratio of heat capacities $\gamma = c_p / c_v$:
\begin{equation}
\Gamma = \frac{1}{\rho} \left. \frac{\partial p }{\partial e }  \right| _\rho , \label{gruneisen}
\end{equation}
where $e$ is the specific internal energy. Using the definition of $c_v$ and the triple product identity, the Gr\"uneisen parameter can be written $\Gamma = \alpha / ( c_v \left. \partial \rho / \partial p \right| _T ) $. 
Then using Mayer's relation, we obtain:
\begin{equation}
\Gamma = \frac{\gamma -1}{\alpha T} = \frac{ \gamma \mathcal{D} }{\hat{\alpha} \widetilde{\mathcal{D}}} . \label{gruneisen2}
\end{equation}
For condensed matter, theoretical reasons, and more importantly experimental measurements for a range of materials, pressure and temperature, converge towards values of $\Gamma$ comprised between $1$ and $2$ \citep{aio1992} while Mayer's relation leads to $\gamma \simeq 1$. This implies that choosing a small value for the product $\alpha T$ should imply that the ratio of specific heat capacities should be chosen accordingly $\gamma - 1 \simeq \alpha T$, {\it{i.e.}} $\hat{\alpha} \simeq \mathcal{D} / \widetilde{\mathcal{D}}$. A decrease of $\hat{\alpha}$ implies an increase of $\widetilde{\mathcal{D}}$ for a given dissipation number $\mathcal{D}$. So, that small values of $\hat{\alpha}$ will be completely (for the quasi-Boussinesq difference) or partly (for the quasi-ALA difference) compensated by an increase in $\widetilde{\mathcal{D}}$. If the coefficient $F$ is of order unity, and the Gr\"uneisen parameter of order unity $\Gamma \simeq 1$, we may rewrite (\ref{deltaRaALAgeneric}) as 
\begin{equation}
\delta Ra_{SA}^{ALA} \propto \hat{\alpha} a \mathcal{D} . \label{deltaRaALAgenericApp}
\end{equation}
This does not apply to ideal gases. They can have a Gr\"uneisen number smaller than unity, with $\hat{\alpha}=1$ and $\gamma -1 << 1$ (polyatomic gases), so that the quasi-ALA may still be a good approximation for them: an anelastic liquid approximation is indeed an accurate approximation for a gas with molecules constituted by many atoms.

\begin{figure}
\begin{center}
\includegraphics[width=12 cm, keepaspectratio]{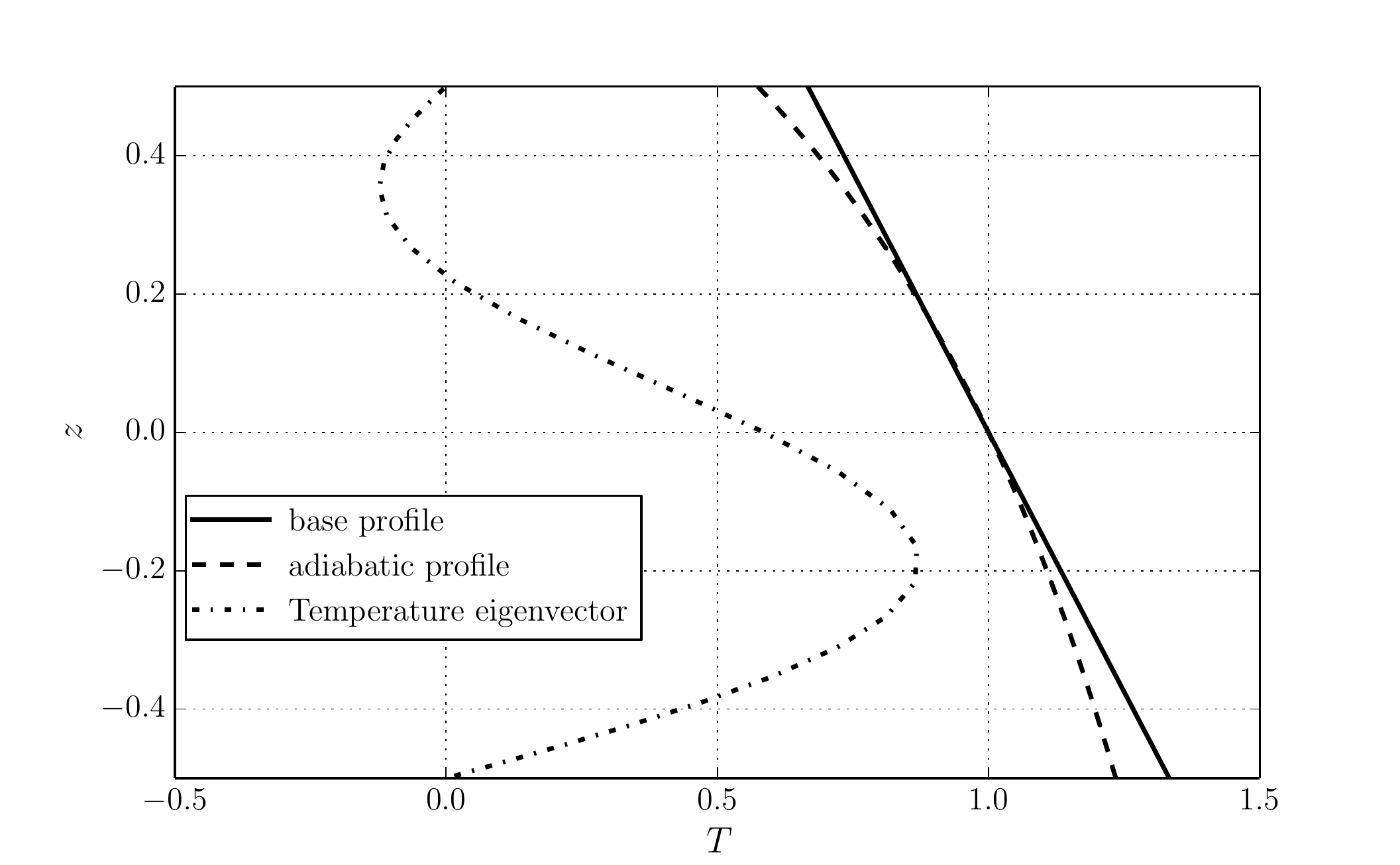}
\caption{Profiles of the base temperature and adiabatic profile, for $\mathcal{D} = 0.6$, $r=2$, $\alpha _0 T_0 = 0.03$, $\gamma _0 = 1.03$, $n=3$ with Murnaghan EoS. The eigenvector for temperature is also plotted.}
\label{figcore}
\end{center}
\end{figure}

Let us consider typical results relevant to the mantle and core of the Earth. For the mantle, we may consider typical values of $\alpha _0 T_0 = 0.03$, $\gamma =1.03$, $\mathcal{D} = 0.5$ and a temperature ratio of 10 between the bottom of the mantle (CMB, core mantle boundary) and the surface of the solid Earth. With a Murnaghan EoS with $n=3$, we obtain the following critical superadiabatic Rayleigh numbers: 
\begin{equation}
{Ra_{SA}^x = 645.04, \hspace*{1 cm} Ra_{SA}^B = 630.84, \hspace*{1 cm} Ra_{SA}^{ALA} = 642.87.} \label{typMantle}
\end{equation}
Although the adiabatic temperature difference is only half the total temperature difference, the quasi-ALA approximation is closer to the exact result than the quasi-Boussinesq approximation by a factor 10. For the Earth's core (assuming that a free-free top and bottom boundary conditions are appropriate), the adiabatic temperature difference is very close to the total temperature difference: we choose $r=2$ and $\mathcal{D}=0.6$. Otherwise, we use the same parameters as for the typical mantle above. The results are the following:
\begin{equation}
{Ra_{SA}^x = 90.885, \hspace*{1 cm} Ra_{SA}^B = 88.726, \hspace*{1 cm} Ra_{SA}^{ALA} = 90.780.} \label{typCore}
\end{equation}
The quasi-ALA is about 20 times closer to the exact result $Ra_{SA}^x$ than the quasi-Boussinesq approximation. Note that the small values of the superadiabatic Rayleigh numbers are due to the non-linearity of the adiabatic gradient, so that the layer is made of a stable region superimposed on top of an unstable one. Figure \ref{figcore} shows the conductive base temperature profile and the adiabatic profile. The base temperature gradient exceeds the adiabatic gradient only in the lower half of the layer: the temperature eigenvector (see Fig.~\ref{figcore}) is thus mainly restricted to this region.   

\section{Conclusions}
\label{conclusion}

We have made a contribution to the study of the convection stability beyond that of Jeffreys: using an approximate analysis based on two functions ($\cos ( \upi z)$ and $ \sin ( 2 \upi z )$), we have shown that the critical superadiabatic Rayleigh number can be expressed as the sum of the Boussinesq value $27 \upi ^4 /4$ and a quadratic function of the dimensionless temperature gradient $a$ and the dissipation number $\mathcal{D}$. That quadratic function is entirely dependent on the choice of an equation of state.
Rayleigh number may be split into an adiabatic part (based on the adiabatic gradient) and a superadiabatic part:
\begin{equation}
Ra = Ra_{ad} + Ra_{SA} , \label{Rasplit}
\end{equation} 
Noting $\Delta T _{ad}$ the adiabatic temperature difference between bottom and top, and $\Delta T$ the imposed temperature difference ($T_{bottom} - T_{top}$), we have $Ra_{ad} = Ra \Delta T _{ad}/\Delta T$ and equation (\ref{Rasplit}) can be written:
\begin{equation}
Ra = \frac{ Ra_{SA} }{1 - \frac{\Delta T _{ad}}{\Delta T} }, \label{Rasplit2}
\end{equation}
In dimensionless terms, $\Delta T = a$ exactly and $\Delta T _{ad} = \mathcal{D} + \mathcal{O}(\mathcal{D}^3 ) $, as argued in section \ref{estimate} for small values of $\mathcal{D}$. 
Hence the critical Rayleigh number can be expressed as:
\begin{equation}
Ra_c \simeq \frac{\frac{27 \upi ^4 }{4} + dRa_{SA} }{1 - \frac{\mathcal{D}}{a} + \frac{1}{24} \frac{\mathrm{d}^3 T_a }{ \mathrm{d} z^3} (z=0)} , \label{Rasplitquad}
\end{equation}
where $dRa_{SA}$ is evaluated correctly up to the second order in the parameters measuring the distance to the Boussinesq limit, $a$ and $\mathcal{D}$. Note that, because of the singularity in $\mathcal{D}=a$, the departure of the superadiabatic Rayleigh number $dRa_{SA}$ is not always a quadratic polynomial in  $a$ and $\mathcal{D}$. However $dRa_{SA}$ is always an homogeneous function of degree $2$ in $a$ and $\mathcal{D}$: when both parameters are multiplied by a real constant $\xi$, $dRa_{SA}$ is multiplied by $\xi ^2$. This is the case when $dRa_{SA}$ is the ratio between a polynomial of degree $4$ in $a$ and $\mathcal{D}$, divided by a polynomial of degree $2$ (see equation (\ref{sol_dRaSA}), along with table \ref{tableMurnaghan} or \ref{tableGeneric}). 

\begin{figure}
\begin{center}
\includegraphics[width=14 cm, keepaspectratio]{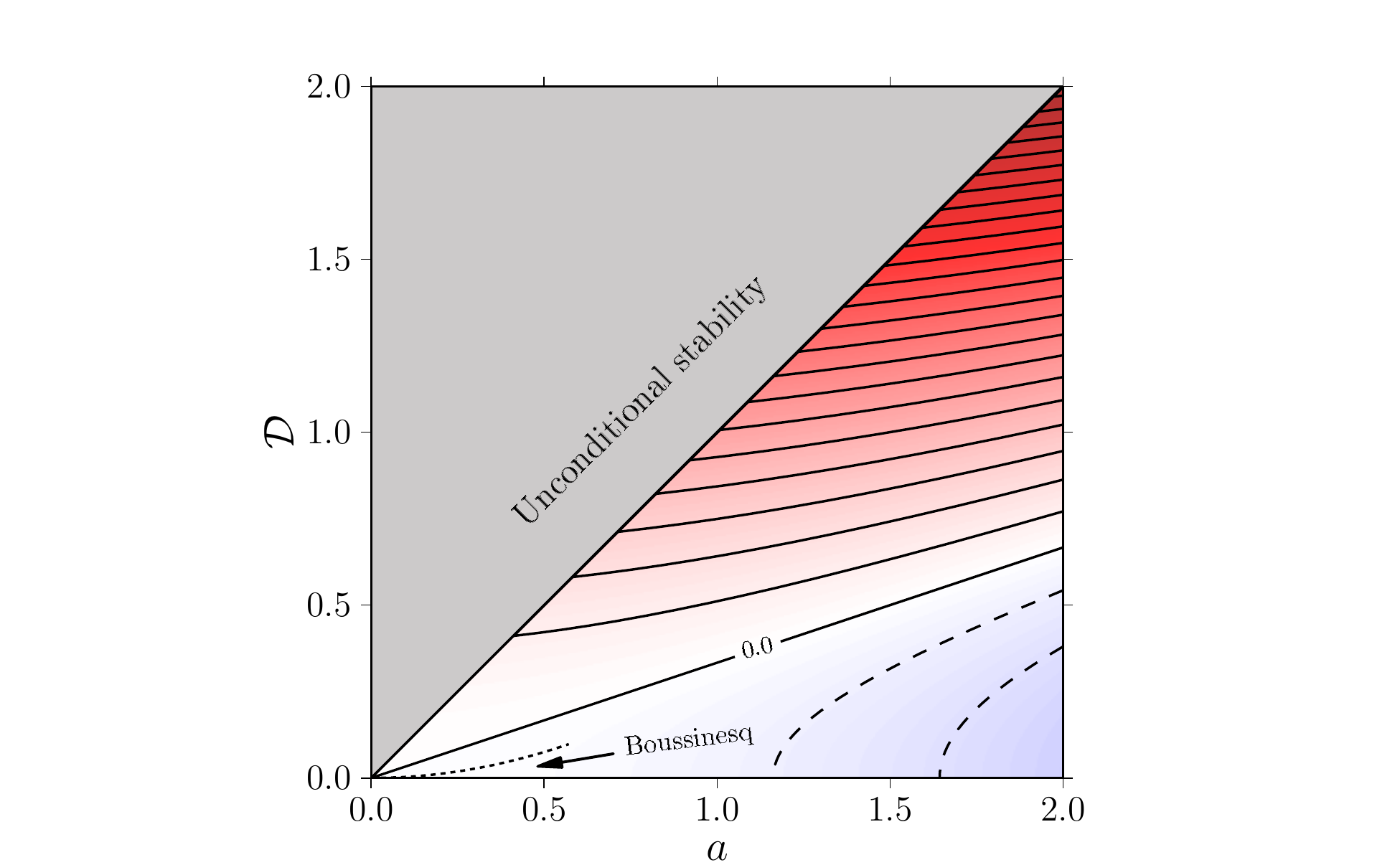}
\caption{Typical representation of the quadratic $dRa_{SA}$ in the plane ($a$, $\mathcal{D}$). The value of $dRa_{SA}$ is related to the background color (red is positive, blue negative). Solid lines are contours of constant positive values, while dashed lines are contours of constant negative values. In the limit of small $a$ and $\mathcal{D}$,  $dRa_{SA}$ vanishes and the superadiabatic critical Rayleigh number is equal to the traditional value $27 \upi ^4 / 4$ \citep{jeffreys}. In addition, the strict Boussinesq limit requires that the superadiabatic Rayleigh number and the Rayleigh number coincide, corresponding to the additional constraint $\mathcal{D} << a$. Above $\mathcal{D} = a$, the configuration is unconditionally stable \citep{schwarzschild}.}
\label{schema}
\end{center}
\end{figure}

A typical representation of the departure of the critical superadiabatic Rayleigh number is shown on Fig.~\ref{schema} which serves here as a reminder for important features of compressible convection. In the plane ($a$, $\mathcal{D}$), the Schawrzschild criterion of stability corresponds to $\mathcal{D}<a$, Jeffreys limit to small $a$ and $\mathcal{D}$, Boussinesq limit to the additional requirement $\mathcal{D} << a$.

We have also studied two variants of the stability problem (quasi-Boussinesq and quasi-ALA models), which are in the spirit of the Boussinesq and of the anelastic liquid models. Approximate analytical expressions have been obtained for the discrepancy of the critical superadiabatic Rayleigh number obtained with these two models (see the general expressions (\ref{deltaRaBgeneric}) and (\ref{deltaRaALAgeneric})). Although our study does not provide any indication concerning the 
quality of the Boussinesq or anelastic liquid approximations for developed convection, 
we have assessed them in terms of critical threshold for convection: the quasi-ALA approximation is in general better than the quasi-Boussinesq approximation, except for very small values of the dissipation parameter $\mathcal{D}$. This tendency is even more pronounced as $\gamma _0$ is closer to unity.  

{Besides providing accurate estimates for the superadiabatic Rayleigh threshold, we have used a two-modes analysis to obtain analytical expressions for the superadiabatic critical Rayleigh number, depending explicitly on the governing physical parameters. We have combined the two-mode analysis to a generic equation of state (\ref{genericEoSadim}) to prove that a cubic expansion of density (or specific volume) in terms of pressure and temperature is needed for the evaluation of the quadratic departure, in terms of $a$ and $\mathcal{D}$, of the superadiabatic critical Rayleigh number beyong the Boussinesq limit. 
The first derivatives of density (or specific volume) with respect to temperature and pressure are prescribed through the two dimensionless parameters $\hat{\alpha}$ and $\widetilde{\mathcal{D}}$. The second derivatives are specified with the introduction of three dimensionless parameters ($E$, $F$ and $G$), while the third order derivatives are defined with four dimensionless parameters ($J$, $K$, $L$ and $M$). 
We also needed to expand the temperature dependence of the heat capacity $c_p$ up to degree two (\ref{cpgeneric1}): dimensionless coefficients $A$ and $B$ specify the linear and quadratic temperature dependence. We have shown that only $M$ (related to $\left. \partial ^3 \rho / \partial p^3 \right| _T$) does not affect the superadiabatic critical Rayleigh number. The superadiabatic Rayleigh number thus depends on eleven parameters: $\hat{\alpha}$, $\mathcal{D}$, $\widetilde{\mathcal{D}}$, $E$, $F$, $G$, $J$, $K$, $L$, $A$ and $B$.   
The differences in critical suparadiabatic Rayleigh numbers induced in the quasi-Boussinesq and quasi-ALA approximations have been found to depend on fewer parameters $\hat{\alpha}$, $\mathcal{D}$, $\widetilde{\mathcal{D}}$, $E$, $F$ and $G$, in effect on the expansion of the specific volume up to degree two in temperature and pressure. }

Our results are in principle valid for any equation of state, hence the introduction of a generic equation of state. We have tested it against the ideal gas equation and Murnaghan's equation of state for condensed matter. Other equations of state might be considered, like those concerning fluids in the vicinity of the critical point, which are the subject of a number of papers devoted to the threshold of convection \citep{adop2010,mk2002}.

{
A feature of our two-mode analysis is that we have treated the equations of thermodynamics as rigorously as those of fluid mechanics. There are thermodynamic relations between $\alpha$, $c_p$, $\gamma$ and other parameters \citep{ar13,ar13c}, so that it is not exact to assume independent expansions of all parameters in terms of temperature and pressure. Our analysis is based on the general form of an equation of state with coherent associated expressions for the heat capacities.  }

\vspace*{1 cm}
Acknowledgements are due to the Labex Lyon Institute of Origins (ANR-10-LABX-0066) and its financial support (ANR-11-IDEX-0007), to the CrysCore project (ANR-08-BLAN-0234-01), to the program PNP of INSU (CNRS), for financial support, and to Fr\'ed\'eric Chambat for fruitfull discussions.

\bibliography{bib}

\end{document}